\documentclass[useAMS,usenatbib]{mn2e}

\usepackage{amssymb,graphicsx} % for resolution symbol and \blacktriangledown
\usepackage{longtable}
\usepackage{booktabs,multirow}
\usepackage{wasysym} % for \astrosun

\newcommand\nodata{ ~$\cdots$~ }

\title[The Aquarius Co-Moving Group is Not a Disrupted Classical Globular Cluster]{The Aquarius Co-Moving Group is Not a Disrupted Classical Globular Cluster$^{1}$}

\author[A. R. Casey et al.]
{\parbox{\textwidth}{A. R. Casey$^{2,3}$\thanks{andrew.casey@anu.edu.au}, S. C. Keller$^{2}$, A. Alves-Brito$^{2,4}$, A. Frebel$^{3}$, G. Da Costa$^{2}$,\\A. Karakas$^{2}$, D. Yong$^{2}$, K. C. Schlaufman$^{3}$, H. R. Jacobson$^{3}$, Q. Yu$^{3}$ and\\C. Fishlock$^{2}$}\vspace{0.4cm}\\
\parbox{\textwidth}{$^{1}$This paper includes data gathered with the 6.5 meter Magellan Telescopes located at Las Campanas Observatory, Chile.\\
$^{2}$Research School of Astronomy and Astrophysics, Australian National University, Canberra, ACT 2611, Australia\\
$^{3}$Massachusetts Institute of Technology, Kavli Institute for Astrophysics and Space Research, 77 Massachusetts Avenue,\\Cambridge, MA 02139, USA\\
$^{4}$Departamento de Astronom\'ia y Astrof\'isica, Ponticia Universidad Cat\'olica de Chile, Av. Vicu\~na Mackenna 4860,\\Macul, 782-0436 Santiago, Chile}}

\begin{document}

\pagerange{\pageref{firstpage}--\pageref{lastpage}} \pubyear{2013}

\maketitle
\label{firstpage}

\begin{abstract}
We present a detailed analysis of high-resolution, high $S/N$ spectra for 5 Aquarius stream stars observed with the MIKE spectrograph on the Magellan Clay telescope. Our sample represents one third of the 15 known members in the stream. We find the stream is not mono-metallic: the metallicity ranges from [Fe/H] = $-$0.63 to $-$1.58. No anti-correlation in Na--O abundances is present, and we find a strong positive Mg--Al relationship, similar to that observed in the thick disk. We find no evidence that the stream is a result of a disrupted classical globular cluster, contrary to a previously published claim. High [(Na, Ni, $\alpha$)/Fe] and low [Ba/Y] abundance ratios in the stream suggests it is not a tidal tail from a disrupted dwarf galaxy, either. The stream is chemically indistinguishable from Milky Way field stars with the exception of one candidate, C222531-145437. From its position, velocity, and detailed chemical abundances, C222531-145437 is likely a star that was tidally disrupted from $\omega$-Centauri. We propose the Aquarius stream is Galactic in origin, and could be the result from a disk-satellite perturbation in the Milky Way thick disk on the order of a few Gyr ago: derived orbits, $UVW$ velocities, and angular momenta of the Aquarius members offer qualitative support for our hypothesis. Assuming C222531-145437 is a tidally disrupted member of $\omega$-Centauri, this system is the most likely disk perturber. In the absence of compelling chemical and/or dynamical evidence that the Aquarius stream is the tidal tail of a disrupted satellite, we advocate the ``Aquarius group'' as a more appropriate description. Like the Canis Major over-density, as well as the Hercules and Monoceros groups, the Aquarius group joins the list of kinematically-identified substructures that are not actually accreted material: they are simply part of the rich complexity of the Milky Way structure.
\end{abstract}


\section{Introduction}
Galaxies are formed hierarchically through chaotic mergers of smaller systems, and the Milky Way is no exception \citep{searle_zinn,bullock;johnston_2005,helmi_2008}. The accumulating stellar debris in our own Galactic halo provides ongoing evidence for such merging events \citep[e.g., ][]{bell;et-al_2008}. As satellites fall towards the Galaxy, tidal forces disrupt the system, hurtling stars in leading and trailing directions. The position and velocities of stars within these ``stellar streams'' are sensitive to the Galactic potential. As such, their phase-space information can collectively constrain the fraction and distribution of accreted matter in the galaxy, the sub-halo mass function, as well as the shape and extent of the Milky Way's dark matter halo. Additionally, individual chemical abundances can trace the chemical evolution of the Galaxy and its satellite systems.

Wide-field deep imaging surveys have proved excellent sources for finding stellar streams \citep[e.g., ][]{belokurov;et-al_2007}. Dozens of streams have been identified through careful photometric selections and matched-filtering techniques, with some to a Galactocentric distance of 100\,kpc \citep[e.g., see][]{drake;et-al_2013}. This suggests that a large fraction of the stellar halo has been built up by accretion. However, as \citet{helmi;white_1999} point out, these detection strategies are most successful for identifying streams that are sufficiently distant from the solar neighborhood. A nearby stream, within $\sim$10\,kpc, will not appear as a photometric over-density because the stars would be sparsely positioned across the sky. Such substructures would only be detectable by their kinematics, or perhaps with precise elemental abundances through a ``chemical tagging'' approach (e.g., see \citealt[][]{freeman;bland-hawthorn_2002}). The confirmation of such substructures would serve to substantially increase the fraction of the known accreted material in the Galaxy.

It is therefore necessary to spectroscopically survey stars in the solar neighbourhood to reveal any nearby substructures. The Radial Velocity Experiment (RAVE) team began such a survey in 2003 and has taken spectra of over 500,000 stars across 17,000\,deg$^{2}$ \citep{steinmetz;et-al_2006}. The primary goal of RAVE is to obtain radial velocities for stars in the solar neighbourhood and beyond. In an attempt to remain kinematically unbiased, RAVE candidates were selected solely by their apparent magnitude ($9 < I < 13$). Almost all have radial velocities published in the RAVE data releases \citep{steinmetz;et-al_2006}, and for a subset of stars with a sufficient signal-to-noise ($S/N$) ratio, stellar parameters have been derived by a $\chi^2$-minimisation technique \citep{zwitter;et-al_2008, siebert;et-al_2011}. 

Using these data, \citet{williams;et-al_2011} identified a co-moving group of nearby (${0.5\,\mbox{kpc} \lesssim D \lesssim 10\,\mbox{kpc}}$) stars near ${(l, b) = (60^\circ, -55^\circ)}$, in the vicinity of the Aquarius constellation. Thus, the co-moving group was named the Aquarius stream. The stream is most apparent when examining heliocentric velocities against Galactic latitude for stars within $-70^\circ < b < -50^\circ$. \citet{williams;et-al_2011} employed a selection criteria of $-250\,\mbox{km s}^{-1} < V_{\rm hel} < -150\,\mbox{km s}^{-1}$, $30^\circ < l < 75^\circ$ and $J > 10.3$ to maximize the contrast between the stream and stellar background, identifying 15 stars in the process. The average heliocentric velocity of these members was found to be $V_{\rm hel} = -199\,\mbox{km s}^{-1}$, with a dispersion of 27\,km s$^{-1}$. The radial velocity uncertainties provided by the RAVE catalog are described to be $\sim$2\,km s$^{-1}$, so the stream's wide velocity distribution appears to be real.

Through a statistical comparison with predictions of stellar positions and kinematics from the Galaxia \citep{sharma;et-al_2011} and Besan\c{c}on \citep{robin;et-al_2003} models of the Milky Way, \citet{williams;et-al_2011} found the stream \ to be statistically significant ($>$4$\sigma$). The choice of model, cell dimension, or extinction rate made no substantial difference to the detection significance. The authors concluded the over-density was genuine, and inferred that the co-moving group is a stellar stream. Based on the phase space information available, \citet{williams;et-al_2011} concluded that the newly discovered stream could not be positively associated with the Sagittarius or Monoceros stream, the Hercules-Aquila cloud, or either the Canis Major or Virgo over-densities. 

RAVE data suggest the Aquarius stream has a metallicity of [Fe/H] $= -1.0 \pm 0.4$\,dex\footnote{\citet{williams;et-al_2011} formally quote [M/H], but for the sake of a consistent discussion we assume [M/H] $\equiv$ [Fe/H] throughout this study.}, whereas field stars at the same distance show ${\mbox{[Fe/H]} = -1.1 \pm 0.6}$\,dex after the same selection cuts had been employed. Of the 15 Aquarius stream stars in the \citet{williams;et-al_2011} discovery sample, the metallicity range determined from medium-resolution spectroscopy is wide: from [Fe/H] = --2.02 to --0.33. High-resolution spectra with high $S/N$ are necessary to accurately characterise the stream's metallicity distribution function (MDF).

To this end, \citet{wylie-de-boer;et-al_2012} obtained high-resolution ($\mathcal{R} = 25,000$) spectra with a modest signal-to-noise ($S/N$) ratio of $\sim$30 for six Aquarius stream stars using the echelle spectrograph on the Australian National University's 2.3m telescope. Their data indicate a surprisingly narrow spread in metallicity compared to previous work: ${\mbox{[Fe/H]} = -1.09 \pm 0.10}$\,dex, with a range extending only from $-1.25$ to {--0.98\,dex}. Samples with such small dispersions in metallicity are typically observed in mono-metallic environments (e.g., globular or open clusters). 

In addition to ascertaining stellar parameters, \citet{wylie-de-boer;et-al_2012} measured elemental abundances for the Aquarius stream stars -- the only study to date to do so. The authors primarily focussed on Na, O, Mg, Al, and Ni. These elements have been extensively studied in globular cluster stars, where unique abundance patterns are observed. Specifically, an anti-correlation between sodium and oxygen content appears ubiquitous to stars in globular clusters \citep{carretta;et-al_2009a}. \citet{wylie-de-boer;et-al_2012} identified two stream stars with slightly higher [Na/Fe] abundance ratios than halo stars of the same metallicity. No strong oxygen depletion was evident in the data, and no overall {Na-O} anti-correlation was present. \citet{wylie-de-boer;et-al_2012} also found [Ni/Fe] abundance ratios similar to thick disk/globular cluster stars, markedly higher than those reported for the Fornax dwarf spheroidal (dSph) galaxy, which has a comparable mean metallicity to the Aquarius stream.

Combined with the low level of [Fe/H] scatter present in their sample, these chemical abundances led \citet{wylie-de-boer;et-al_2012} to conclude that the Aquarius stream is the result of a tidally disrupted globular cluster. We note, though, that \citet{williams;et-al_2011} previously excluded this scenario after modelling an Aquarius-like progenitor falling towards the Milky Way. The predicted positions and velocities from their simulations could not be reconciled with any known globular cluster, except for $\omega$-Centauri, although no explicit link was argued. Alternatively, any parent cluster may have been totally disrupted, leaving no identifiable remnant for discovery. 

We seek to investigate the nature of the Aquarius stream, specifically the  globular cluster origin claimed by \citet{wylie-de-boer;et-al_2012}. Details of the observations and data reduction  are outlined in the following section. The bulk of our analysis is presented in Section \ref{sec:analysis} and our chemical abundance analysis is chronicled separately in Section \ref{sec:chemical-abundances}. A detailed discussion of our results is made in Section \ref{sec:discussion}, and we conclude in Section \ref{sec:conclusions} with a summary of our findings and critical interpretations.

\begin{table*}
\caption{Observations and radial velocities\label{tab:observations}}
%\resizebox{\textwidth}{!}{%
\begin{tabular}{lccccccccc}
\hline
\hline
Designation & $\alpha$ & $\delta$ & $V$   & $B-V$ & UT   & UT   & $t_{\rm exp}$ & $S/N$$^{a}$ & $V_{\rm hel}$ \\
            & (J2000)  & (J2000)  & (mag) &       & Date & Time & (secs)        & (px$^{-1}$) & (km s$^{-1}$) \\
\hline
\\
\multicolumn{10}{c}{\textbf{Standard Stars}} \\
\hline
HD 41667 & 06:05:03.7 & $-$32:59:36.8	& 8.52 & 0.76 & 2011-03-13	& 23:40 & 90 	& 340 & 297.1 		\\
HD 44007 & 06:18:48.6 & $-$14:50:44.2 & 8.06 & 0.79 & 2011-03-13	& 23:52 & 120	& 280 & 161.8 		\\
HD 76932 & 08:58:44.2 & $-$16:07:54.2	& 5.86 & 0.53 & 2011-03-14	& 00:16 & 25	& 330 & 117.8		\\
HD 136316 & 15:22:17.2 & $-$53:14:13.9	& 8.77 & 1.12 & 2011-03-14	& 09:37 & 120 	& 400 & $-$38.8 	\\
HD 141531& 15:49:16.9 & $+$09:36:42.5	& 9.28 & 1.03 & 2011-03-14	& 09:52 & 120	& 350 & 2.8		\\
HD 142948	 & 16:00:01.6 & $-$53:51:04.1	& 9.27 & 0.60 & 2011-03-14	& 09:45 & 90 	& 320 & 29.9 		\\
\hline
\\
\multicolumn{10}{c}{\textbf{Program Stars}} \\
\hline
C222531-145437 & 22:25:31.7 & $-$14:54:39.6	& 12.49 & 1.20 & 2011-07-30	& 06:52 & 650 & 135	& $-$156.4 \\
C230626-085103 & 23:06:26.6 & $-$08:51:04.8	& 12.60 & 1.28 & 2011-07-30	& 08:15 & 650 & 100	& $-$221.1 \\
J221821-183424 & 22:18:21.2 & $-$18:34:28.3	& 12.12 & 0.96 & 2011-07-30	& 05:58 & 650 & 115	& $-$159.5 \\
J223504-152834 & 22:35:04.5 & $-$15:28:34.9	& 12.26 & 1.02 & 2011-07-30 & 07:34 & 650 & 130	& $-$169.7 \\
J223811-104126 & 22:38:11.6 & $-$10:41:29.4	& 11.93 & 0.79 & 2011-07-30	& 08:57 & 650 & 115	& $-$235.7  \\
\hline
\multicolumn{10}{l}{$^{a}S/N$ measured per pixel ($\sim$0.09\,{\AA} px$^{-1}$) at 600\,nm for each target.} \\
\end{tabular}
\end{table*}

\section{Observations \& Data Analysis}

The most complete sample of Aquarius stream stars is presented in the discovery paper of \citet{williams;et-al_2011}. We have obtained high-resolution, high $S/N$ spectra for 5 Aquarius stream candidates using the Magellan Inamori Kyocera Echelle (MIKE) spectrograph \citep{bernstein;et-al_2003} on the Magellan Clay telescope. Although these observations were carried out independently of the \citet{wylie-de-boer;et-al_2012} study, by chance there are four stars common to both samples. The additional star in this sample, C2306265-085103, was observed by the RAVE survey but had a $S/N$ ratio too low for stellar parameters to be accurately determined. All program stars were observed in July 2011 in $\sim$1\arcsec{} seeing at low airmass (Table \ref{tab:observations}), and six standard stars were observed in March 2011. All observations were taken using a 1.0\arcsec{} slit without  spectral or spatial binning, providing a spectral resolution in excess of $\mathcal{R} = 28,000$ in the blue arm and $\mathcal{R} = 25,000$ in the red arm. The exposure time for our program stars was 650\,seconds per star in order to ensure a $S/N$ ratio in excess of 100\,pixel$^{-1}$ at 600\,nm. 

Calibration frames were taken at the start of each night, including 20 flat-field frames (10 quartz lamp, 10 diffuse flats) and 10 Th-Ar arc lamp exposures. The data were reduced using the CarPy pipeline\footnote{http://code.obs.carnegiescience.edu/mike}. For comparison purposes one of the standard stars, HD 41667, was also reduced using standard extraction and calibration methods in \textsc{iraf}. The resultant spectra from both approaches were compared for residual fringing, $S/N$, and wavelength calibration. No noteworthy differences were present, and the CarPy pipeline was utilized for the remainder of the data reduction. Each reduced echelle order was carefully normalized using a cubic spline with defined knot spacing. Normalized orders were stitched together to provide a single one-dimensional spectrum from 333 to 916\,nm. A portion of normalised spectra for the program stars is shown in Figure \ref{fig:spectra}.
 
The white dwarf HR 6141 was observed in March 2011 as a telluric standard. The $S/N$ ratio for HR 6141 exceeds that of any of our standard or program stars. Although the atmospheric conditions at Las Campanas Observatory are certain to change throughout the night and between observing runs, we are primarily using this spectrum to identify stellar absorption lines that are potentially affected by telluric absorption. 

\section{Analysis}
\label{sec:analysis}

\subsection{Radial Velocities}
\label{sec:radial-velocities}

The radial velocity for each star was determined in a two step process. An initial estimate of the radial velocity was ascertained by cross-correlation with a synthetic spectrum of a giant star with $T_{\rm eff} = 4500$\,K, $\log{g} = 1.5$, and ${\mbox{[Fe/H]} = -1.0}$ across the wavelength range 845 to 870\,nm. The observed spectrum was shifted to the pseudo-rest frame using this initial velocity estimate. Equivalent widths (EWs) were measured for $\sim$160 atomic transitions by integrating fitted Gaussian profiles (see Section \ref{sec:line-measurements}). In each case a residual line velocity was calculated from the expected rest wavelength and the measured wavelength. The mean residual velocity offset correction is small in all cases ($<$1\,km s$^{-1}$), and this residual correction is applied to the initial velocity measurement from cross-correlation. The final heliocentric velocities are listed in Table \ref{tab:observations}, where the typical uncertainty is $\pm0.1$\,km s$^{-1}$. These velocities agree quite well with those compiled by \citet{williams;et-al_2011} as part of the RAVE survey: the mean offset of 2.5 km s$^{-1}$ with a standard deviation of 2.7 km s$^{-1}$.

\subsection{Line Measurements}
\label{sec:line-measurements}

\begin{table*}
\caption{List of Atomic Transitions and Equivalent Width Measurements for Program and Standard Stars \label{tab:equivalent-widths}}
\resizebox{\textwidth}{!}{%
\begin{tabular}{lccccccccccccccc}
\hline
\hline
& & & & & \multicolumn{5}{c}{Equivalent Width} \\
\cline{5-10} \\
Wavelength & Species & $\chi$ & $\log{gf}$ & C222531-145437 & C2306265-085103 & J221821-183424 & J223504-152834 & J223811-104126 & (cont..) \\
({\AA}) & & (eV) & & (m{\AA}) & (m{\AA}) & (m{\AA}) & (m{\AA}) & (m{\AA}) \\
\hline
 6300.30 &  O I &      0.00 &    --9.72 &      45.4 &      66.9 &      38.5 &      32.8 &      17.9 \\
 6363.78 &  O I &      0.02 &   --10.19 &      19.5 &      28.3 &      13.0 &      18.8 &   \nodata \\
 5688.19 & Na I &      2.11 &    --0.42 &   \nodata &   \nodata &      49.0 &     131.5 &      38.4 \\
 6154.23 & Na I &      2.10 &    --1.53 &      24.1 &      38.9 &   \nodata &      48.5 &   \nodata \\
 6160.75 & Na I &      2.10 &    --1.23 &      37.7 &      58.5 &   \nodata &      65.7 &   \nodata \\
 6318.72 & Mg I &      5.11 &    --1.97 &   \nodata &      47.9 &      14.7 &      62.8 &       8.9 \\
 6319.24 & Mg I &      5.11 &    --2.22 &      30.8 &   \nodata &       5.5 &   \nodata &   \nodata \\
 6965.41 & Mg I &      5.75 &    --1.51 &   \nodata &   \nodata &   \nodata &      59.2 &   \nodata \\
\hline
\end{tabular}}
\parbox{\textwidth}{Table \ref{tab:equivalent-widths} is published in its entirety in the electronic edition. A portion is shown here for guidance regarding its form and content.}
\end{table*}

For the measurement of atomic absorption lines, we employed the line list of \citet{yong;et-al_2005} with additional transitions of Cr, Sc, Zn, and Sr from \citet{roederer;et-al_2010}. The list has been augmented with molecular CH data from \citet{plez;et-al_2008}. For molecular features (e.g., CH), or lines with hyperfine and/or isotopic splitting (Sc, V, Mn, Co, Cu, Ba, La, Eu), we determined the abundance using spectral synthesis with the relevant data included. Isotopic and hyperfine splitting data was taken from \citet{sneden} for Sc, V, Mn, Co and Cu, \citet{biemont} for Ba, \citet{Lawler;et-al_2001} for La, and \citet{lawler_eu} for Eu. For all other transitions, abundances were obtained using the measured EWs.

\begin{figure}
	\includegraphics[width=\columnwidth]{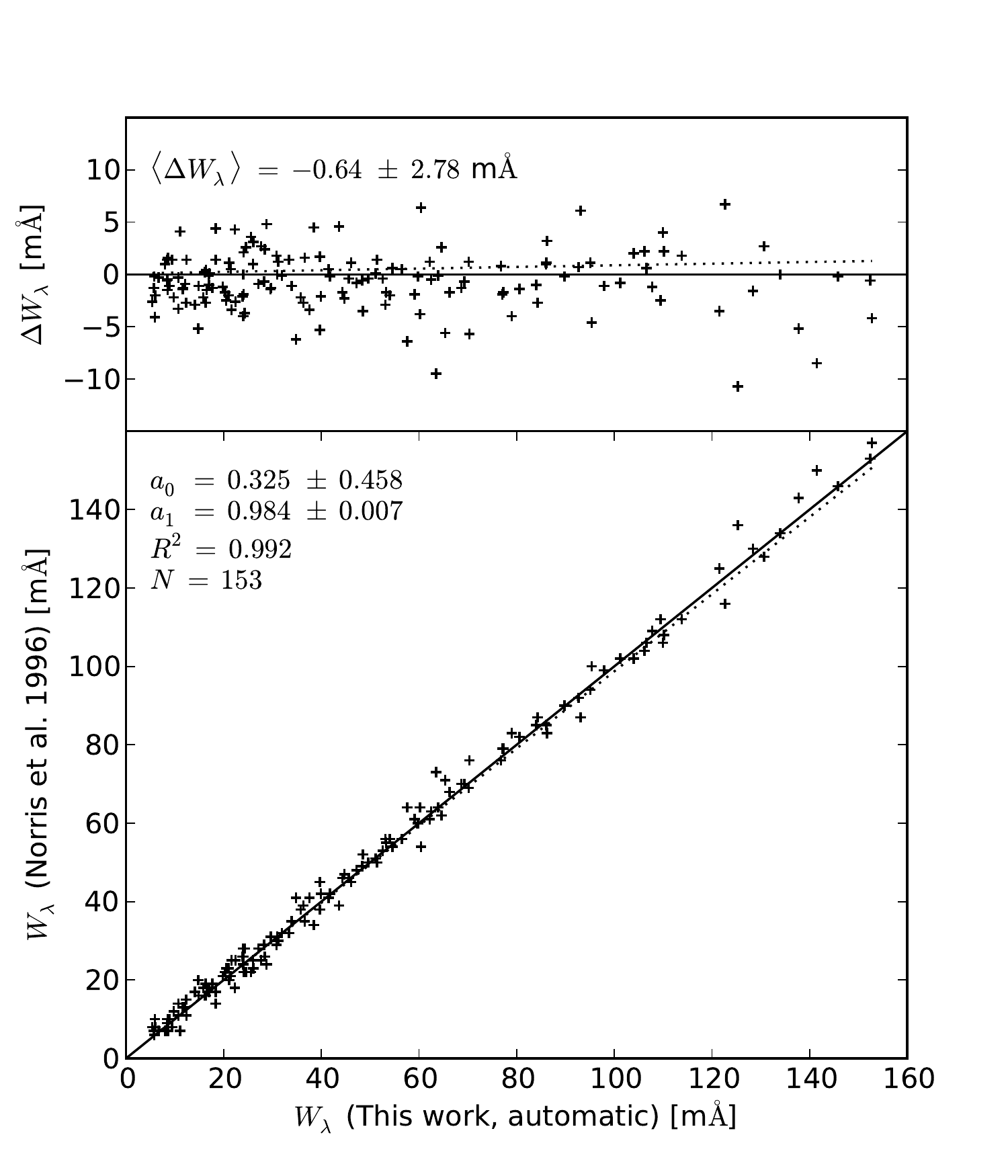}
	\label{fig:ew-compare}
	\caption{Comparison showing equivalent widths measured for HD\,140283 using our automatic routine (see \S\ref{sec:line-measurements}), and manual measurements by \citet{norris;et-al_1996}. No systematic trend is present, and the mean difference between these studies is ${\langle\Delta{}W_\lambda\rangle = -0.64 \pm 2.78}$\,m\AA{}. The offset ($a_0$) and the slope ($a_1$) of the fit are shown.}
\end{figure}

\begin{figure*}
	\includegraphics[width=\textwidth]{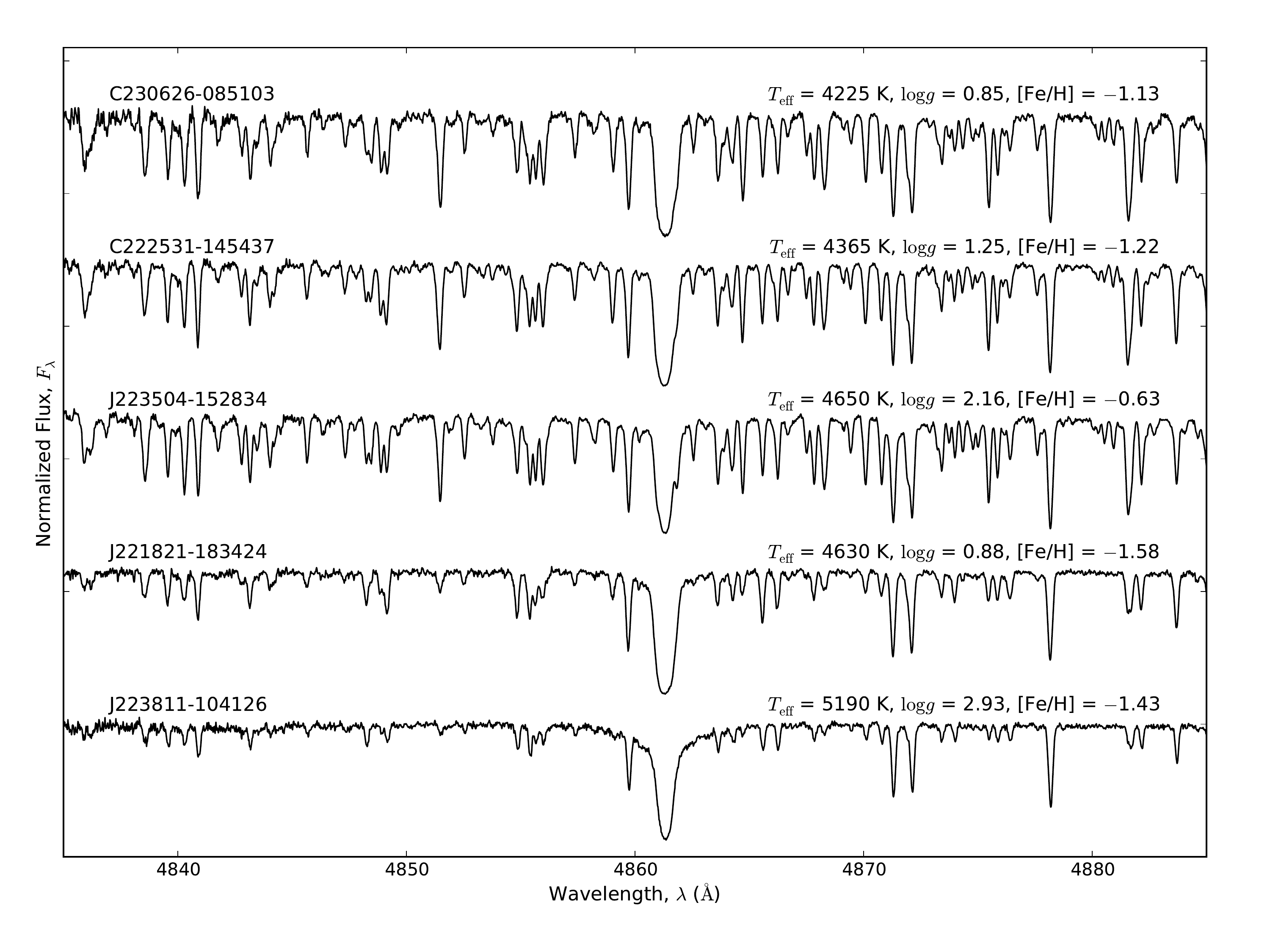}
	\caption{Normalized rest-frame spectra surrounding the H-$\beta$ absorption line for all Aquarius stream candidates with offset fluxes. The effective temperature, surface gravity and metallicity is shown for all stars.}
	\label{fig:spectra}
\end{figure*}

The EWs for all absorption lines were measured automatically using software written during this study. The local continuum surrounding every atomic transition is determined, and a Gaussian profile is iteratively fit to the absorption feature of interest. Our algorithm accounts for crowded or blended regions by weighting pixels as a function of difference to the rest wavelength. These algorithms and software will be fully outlined in a future contribution (Casey et al., in preparation). For this study we have verified our approach by comparing EWs of 156 lines measured by hand and tabulated in \citet{norris;et-al_1996}. We only included measurements in the \citet{norris;et-al_1996} study that were not marked by \citet{norris;et-al_1996} to have questionable line quality parameters. Excellent agreement is found between the two studies, which is illustrated in Figure \ref{fig:ew-compare}. The mean difference is a negligible ${-0.64 \pm 2.78}$\,m\AA{}, and no systematic trend is present. The scatter can be attributed to the lack of significant digits in the \citet{norris;et-al_1996} study, as well as the $S/N$ of the data. Other studies \citep[e.g.][]{frebel;et-al_2013} using the same algorithm used here find better agreement for spectra with higher $S/N$ ratios: $0.20 \pm 0.16$\,m{\AA} when we compare our results with manual measurements by \citet{aoki;et-al_2007}, and a difference of $0.25 \pm 0.28$\,m{\AA} is found between manual measurements by \citet{cayrel;et-al_2004} and our automatic results. Although we are extremely confident in our EW measurements, \textit{every} absorption profile was repeatedly examined by eye for quality, and spurious measurements were removed.

We list the atomic data and measured EWs in Table \ref{tab:equivalent-widths}. Transitions near the flat portion of the curve-of-growth have been excluded by removing measurements with reduced equivalent widths (REW), ${\log_{10}{(EW/\lambda)} > -4.5}$. A minimum detectable EW was calculated as a function of wavelength, $S/N$ and spectral resolution following \citet{norris;et-al_2001},

\begin{equation}
EW_{\rm min} \approx \left(\frac{S}{N}\right)^{-1}\sqrt{1.5\times{}FWHM\times\delta\lambda}
\label{eq:min_ew}
\end{equation}

\noindent{}where $FWHM$ is the minimum detectable line profile limited by instrumental broadening and $\delta\lambda$ is the pixel size. Only lines that exceeded a 3$\sigma$ detection significance were included for this analysis.

\subsection{Model Atmospheres}
We have employed the ATLAS9 plane-parallel stellar atmospheres of \citet{castelli;kurucz_2003}. These one-dimensional models ignore any center-to-limb spatial variations, assume hydrostatic equilibrium and no convective overshoot from the photosphere. The stellar parameter spacing between models is 250\,K in temperature, 0.5\,dex in surface gravity, 0.5\,dex in [M/H] and 0.4\,dex in [$\alpha$/Fe]. We interpolated the temperature, gas and radiative pressure, electron density and opacities between atmosphere models using the Quickhull algorithm \citep{barber;et-al_1996}. Quickhull is reliant on Delaunay tessellation, which suffers from extremely skewed cells when the grid points vary in size by orders of magnitude -- as $T_{\rm eff}$ values do compared to $\log{g}$ or [(M,$\alpha$)/H]. If unaccounted for, performing interpolation using such asymmetric cells can result in significant errors in atmospheric properties across all photospheric depths. We scaled each stellar parameter between zero and unity before interpolation to minimise these interpolation errors. 

\begin{table*}
\caption{Stellar parameters for standard and program stars\label{tab:stellar-parameters}}
\resizebox{\textwidth}{!}{%
\begin{tabular}{lccccccccccccccc}
\hline
\hline
& \multicolumn{4}{c}{{\bf This Study}} && \multicolumn{5}{c}{{\bf Literature}} \\
\cline{2-5} \cline{7-11}
Designation & $T_{\rm eff}$ & $\log{g}$ & $\xi_t$ & [Fe/H] & & $T_{\rm eff}$ & $\log{g}$ & $\xi_t$ & [Fe/H] & Reference \\
& (K) & (dex) & (km\,s$^{-1}$) & (dex) && (K) & (dex) & (km s$^{-1}$) & (dex) \\
\hline
\\
\multicolumn{11}{c}{\textbf{Standard Stars}} \\
\hline
HD\,41667		& 4660	& 1.71	& 1.84 	& $-$1.20&
				& 4605	& 1.88	& 1.44	& $-$1.16
				& \citet{gratton;et-al_2000} \\
HD\,44007		& 4835	& 1.78	& 1.95	& $-$1.77&
				& 4850	& 2.00 	& 2.20	& $-$1.71
				& \citet{fulbright_2000} \\
HD\,76932		& 5800 & 3.88 & 1.65 & $-$1.05&
				& 5849 & 4.11 & \nodata{} & $-$0.88 
				& \citet{nissen;et-al_2000} \\
HD\,136316		& 4355 & 0.58 & 2.06 & $-$1.93&
				& 4414 & 0.94 & 1.70 & $-$1.90 
				& \citet{gratton_sneden_1991} \\
HD\,141531		& 4345 & 0.63 & 2.07 & $-$1.69 &
				& 4280 & 0.70 & 1.60 & $-$1.68
				& \citet{shetrone_1996} \\
HD\,142948		& 5025	& 2.25	& 2.05	& $-$0.74&
				& 4713 	& 2.17 	& 1.38	& $-$0.77
				& \citet{gratton;et-al_2000} \\
\hline \\

\multicolumn{11}{c}{\textbf{Program Stars}} \\
\hline 

C222531-145437	& 4365				& 1.25				& 1.94				& $-$1.22	&
				& 4235\,$\pm$\,118 	& 1.45\,$\pm$\,0.21	& 1.96\,$\pm$\,0.11 	& $-$1.20\,$\pm$\,0.14 
				& \citet{wylie-de-boer;et-al_2012} \\
C230626-085103	& 4225	& 0.85	& 1.92 	& $-$1.13&
				& \dots	& \dots	& \dots	& \dots
				& \dots \\	
J221821-183424	& 4630				& 0.88				& 2.16				& $-$1.58&
				& 4395\,$\pm$\,205 	& 1.45\,$\pm$\,0.35 	& 1.96\,$\pm$\,0.18 	& $-$1.15\,$\pm$\,0.21
				& \citet{wylie-de-boer;et-al_2012} \\
J223504-152834	& 4650				& 2.16 				& 1.55				& $-$0.63&
				& 4597\,$\pm$\,158 	& 2.40\,$\pm$\,0.14 	& 1.47\,$\pm$\,0.07 	& $-$0.98\,$\pm$\,0.17
				& \citet{wylie-de-boer;et-al_2012} \\
J223811-104126	& 5190				& 2.93				& 1.62				& $-$1.43&
				& 5646\,$\pm$\,147 	& 4.60\,$\pm$\,0.15 	& 1.09\,$\pm$\,0.11 	& $-$1.20\,$\pm$\,0.20
				& \citet{wylie-de-boer;et-al_2012} \\
\hline
\end{tabular}}
\end{table*}

\subsection{Stellar Parameters}
\label{sec:stellar-parameter-derivation}
The May 2011 version of the \textsc{moog}  \citep{sneden;et-al_1973} spectral synthesis code has been used to derive individual line abundances and stellar parameters. This version employs Rayleigh scattering \citep{sobeck;et-al_2011} instead of treating scattering as true absorption, which is particularly important for transitions blue-ward of 450\,nm. This is noteworthy, but is less relevant for these analyses as most of the atomic transitions utilized here are red-ward of 450\,nm.

\subsubsection{Effective Temperature}
\label{sec:effective-teffs}

The effective temperature, $T_{\rm eff}$, for each star was found by demanding a zero-trend in excitation potential and line abundance for measurable Fe\,\textsc{i} transitions. The data were fitted with a linear slope, and gradients less than $|10^{-3}|$\,dex eV$^{-1}$ were considered to be converged. For comparison, photometric temperatures were calculated after our spectroscopic temperatures had been derived, and these are discussed in Section \ref{sec:photometric-temperatures}.

\subsubsection{Microturbulence and Surface Gravity}

The microturbulence for each star was found by forcing a zero-trend in the REW and abundance for Fe\,\textsc{i} lines. Similar to the effective temperature, linear slopes in REW and abundance of less than $|10^{-3}|$\,dex were considered converged. The surface gravity for all stars was found by forcing the mean Fe\,\textsc{i} and Fe\,\textsc{ii} abundances to be equal. A tolerance of $|\langle$Fe\,\textsc{i}$\rangle - \langle$Fe\,\textsc{ii}$\rangle| \leq 0.05$ was deemed acceptable. The process is iterative: a zero trend with the excitation potential, REW and abundances must be maintained. A solution was only adopted when the all criteria were simultaneously satisfied. \\ % for nice formatting

\subsubsection{Metallicity}
The model atmosphere metallicity was exactly matched to that of our mean Fe\,\textsc{i} abundance. Individual Fe line abundances that were unusually deviant (e.g., $>$3$\sigma$) from the mean abundance were removed. The largest number of outlier measurements removed for any observation was nine for C222531-145437. These were transitions near the flat part of the curve-of-growth with REWs $\sim -4.5$, leaving 60 Fe\,\textsc{i} and 10 Fe\,\textsc{ii} lines for the analysis of C222531-145437. Usually only one outlier measurement was removed for the other candidates. The minimum number of Fe transitions employed for stellar parameter determination was 42 lines (33 Fe\,\textsc{i} and 9 Fe\,\textsc{ii}), which occurred for our hottest star, J223811-104126.

\subsubsection{Photometric Effective Temperatures}
\label{sec:photometric-temperatures}
 
As a consistency check for our spectroscopic temperatures, we have estimated effective temperatures using the colour-$T_{\rm eff}$ empirical relationship for giant stars from \citet{ramirez;melendez_2005}. The $V-K$ colour has been employed as its calibration has the lowest residual fit. This relationship has a slight dependence on metallicity, and as such we have adopted the spectroscopic [Fe/H] values in Table \ref{tab:stellar-parameters} for these calculations. Optical $V$-band magnitudes from the APASS catalogue \citep{henden;et-al_2012} have been employed, and $K$-band magnitudes have been sourced from the 2MASS catalog \citep{skrutskie;et-al_2006}. The reddening maps of \citet{schlegel;et-al_1998} estimate that the extinction for our stars varies between ${E(B-V) = 0.03}$ to {0.07\,mags}, and these values have been used to de-redden our $V-K$ colour.

\setlength{\tabcolsep}{4pt}
\begin{table}
\centering
\caption{Reddening \& photometric temperatures for program stars\label{tab:photometric-temperatures}}
%\resizebox{\textwidth}{!}{%
\begin{tabular}{lccccccc}
\hline
\hline
Designation & $E(B-V)$ & $(V-K)_0$ & $T_{\rm phot}$ & $T_{\rm spec}$ & $\Delta{}T$ \\
	& (mag) & (mag) & (K) & (K) & (K) \\
\hline
C222531-145437 	& 0.03	& 	2.86 & 4285	& 4365 & $-$80 \\	
C230626-085103		& 0.05	&	3.00 & 4196	& 4225 & $-$29 \\	
J221821-183424 	& 0.03	&	2.36 & 4685	& 4630 &   +55  \\	
J223504-152834 	& 0.04	&	2.50 & 4557	& 4650 & $-$93 \\	
J223811-104126 	& 0.07	&	1.84 & 5240	& 5190 &   +50  \\
\hline
\end{tabular}%}
\end{table}
\setlength{\tabcolsep}{6pt}

Calculated photometric temperatures are listed in Table \ref{tab:photometric-temperatures}. The mean difference between the photometric temperatures and those found by excitation balance is $-$19\,K, where the largest variation is $-$93\,K for J223504-152834. While these photometric temperatures serve as a confirmation for our spectroscopically-derived values, for the remainder of this analysis we have employed effective temperatures determined by excitation balance.

\subsection{Uncertainties in Stellar Parameters}
\label{sec:uncertainties-in-stellar-parameters}
Due to scatter in neutral iron lines measurements, there is a formal uncertainty in our calculated trend line between excitation potential and abundance, as well as between the reduced equivalent width and abundance. We have calculated 1$\sigma$ uncertainties in effective temperature and microturbulence by independently varying each stellar parameter until the relevant slope matches that formal uncertainty. This process is repeated for positive and negative offsets in temperature and microturbulence to allow for asymmetric uncertainties. The largest absolute offset is taken as the $1\sigma$ uncertainty. For surface gravity, the uncertainty has been calculated by varying $\log{g}$ until the difference in mean Fe\,\textsc{i} - Fe\,\textsc{ii} abundance matches the standard error about the mean for Fe\,\textsc{i} and Fe\,\textsc{ii} in quadrature. The calculated uncertainties are tabulated in Table \ref{tab:stellar-parameter-uncertainties}. 

These uncertainties ignore any correlations between stellar parameters, and therefore are likely to be under-estimated. As such, we have assumed the total uncertainty in stellar parameters to be $\sigma(T_{\rm eff}) = \pm125$\,K, $\sigma(\log{g}) = \pm0.30$\,dex, and $\sigma(\xi_t) = \pm0.20$ km s$^{-1}$. These adopted uncertainties are higher than those listed in Table \ref{tab:stellar-parameter-uncertainties}, and can be regarded as extremely conservative.

\begin{table}
\centering
\caption{Uncorrelated uncertainties in stellar parameters for standard and program stars\label{tab:stellar-parameter-uncertainties}}
%\resizebox{\textwidth}{!}{%
\begin{tabular}{lccc}
\hline
\hline
Designation & $\sigma(T_{\rm eff})$ & $\sigma(\xi_t)$ & $\sigma(\log{g})$ \\
 & (K) & (km s$^{-1}$) & (dex) \\
\hline
HD\,41667  	&    53   	&   0.09   &   0.13   \\
HD\,44007 		&    81   	&   0.29   &   0.09   \\
HD\,76932 		&   107  	&   0.08   &   0.19   \\
HD\,136316		&    33   	&   0.15   &   0.12   \\
HD\,141531		&    25   	&   0.05   &   0.13   \\
HD\,142948		&    47   	&   0.10   &   0.11   \\
J221821-183424 	&    42   	&   0.11   &   0.09   \\
C222531-145437  	&    46   	&   0.05   &   0.12   \\
J223504-152834  	&    61   	&   0.08   &   0.05   \\
J223811-104126  	&    49   	&   0.16   &   0.08   \\
C230626-085103  	&    52   	&   0.05   &   0.04   \\
\hline
\end{tabular}
\end{table}

\subsection{Distances}
\label{sec:distances}

\begin{table*}
\caption{Parameters and uncertainties for Monte-Carlo realisations\label{tab:monte-carlo-data}}
\resizebox{\textwidth}{!}{%
\begin{tabular}{lccccccccc}
\hline
\hline
& \multicolumn{7}{c}{{\bf Input Parameters for Monte-Carlo Simulation}} && {\bf Output} \\
\cline{2-8} \cline{10-10}
Designation & $T_{\rm eff}$ & $\log{g}$ & $J$ & $E(B-V)$ & $V_{\rm hel}$ & $\mu_\alpha$ & $\mu_\delta$ && $D$ \\
& (K) & (dex) & (mag) & (mag) & (km s$^{-1}$) & (mas yr$^{-1}$) & (mas yr$^{-1}$) & &(kpc) \\
\hline
C222531-145437 & $4365 \pm 125$ & $1.25 \pm 0.20$ & $10.341 \pm 0.022$ & $0.03 \pm 0.01$ & $-156.4 \pm 0.1$ & \,$3.5 \pm 2.1$ & $-14.7 \pm 2.2$ &\,& $5.1^{+1.1}_{-0.8}$\\
C230626-085103 & $4225 \pm 125$ & $0.85 \pm 0.20$ & $10.312 \pm 0.025$ & $0.05 \pm 0.01$ & $-221.1 \pm 0.1$ & $-2.5 \pm 2.8$ & $-15.4 \pm 2.7$ &\,& $6.5^{+1.4}_{-1.1}$\\
J221821-183424 & $4630 \pm 125$ & $0.88 \pm 0.20$ & $10.340 \pm 0.021$ & $0.03 \pm 0.01$ & $-159.5 \pm 0.1$ & $-10.6 \pm 2.5$ & $-19.3 \pm 2.5$ &\,& $5.6^{+1.3}_{-0.9}$ \\
J223504-152834 & $4650 \pm 125$ & $2.16 \pm 0.20$ & $10.363 \pm 0.025$ & $0.04 \pm 0.01$ & $-169.7 \pm 0.1$ & \,$15.9 \pm 2.2$ & $-12.8 \pm 2.2$ &\,& $1.9^{+0.5}_{-0.4}$ \\
J223811-104126 & $5190 \pm 125$ & $2.93 \pm 0.20$ & $10.420 \pm 0.018$ & $0.07 \pm 0.01$ & $-235.7 \pm 0.1$ & $-25.3 \pm 2.1$ & $-99.5 \pm 2.1$ &\,& $1.1^{+0.3}_{-0.2}$ \\
\hline
\end{tabular}}
\end{table*}

Distances to the Aquarius stars are necessary for understanding the dynamical history of the parent cluster. Many groups have determined distances for stars in the RAVE survey catalog, which includes all Aquarius stream members. \citet{williams;et-al_2011} tabulated a range of distances inferred by different techniques. Not every measurement technique was applicable to all Aquarius stars. The reduced proper motion distance technique was the only method to estimate distances for all Aquarius candidates. The variations between distance measurements are large. In particular, the distance for {C222531-146537} ranged from $1.4 \pm 0.1$\,kpc \citep{burnett_binney_2010} to $10.3 \pm 2.4$\,kpc \citep{breddels;et-al_2010}, where both groups claim to have the ``most likely'' distances.

\begin{figure}
	\includegraphics[width=\columnwidth]{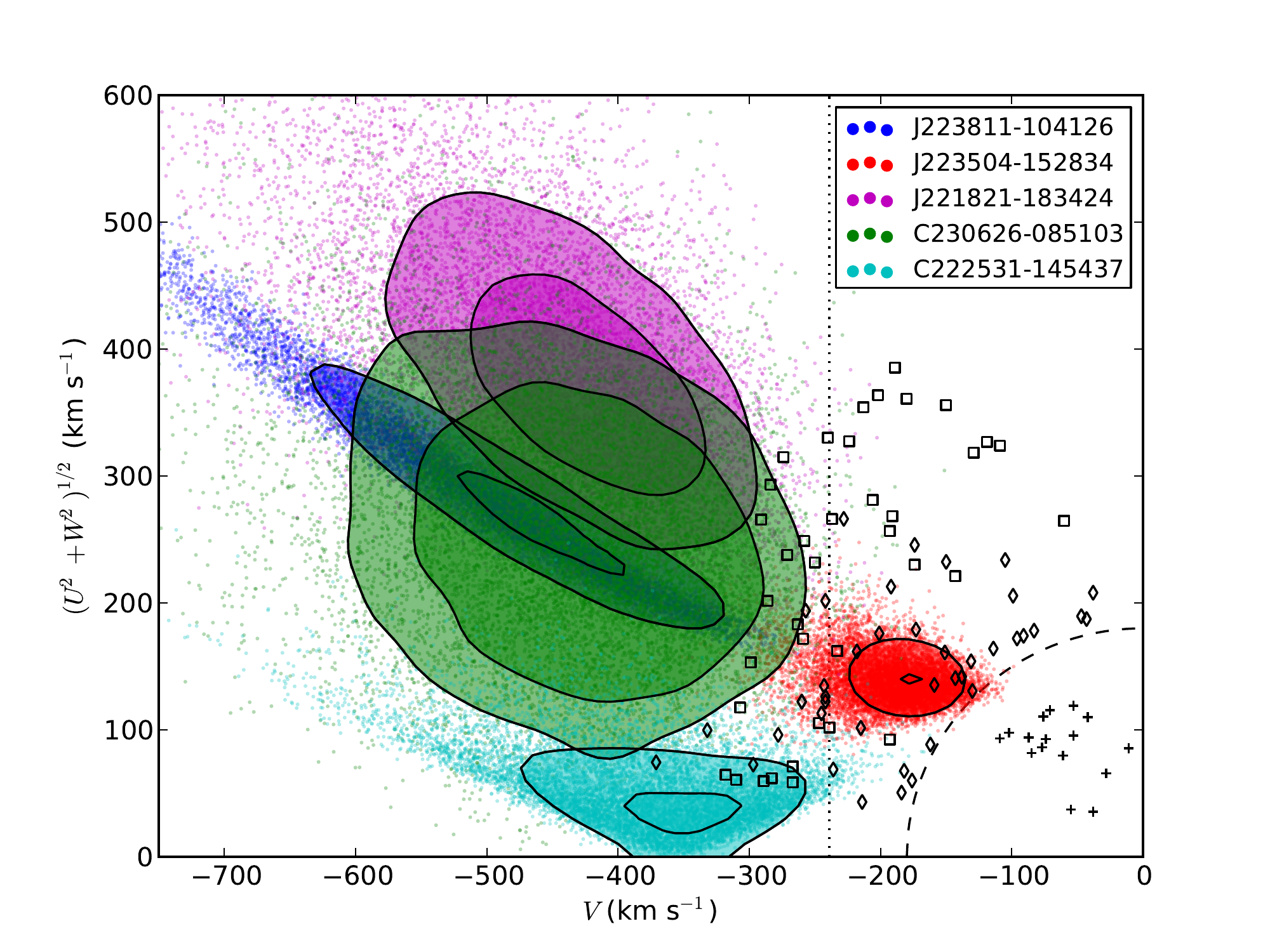}
	\caption{Galactic plane rotational velocities versus out-of-plane total velocities. The contours of each star represent the 68\% and 95\% confidence intervals from 10,000 Monte-Carlo realisations of the parameter distributions shown in Table \ref{tab:monte-carlo-data}. A sample of thick disk data from \citet{nissen;schuster_2010} is shown ($+$), as well as their high- and low-alpha halo populations ($\diamond$ and $\circ$ respectively).}
	\label{fig:toombre}
\end{figure}

Using the stellar parameters tabulated in Table \ref{tab:stellar-parameters}, we have calculated distances by isochrone fitting. The \citet{dotter;et-al_2008} $\alpha$-enhanced isochrones were used for these calculations, and an age of 10\,Gyr was assumed for all stars \citep{williams;et-al_2011,wylie-de-boer;et-al_2012}. The closest point to the isochrone was found by taking the uncertainties in $T_{\rm eff}$ and $\log{g}$ (see Section \ref{sec:uncertainties-in-stellar-parameters}) into account and measuring the distance modulus in the $J$ band. Given the (i) number of uncertain measurements involved in calculating distances ($T_{\rm eff}$, $\log{g}$, $E(B-V)$, $J$) and (ii) the resultant asymmetric uncertainties, distances were determined from 10,000 Monte-Carlo realisations. Table \ref{tab:monte-carlo-data} lists the input parameters and uncertainties adopted for the Monte-Carlo realisations, as well as the emergent distances and uncertainties. Uncertainties in input parameters were assumed to be normally distributed. Of the distance scales collated in \citet{williams;et-al_2011}, our distances are in most agreement with the \citet{zwitter;et-al_2010} system. In fact, we find the best agreement with the mean of all the distance scales tabulated in \citet{williams;et-al_2011}. The uncertainties in our distance determinations are on the order of twenty per cent.

\subsection{Dynamics}

Velocity vectors and Galactic orbits have been determined in the same Monte-Carlo realisations outlined in Section \ref{sec:distances}, which includes uncertainties in distances, proper motions\footnote{The proper motions in Table 1 of \citet{williams;et-al_2011} are erroneous in that they are associated with the wrong stars. The error was typographical and did not affect the transverse velocity calculations (M.E.K. Williams, private communication). The proper motions listed in our Table \ref{tab:monte-carlo-data} are correct.} and heliocentric velocities. We assume no uncertainty in on-sky position $(\alpha, \delta)$. Orbital energy calculations have assumed a three-component (bulge, disk, halo) model of the Galactic potential that reasonably reproduces the Galactic rotation curve. The bulge is represented by a Hernquist potential:

\begin{equation}
	\Phi_{\rm bulge}(x, y, z) = \frac{GM_b}{r + a}
\end{equation}

\noindent where $a = 0.6$\,kpc. The disk is modelled as a Miyamoto-Nagai potential \citep{miyamoto;nagai_1975} where:

\begin{equation}
	\Phi_{\rm disk}(x, y, z) = \frac{GM_{\rm disk}}{\sqrt{x^2 + y^2 + (b + \sqrt{z^2 + c^2})^2}}
\end{equation}

\noindent with $b = 4.5$\,kpc and $c = 0.25$\,kpc and the Galactic halo is represented by a Navarro-Frenk-White dark matter halo \citep{navarro;et-al_1997}:

\begin{equation}
	\Phi_{\rm halo} = -\frac{GM_{\rm vir}}{r\left[\log{(1 + c)} - c/(1 + c)\right]}\log{\left(1 + \frac{r}{r_s}\right)}
\end{equation}

\noindent{}with the three components scaled such that the disk provides 85\% of the radial force at $R_{\rm GC}$, in order to yield a flat circular-speed curve at $R_{\rm GC}$. The solar motion of \citet{schonrich;et-al_2012} has been adopted, where $R_{\rm GC}$ = 8.27\,kpc and a circular velocity speed $V_c = 238$\,km s$^{-1}$.

\begin{figure}
	\includegraphics[width=\columnwidth]{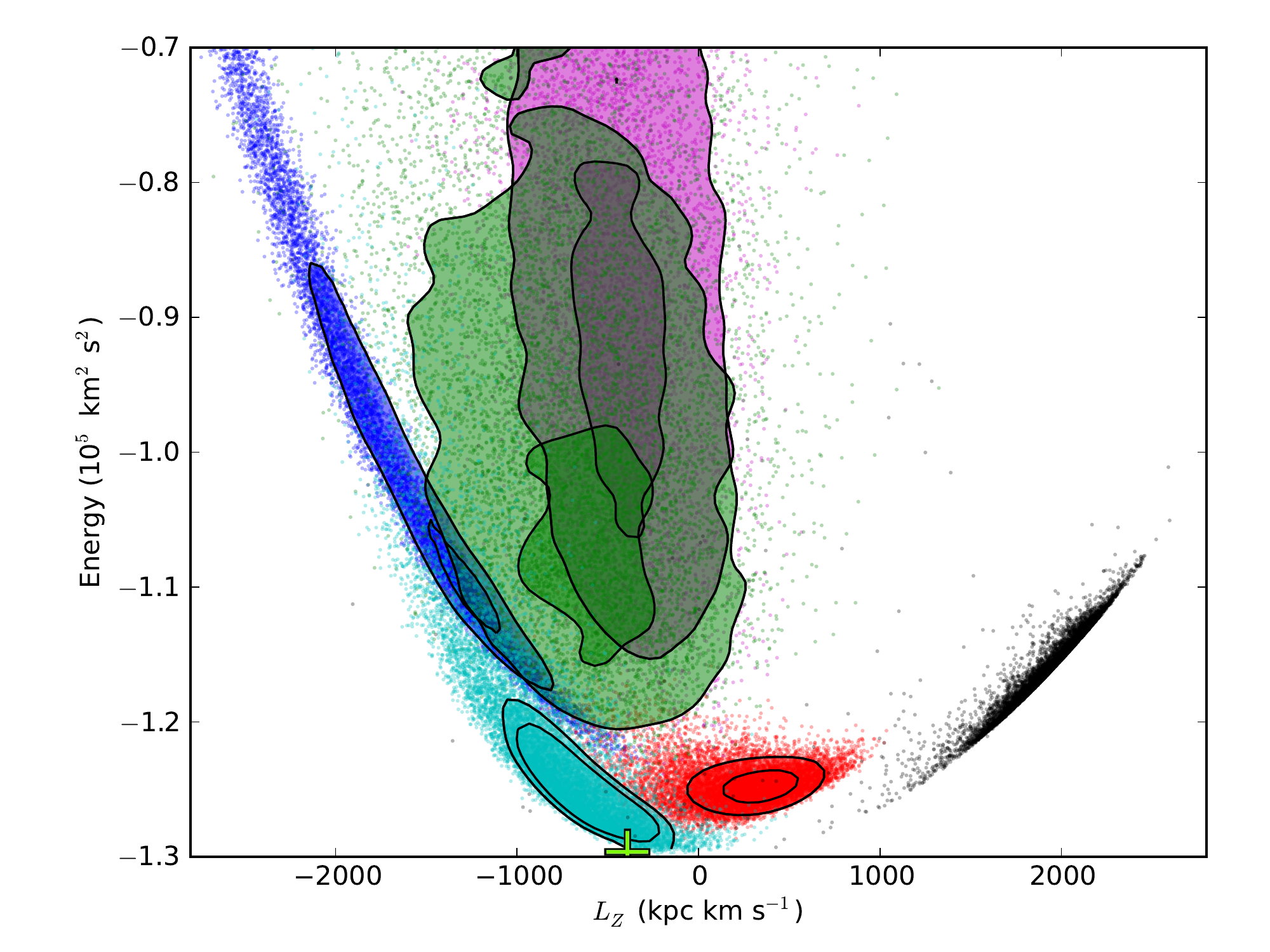}
	\caption{A Linblad ($L_Z-E$) diagram showing angular momenta and orbital energies after 10,000 Monte-Carlo realisations for each Aquarius stream star. Iso-contours represent the 68\% and 95\% confidence intervals. $\omega$-Centauri is shown as a lime green marker \citep{wylie-de-boer;et-al_2010}. The black points without contours are from the Geneva-Cophenagen Survey sample \citep{nordstrom;et-al_2004}, which primarily consists of nearby disk stars and serves as a validation of our orbital energy calculations. Colors are the same as in Figure \ref{fig:toombre}.}
	\label{fig:orbits}
\end{figure}

The Aquarius stream members have bound orbits, all of which are probably retrograde except for {J223504-152834} (Figure \ref{fig:toombre}). Orbital energies and angular momenta from Monte-Carlo simulations are illustrated in Figure \ref{fig:orbits}. The 16,686 stars from the Geneva-Cophenhagen Survey sample \citep{nordstrom;et-al_2004} are also shown as a reference, which primarily consists of nearby disk stars.

\section{Chemical Abundances}
\label{sec:chemical-abundances}
We have scaled our chemical abundances to Solar values using the chemical composition described in \citet{asplund;et-al_2009}. The abundances for the standard and program stars are shown in Tables \ref{tab:standard-star-abundances} and \ref{tab:program-star-abundances}, respectively. The discussion of comparable elements are grouped accordingly. \newpage % for nice formatting

\subsection{Carbon}
Carbon is produced by the triple-$\alpha$ process and ejected through supernovae events, or by mass-loss from asymptotic giant branch (AGB) stars \citep{kobayashi;et-al_2011}. 

\begin{figure}
	\includegraphics[width=\columnwidth]{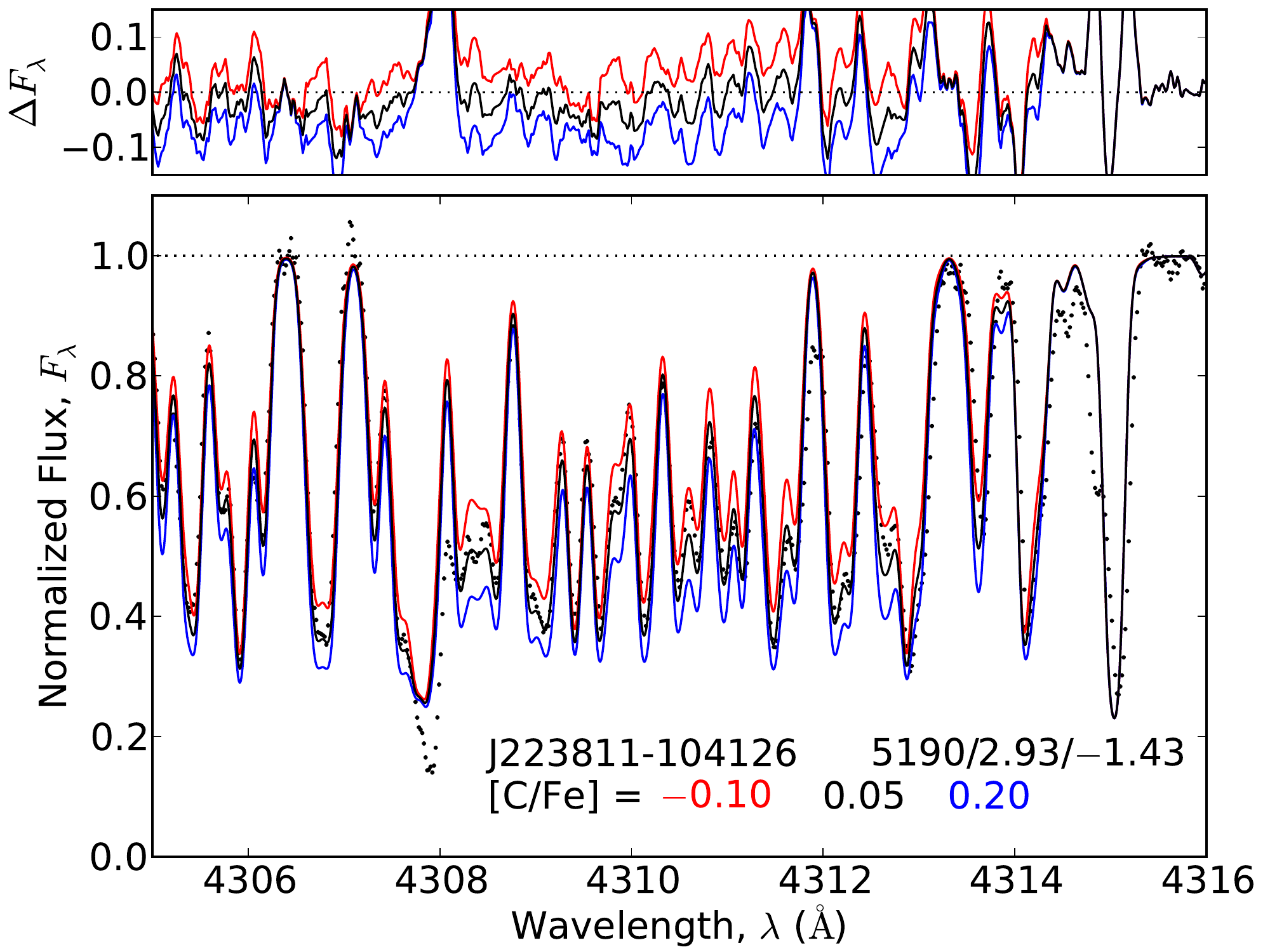}
	\caption{The carbon CH feature near 4313\,{\AA} in program star J223811-104126. The best-fit synthetic spectra is shown, with synthetic spectra for $\pm0.15$\,dex about the best-fitting abundance.}
	\label{fig:carbon}
\end{figure}

We have measured carbon abundances for all stars from the G-band head near 4313\,{\AA} and the CH molecular feature at 4323\,{\AA}, by comparing observed spectra with synthetic spectra for different carbon abundances. The synthetic spectra were convolved with a Gaussian kernel where the width was determined from nearby atomic lines with known abundances. Carbon was measured separately for both features, and in all stars the two measurements agree within 0.10\,dex. An example fit to this spectral region for J223811-104126 is shown in Figure \ref{fig:carbon}.

Carbon abundances in our standard stars agree well with the literature. For HD\,136316 we find {${\rm [C/Fe]} = -0.50 \pm 0.15$}, where \citet{gratton;et-al_2000} find {[C/Fe] $= -0.66$}. Our ${{\rm [C/Fe]} = -0.48}$ measurement for HD\,141531 agrees with \citet{gratton;et-al_2000} to within 0.06\,dex. Most program stars have near-solar carbon abundances, ranging from ${{\rm [C/Fe]} = -0.30}$ for {J221821-183424}, and +0.05 for {J223811-104126}.

\subsection{Sodium and Aluminium}
\label{sec:sodium-abundances}
Our line list includes three clean, unblended sodium lines at $\lambda$5688, $\lambda$6154 and $\lambda$6161. Not all three of these lines were detectable in each star. In the hottest and most metal-poor stars, J223811-104126 and J221821-183424 respectively, only the $\lambda$5688 line was measurable. For stars where multiple sodium lines were available, the line-to-line scatter is usually around 0.04\,dex with a maximum of 0.09\,dex in HD\,41667. However, in calculating total abundance uncertainties (see Section \ref{sec:chemical-abundance-uncertainties}) we have conservatively assumed a minimum random scatter of $\pm0.10$\,dex for all stars.

Our [Na/Fe] abundances appear systematically higher than values found in the literature by $\sim$0.10\,dex. For HD\,142948 we find {[Na/Fe] = 0.22}, which is +0.10\,dex higher than that found by \citet{gratton;et-al_2000}, and similarly we find HD\,76932 to be +0.10\,dex higher than reported by \citet{fulbright_2000}. \citet{gratton;et-al_2000} also found HD\,136316 to have {[Na/Fe] = $-$0.29}, where we find {[Na/Fe] = $-0.14$}, yet excellent agreement is found in the stellar parameters in \citet{gratton;et-al_2000} and this study. Different solar compositions employed between this study and earlier work can account for $\sim$0.08\,dex of this effect, leaving the residual difference well within the observational uncertainties. However, it is important to note that the [Na/Fe] abundance ratios presented in this study may be slightly higher compared to previous studies. While a systematic offset may be present, no intrinsic abundance dispersion in [Na/Fe] is present in the Aquarius sample.

There are six aluminium transitions in our optical spectra. The strongest of these lines occur at $\lambda$3944 and $\lambda$3961 and are visible in all of our stars. However this is a particularly crowded spectral region: the lines fall between the strong Ca H and K lines, with the $\lambda$3961 transition clearly located in the wing of the Ca H line. Additionally, the $\lambda$3944 and $\lambda$3961 lines have appreciable departures from the assumption of LTE, resulting in under-estimated abundances by up to $\sim$0.6\,dex \citep{baumueller;gehren_1997}. Instead, we have measured Al abundances from other available transitions: the {Al\,\textsc{i}} lines at $\lambda$6696, $\lambda$6698, $\lambda$7835 and $\lambda$7836. Generally the four {Al\,\textsc{i}} lines are in reasonable agreement with one another, yielding random scatter of less than 0.05\,dex. 

\subsection{$\alpha$-elements (O, Mg, Si, Ca and Ti)}
\label{sec:alpha-elements}

The $\alpha$-elements (O, Mg, Si, Ca and Ti) are forged through $\alpha$-particle capture during hydrostatic burning of carbon, neon and silicon. Material enriched in $\alpha$-elements is eventually dispersed into the interstellar medium following Type II core-collapse supernovae (SN). 

Oxygen can be a particularly difficult element to measure. There are only a handful of lines available in an optical spectrum: the forbidden {[O\,\textsc{i}]} lines at $\lambda$6300 and $\lambda$6363 and the O\,\textsc{i} triplet lines at $\sim$7775\,{\AA}. The forbidden lines are very weak and become difficult to measure in hot and/or metal-poor stars (${\mbox{[Fe/H]} \lesssim -1.5}$\,dex). When they are present, depending on the radial velocity of the star, the {[O\,\textsc{i}]} lines can be significantly affected by telluric absorption. The $\lambda$6363 line is intrinsically weak, blended with CN, and it falls in the wing of a strong Ca I auto-ionization feature. Because of these properties it is rarely used in abundance studies. Moreover, the $\lambda$6300 line is blended with a Ni\,\textsc{i} absorption line \citep{allende-prieto;et-al_2001}. Hence the region requires careful consideration. Although the O\,\textsc{i} triplet lines at $\sim$7775\,{\AA} are stronger than the forbidden lines, they are extremely susceptible to non-LTE effects, surface granulation \citep{asplund;perez_2001}, and are sensitive to changes in microturbulence. Our forbidden {[O\,\textsc{i}]} abundances for HD\,136316 agree well with those from \citet{gratton;et-al_2000} -- the difference is only 0.07\,dex.

The {[O\,\textsc{i}]} lines were measurable in four of our Aquarius stream candidates. The $\lambda$6300 line in one of our candidates, {C2306265-085103}, was sufficiently affected by telluric absorption such that we deemed the line unrecoverable. Thus, only the $\lambda$6363 transition was used to derive an oxygen abundance for {C2306265-085103}. In our hottest star, {J223811-104126}, the forbidden oxygen lines were not detected above a 3$\sigma$ significance. After synthesising the region, we deduce a very conservative upper limit of ${\mbox{[O/Fe]} < 0.50}$ from the {[O\,\textsc{i}]} lines. This is consistent with the rest of our candidates, with [O/Fe] abundances varying between 0.43 to 0.49\,dex.

\begin{figure}
	\includegraphics[width=\columnwidth]{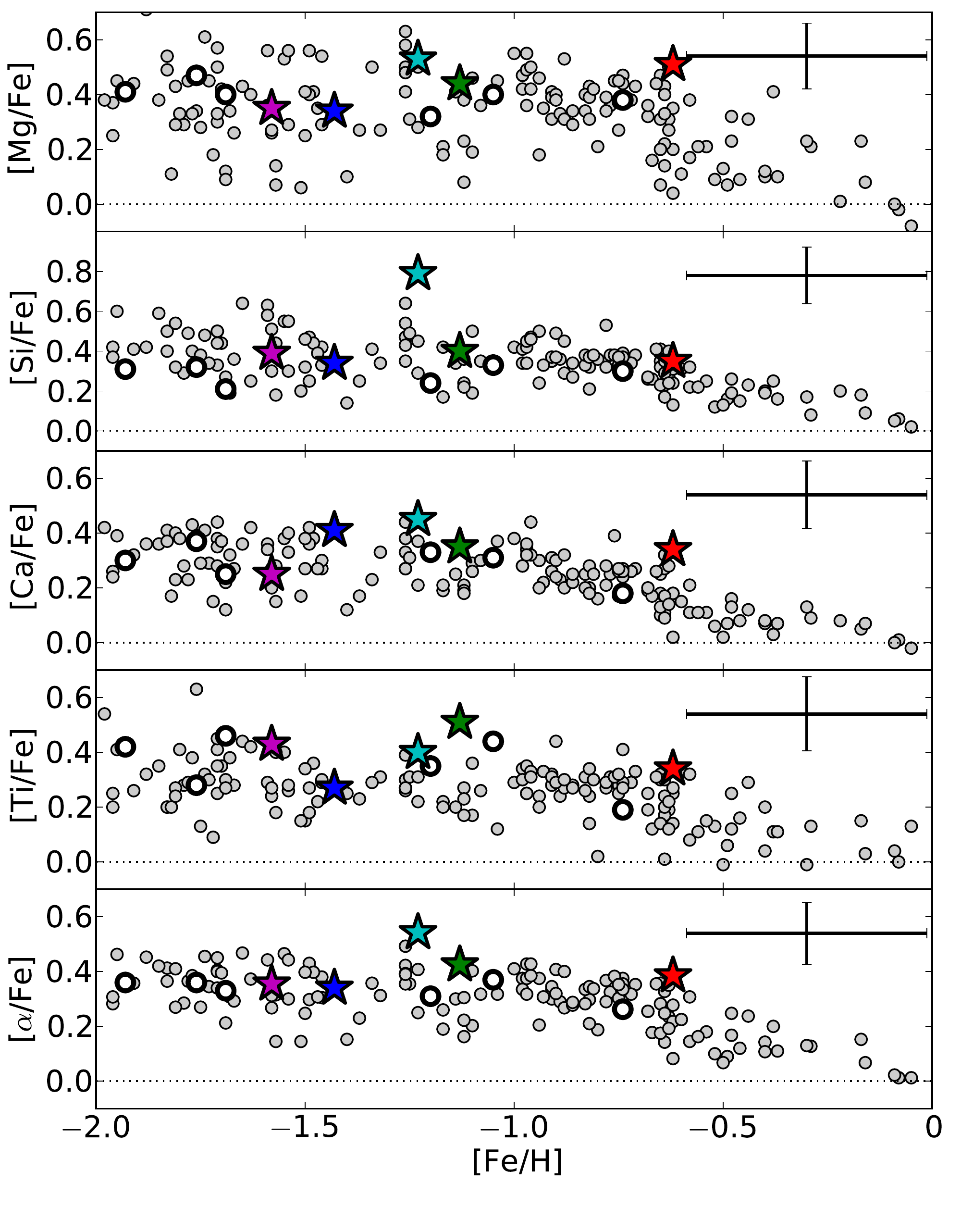}
	\caption{$\alpha$-element abundances with respect to iron content. The mean [$\alpha$/Fe] abundance from these is shown in the bottom panel. The Solar element-to-iron ratio is marked as a dotted line in each panel. Colors are as per Figure \ref{fig:toombre}. Standard stars are shown as open circles. Mean conservative \emph{total uncertainties} (random and systematic) for stars in this study are shown in each panel. Filled circles represent Milky Way field stars from \citet{fulbright_2000}. Oxygen abundances are shown separately in Figure \ref{fig:o-na}.}
	\label{fig:alpha-elements}
\end{figure}

In order to derive an oxygen measurement for {J223811-104126}, we were forced to use the triplet lines at $\sim$7775\,{\AA}. We extended these measurements for all Aquarius stars, and a mean abundance for each candidate was found from the synthesis of the permitted triplet lines. Oxygen abundances inferred from the triplet lines in all other stars were systematically ${\sim}+0.3$\,dex higher than abundances calculated from the {[O\,\textsc{i}]} forbidden lines. \citet{perez;et-al_2006} found the same result from stars with similar stellar parameters: [O/Fe] values based on the {O\,\textsc{i}} permitted triplet lines are on average ${+0.19 \pm 0.07}$\,dex higher than those found from the forbidden lines, which did not include non-LTE corrections of $+$0.08\,dex. Thus, we attribute our ${\sim}+0.3$\,dex offset between measurements of the {[O\,\textsc{i}]} and {O \textsc{i}} triplet lines to non-LTE and 3D effects. \citet{perez;et-al_2006} concluded that the forbidden lines, when not too weak, probably give the most reliable estimate of oxygen abundance. From the permitted O\,\textsc{i} triplet in {J223811-104126}, we derive an oxygen abundance of ${\mbox{[O/Fe]} = 0.42 \pm 0.01}$\,dex (random scatter). This measurement will be systematically higher than the `true' abundance if it were discernible from the [O\,\textsc{i}] lines, on the order of ${\sim}+0.3$\,dex. When we apply this crude offset derived from the rest of our sample, we arrive at a corrected abundance of ${\mbox{[O/Fe]} = 0.15 \pm 0.13}$ (total uncertainty) for {J223811-104126}. This is the most oxygen-deficient star in our sample by a factor of two.

Depending on the radial velocity of the star, some magnesium lines were affected by telluric absorption, particularly the $\lambda$6318 and $\lambda$6965 transitions. Atmospheric absorption was most notable for C222531-145437, where three of the four Mg transitions in our line list suffered some degree of telluric absorption, requiring an attentive correction. Every amended absorption profile was carefully examined, and lines with suspicious profiles were excluded from the final magnesium abundance. All [$\alpha$/Fe] abundance ratios in the standard stars are in excellent agreement with the literature. Typically the difference is 0.01\,dex, with the largest discrepancy of {$\Delta$[Ti/Fe] = +0.13\,dex} for HD\,76932 when compared with \citet{fulbright_2000}.

While \citet{wylie-de-boer;et-al_2012} find almost no scatter {($\pm0.02$\,dex)} in [Mg/Fe] for stars common to both studies, we find {C222531-145437} and {J223504-152834} to be almost {$+0.20$\,dex} higher than the rest of the sample. Of the {Mg\,\textsc{i}} line profiles measured, only two transitions are common to both line lists: $\lambda$6318 and $\lambda$6319. The oscillator strengths differ between studies; in these two lines the $\log{gf}$ differs by $-0.24$ and {$-0.27$\,dex} respectively (our oscillator strengths are lower). This indicates that the difference in oscillator strengths may explain the $\sim{}$0.2\,dex offset in [Mg/Fe] between this study and \citet{wylie-de-boer;et-al_2012}.

Of all the $\alpha$-elements, calcium has the smallest measurement scatter in our stars. The mean was formed from four line measurements in each star, with a typical random scatter of 0.01\,dex. Nevertheless, the aforementioned conservative minimum of 0.10\,dex for random scatter applies, and uncertainties in stellar parameters will contribute to the total error budget. As shown in Figure \ref{fig:alpha-elements}, all Aquarius stream candidates show super-solar [Ca/Fe], ranging between +0.23 to +0.43\,dex, consistent with [(Mg,Si,Ti)/Fe] measurements. 

{C222531-145437} has an unusually high silicon abundance (${\mbox{[Si/Fe]} = 0.79}$), well outside the uncertainties of the rest of our sample. The 5 silicon line abundances in this star are in relatively good agreement with each another. If we exclude the highest measurement, then the mean abundance drops only slightly to ${\mbox{[Si/Fe]} = 0.73 \pm 0.04}$ (random scatter). The lowest silicon line abundance for {C222531-145437} is ${\mbox{[Si/Fe]} = 0.61}$, which is still significantly higher than the mean abundance for any other star. With [Si/Fe] = +0.79, star {C222531-145437} lies above the majority of field stars. Examination of Figure \ref{fig:alpha-elements} would indicate that for all other $\alpha$-elements, it remains near the upper envelope defined by the field stars. It is not obvious why this is the case.

Titanium abundance ratios for the stream show typical levels of $\alpha$-enhancement. Our mean titanium abundances are derived from four to seven clean unblended Ti\,\textsc{i} and Ti\,\textsc{ii} lines. In our hottest and most metal-poor stars the mean Ti abundance is found from only Ti\,\textsc{ii} lines, as no suitable Ti\,\textsc{i} transitions were available.

\subsection{Iron-peak Elements}
\label{sec:fe-peak-abundances}

The Fe-peak elements (Sc to Zn) are primarily synthesized by the explosive nucleosynthesis of oxygen, neon, and silicon burning. Ignition can occur from Type II SN explosions of massive stars, or once a white dwarf accretes enough material to exceed the Chandrasekhar mass limit and spontaneously ignite carbon, leading to a Type Ia SN. 

Although not all {Fe-peak} elements are created equally, many Fe-peak elements generally exhibit similar trends with overall metallicity. All exhibit a positive trend with increasing iron abundance, with varying gradients.

\begin{figure}
	\includegraphics[width=\columnwidth]{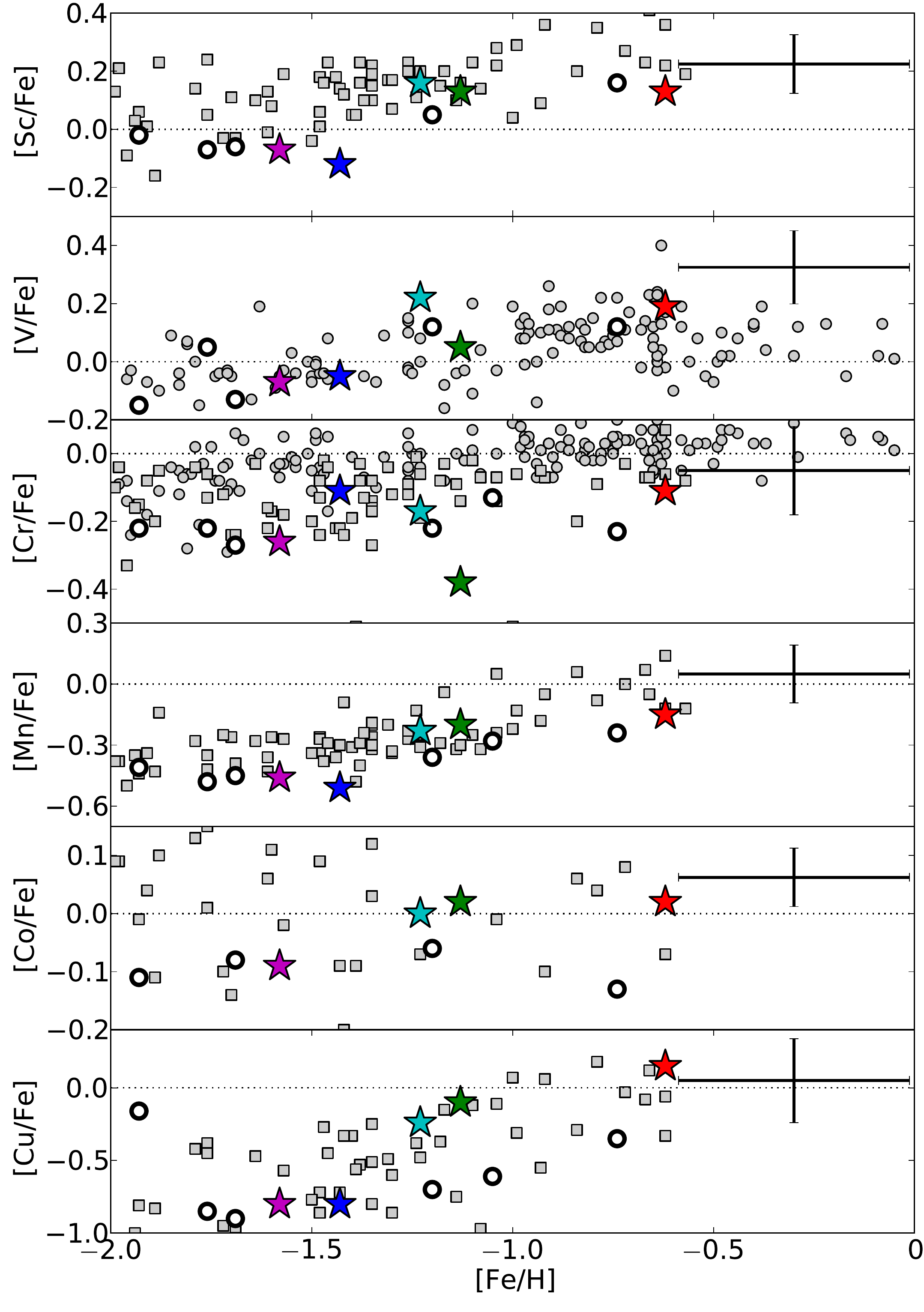}
	\caption{Iron-peak element abundances (Sc, V, Cr, Mn, Co and Cu) with respect to iron for all Aquarius stream stars. Ni, an additional Fe-peak element, is discussed in \S\ref{sec:na-ni-relationship} and shown in Figure \ref{fig:na-ni}. Colors are as per Figure \ref{fig:toombre}. Standard stars are shown as open circles. Mean conservative \emph{total uncertainties} (random and systematic) for this study are shown in each panel. Filled circles and squares represent Milky Way field stars from \citet{fulbright_2000} and \citet{ishigaki;et-al_2013}, respectively. Unlike Figure \ref{fig:alpha-elements}, panels have different $y$-axis ranges to accommodate the data.}
	\label{fig:fe-peak-elements}
\end{figure}

The [Sc/Fe] measurements presented in Figure \ref{fig:fe-peak-elements} are averaged from six clean {Sc\,\textsc{ii}} lines, and there is very little line-to-line scatter, the largest of which is {0.06\,dex}. The number of clean, suitable {Cr\,\textsc{i}} lines available between members fluctuated from three to twelve. Very little line-to-line scatter is present in both {Cr\,\textsc{i}} and Cr\,\textsc{ii}: the random scatter is below {0.04\,dex} for most stars. Chromium abundances are only available for one of the standard stars, where our {[Cr/Fe] = 0.03} is in excellent agreement with \citet{fulbright_2000}, where they find {[Cr/Fe] = 0.04}.

Manganese demonstrates a strong trend with increasing iron abundance (Figure \ref{fig:fe-peak-elements}). A significant source of Mn comes from Type Ia SN, and the strong [Mn/Fe]--[Fe/H] correlation is consistent with chemodynamical simulations \citep{kobayashi;nakasato_2011}, as well as thick disk observations by \citet{reddy;et-al_2006}.  Although Mn is known to demonstrate significant departures from LTE, we have not applied any non-LTE corrections to our abundances. 

Abundances of Co\,\textsc{i} lines were calculated by synthesis, as they demonstrate appreciable broadening due to hyperfine structure. Although they are known to suffer significant departures from LTE \citep{bergemann;et-al_2010}, no corrections have been made for these data. In general, [Co/H] follows [Fe/H] in our candidates.

Most Aquarius stream stars have seven clean {Ni\,\textsc{i}} transitions available. These lines are in excellent agreement, with a typical scatter of {0.03\,dex}. Nickel abundances have been published for two of our standard stars: HD\,76932 \citep{fulbright_2000} and HD\,141531 \citep{shetrone_1996}. In both cases, our [Ni/Fe] abundance ratios are slightly higher by +0.08 and +0.10\,dex respectively. The different solar compositions employed by these studies can only account for 0.01\,dex of this discrepancy, and the differences in oscillator strengths for common Ni\,\textsc{i} lines are negligible. Overall, the [Ni/Fe] abundance ratios in the Aquarius stream stars do not deviate greatly from the Solar ratio.

Hyperfine structure data has been included for the synthesis of Cu abundances. The offsets between EW and synthesis abundances for Cu were significant: $\sim$0.4\,dex higher for some stars without the inclusion of hyperfine structure information. Cu abundances have also been determined by synthesis, and are consistent with the Milky Way trend.

\subsection{Neutron-capture Elements}
Neutron-capture elements (Sr to Eu; ${38 \leqslant Z \leqslant 63}$) can be forged through multiple nucleosynthetic processes. The two primary processes that produce these elements are the rapid ($r$-) process and the slow ($s$-) process. While the $r$-process is theorised to occur in SN explosions, the $s$-process takes place foremost in AGB stars with a significant contribution from massive stars at higher metallicities \citep[e.g.,][]{meyer_1994}, although models of rotating massive stars may change this picture at the very lowest metallicities \citep{frischknecht;et-al_2012}.

\subsubsection{Strontium, Yttrium and Zirconium}
\label{sec:weak-s-process-elements}

These neutron-capture elements belong to the first $s$-process peak, and generally increase in lock-step with each another. {[Y\,\textsc{ii}/Fe]} and {[Zr\,\textsc{i}, Zr \textsc{ii}/Fe]} are in good agreement among all candidates. Strontium was measured by synthesis of the $\lambda$4077 and $\lambda$4215 lines. Although these lines are strong, they are often blended by a wealth of unresolved atomic and molecular features.

The Aquarius stream candidates have Y abundances consistent with halo field stars, with the exception of {C222531-145437}. With {[Y/Fe] = 0.79}, {C222531-145437} is significantly over-abundant in Y for its metallicity \citep[see Figure 4 of][]{travaglio;et-al_2004}. {C222531-145437} is consistently over-abundant in Zr, too. All other program and standard stars have first $n$-capture peak abundances and trends that are consistent with the chemical evolution of the Milky Way.

\subsubsection{Barium and Lanthanum}

Barium and Lanthanum belong to the second $s$-process peak. Ba has appreciable hyperfine and isotopic splitting, and its measurement requires some careful consideration. Solar Ba isotopic ratios have been adopted. Our standard stars have [Ba/Fe] abundances typical of the Milky Way halo. Two standard stars have existing [Ba/Fe] measurements from high-resolution spectra: HD\,44007 and HD\,76932. We find HD\,44007 to have ${\mbox{[Ba/Fe]} = 0.03}$, which is in good agreement with \citet{burris;et-al_2000}, who find {0.05\,dex}. For HD\,76932 our measurement of {$\mbox{[Ba/Fe]} = 0.18$} is in reasonable agreement with the \citet{fulbright_2000} value of $-0.02$\,dex, especially when differences in adopted solar composition are considered.

With one exception, the Aquarius stream candidates have [Ba/Fe] abundance ratios that are indistinguishable from field stars, ranging between ${\mbox{[Ba/Fe]} = -0.10}$ to {0.10\,dex}. The exception is {C222531-145437}, the same star showing enhancements in Y and Zr, which has an anomalously high barium abundance of ${\mbox{[Ba/Fe]} = 0.62}$. This is $\sim$0.60\,dex higher than the Milky Way trend at its given metallicity of ${\mbox{[Fe/H]} = -1.26}$ \citep{ishigaki;et-al_2013}. Our two {Ba\,\textsc{ii}} lines in {C222531-145437} are in excellent agreement with each other: [Ba/Fe] = 0.63, and 0.61. 

Lanthanum abundances have been determined by synthesis of the $\lambda$4558 and $\lambda$5805 lines with hyperfine splitting data included. All stars have La abundances that are consistent with the chemical enrichment of the Galaxy \citep{ishigaki;et-al_2013}, with the exception of the Ba-rich star C222531-145437, where [La/Fe] = 0.64 is observed.

\subsubsection{Cerium, Neodymium and Europium}

Europium is primarily produced by the $r$-process, whereas the production of Ce and Nd is split between $s$- and $r$-process. Europium abundances have been determined by synthesising the $\lambda$6645 Eu\,\textsc{ii} transition with hyperfine splitting data from \citet{sneden}.

We chose not to use the $\lambda$6437 Eu\,\textsc{ii} line as it is appreciably blended by a nearby Si\,\textsc{i} line \citep{Lawler;et-al_2001}, and our measurements were consistent with a hidden blend: the $\lambda$6437 Eu\,\textsc{ii} abundance was systematically higher than the $\lambda$6645 counterpart. One Aquarius stream candidate, {C222531-145437}, appears enhanced in all [$s$-process/Fe] abundance ratios compared to the program and standard sample. However no noteworthy difference in Eu, which is generally considered to be a $r$-process dominated element, was observed.\\

\onecolumn
\begin{longtable}[t!]{lcrcrrclcrcrrcc}
\caption{Standard Star Abundances\label{tab:standard-star-abundances}} \tabularnewline
\cline{1-13}
Species & $N$ & $\log\epsilon(X)$ & $\sigma$ & [X/H] & [X/Fe] && 
Species & $N$ & $\log\epsilon(X)$ & $\sigma$ & [X/H] & [X/Fe] \tabularnewline 
\cline{1-13} \tabularnewline
% This works.
\endhead
\hline
\multicolumn{13}{r}{Continued..}
\endfoot
\hline
\endlastfoot
\\
\multicolumn{6}{c}{HD\,41667} & \, & \multicolumn{6}{c}{HD\,44007} \\
\cline{1-6} \cline{8-13}
   C (CH)       &   2 &    6.95 &    0.20 &  --1.48 &  --0.28 && 
   C (CH)       &   2 &    6.66 &    0.20 &  --1.77 &  --0.01 \\ 
   O \textsc{I} &   2 &    7.95 &    0.06 &  --0.74 &    0.46 && 
   O \textsc{I} &   1 &    7.41 &    0.00 &  --1.28 &    0.48 \\ 
  Na \textsc{I} &   3 &    4.90 &    0.18 &  --1.34 &  --0.14 &&
  Na \textsc{I} &   2 &    4.44 &    0.09 &  --1.80 &  --0.04 \\
  Mg \textsc{I} &   4 &    6.72 &    0.10 &  --0.88 &    0.32 &&
  Mg \textsc{I} &   2 &    6.30 &    0.06 &  --1.29 &    0.47 \\
  Al \textsc{I} &   4 &    5.18 &    0.11 &  --1.27 &  --0.07 &&
  Al \textsc{I} &   1 &   :4.80 & \nodata & :--1.65 &   :0.11 \\
  Si \textsc{I} &   5 &    6.55 &    0.06 &  --0.96 &    0.24 &&
  Si \textsc{I} &   5 &    6.07 &    0.07 &  --1.44 &    0.32 \\
   K \textsc{I} &   1 &    4.64 & \nodata &  --0.39 &    0.81 &&
   K \textsc{I} &   1 &    4.31 & \nodata &  --0.72 &    1.04 \\
  Ca \textsc{I} &   4 &    5.47 &    0.06 &  --0.87 &    0.33 &&
  Ca \textsc{I} &   4 &    4.95 &    0.02 &  --1.39 &    0.37 \\
 Sc \textsc{II} &   5 &    2.00 &    0.12 &  --1.15 &    0.05 &&
 Sc \textsc{II} &   5 &    1.32 &    0.12 &  --1.85 &  --0.07 \\
  Ti \textsc{I} &   4 &    3.96 &    0.04 &  --0.99 &    0.21 &&
  Ti \textsc{I} &   1 &    3.48 & \nodata &  --1.47 &    0.29 \\
 Ti \textsc{II} &   3 &    4.09 &    0.25 &  --0.86 &    0.35 &&
 Ti \textsc{II} &   4 &    3.47 &    0.15 &  --1.48 &    0.28 \\
   V \textsc{I} &   4 &    2.85 &    0.11 &  --1.08 &    0.12 &&
   V \textsc{I} &   1 &    2.22 & \nodata &  --1.72 &    0.05 \\
  Cr \textsc{I} &  10 &    4.22 &    0.08 &  --1.42 &  --0.22 &&
  Cr \textsc{I} &  15 &    3.65 &    0.07 &  --1.99 &  --0.22 \\
 Cr \textsc{II} &   2 &    4.54 &    0.05 &  --1.09 &    0.11 &&
 Cr \textsc{II} &   3 &    4.00 &    0.01 &  --1.64 &    0.12 \\
 Mn \textsc{I}  &   3 &    3.87 &    0.04 &  --1.56 &  --0.36 &&
 Mn \textsc{I}  &   2 &    3.21 &    0.06 &  --2.22 &  --0.48 \\
  Fe \textsc{I} &  61 &    6.30 &    0.12 &  --1.20 &    0.00 &&
  Fe \textsc{I} &  51 &    5.74 &    0.13 &  --1.76 &    0.00 \\
 Fe \textsc{II} &  13 &    6.35 &    0.05 &  --1.15 &    0.05 &&
 Fe \textsc{II} &  15 &    5.74 &    0.10 &  --1.76 &  --0.00 \\
  Co \textsc{I} &   3 &    3.73 &    0.06 &  --1.26 &  --0.06 &&
  Co \textsc{I} &   0 & \nodata & \nodata & \nodata & \nodata \\
  Ni \textsc{I} &   7 &    4.94 &    0.12 &  --1.28 &  --0.08 &&
  Ni \textsc{I} &   4 &    4.47 &    0.05 &  --1.75 &    0.01 \\
  Cu \textsc{I} &   1 &    2.29 & \nodata &  --1.90 &  --0.70 &&
  Cu \textsc{I} &   1 &    1.58 & \nodata &  --2.61 &  --0.85 \\
  Zn \textsc{I} &   2 &    3.36 &    0.08 &  --1.20 &    0.00 &&
  Zn \textsc{I} &   2 &    2.83 &    0.05 &  --1.73 &    0.03 \\
 Sr \textsc{II} &   1 &    1.59 & \nodata &  --1.28  &  --0.08 &&
 Sr \textsc{II} &   2 &    1.13 &    0.09 &  --1.75 &    0.01 \\
  Y \textsc{II} &   5 &    0.97 &    0.19 &  --1.24 &  --0.04 &&
  Y \textsc{II} &   6 &    0.28 &    0.11 &  --1.93 &  --0.16 \\
  Zr \textsc{I} &   2 &    1.42 &    0.05 &  --1.17 &    0.04 &&
  Zr \textsc{I} &   0 & \nodata & \nodata & \nodata & \nodata \\
 Zr \textsc{II} &   1 &    1.28 & \nodata &  --1.30 &  --0.10 &&
 Zr \textsc{II} &   1 &    0.59 & \nodata &  --1.99 &  --0.23 \\
 Ba \textsc{II} &   2 &    0.95 &    0.07 &  --1.23 &  --0.02 &&
 Ba \textsc{II} &   2 &    0.31 &    0.06 &  --1.87 &  --0.11 \\
 La \textsc{II} &   1 &    0.17 & \nodata &  --0.93 &     0.27 &&
 La \textsc{II} &   2 &  --0.57 &    0.05 &  --1.67 &     0.09 \\
 Ce \textsc{II} &   4 &    0.35 &    0.18 &  --1.23 &  --0.02 &&
 Ce \textsc{II} &   3 &  --0.41 &    0.12 &  --1.99 &  --0.23 \\
 Nd \textsc{II} &   9 &    0.54 &    0.10 &  --0.88 &    0.32 &&
 Nd \textsc{II} &   9 &  --0.36 &    0.11 &  --1.78 &  --0.01 \\
 Eu \textsc{II} &   1 &  --0.13 & \nodata &  --0.65 &    0.55 &&
 Eu \textsc{II} &   1 &  --1.16 & \nodata &  --1.68 &    0.08 \\

\cline{1-6} \cline{8-13} \\ \\
\multicolumn{6}{c}{HD\,76932} && \multicolumn{6}{c}{HD\,136316} \\
\cline{1-6} \cline{8-13}
   C (CH)       &   2 &    7.52 &    0.20 &  --0.91 &    0.14 &&
   C (CH)       &   2 &    5.95 &    0.20 &  --2.48 &  --0.50 \\
   O \textsc{I} &   1 &    8.05 & \nodata &  --0.64 &    0.41 &&
   O \textsc{I} &   1 &    7.17 & \nodata &  --1.52 &    0.41 \\
  Na \textsc{I} &   3 &    5.37 &    0.04 &  --0.87 &    0.18 &&
  Na \textsc{I} &   2 &    4.17 &    0.04 &  --2.08 &  --0.14 \\
  Mg \textsc{I} &   3 &    6.95 &    0.20 &  --0.65 &    0.40 &&
  Mg \textsc{I} &   2 &    6.08 &    0.24 &  --1.52 &    0.41 \\
  Al \textsc{I} &   4 &    5.45 &    0.07 &  --1.00 &    0.05 &&
  Al \textsc{I} &   0 & \nodata & \nodata & \nodata & \nodata \\
  Si \textsc{I} &   5 &    6.79 &    0.06 &  --0.72 &    0.33 &&
  Si \textsc{I} &   4 &    5.89 &    0.05 &  --1.62 &    0.31 \\
   K \textsc{I} &   1 &    4.94 & \nodata &  --0.09 &    0.96 &&
   K \textsc{I} &   1 &    3.91 & \nodata &  --1.12 &    0.81 \\
  Ca \textsc{I} &   4 &    5.60 &    0.02 &  --0.74 &    0.31 &&
  Ca \textsc{I} &   4 &    4.71 &    0.02 &  --1.63 &    0.30 \\
 Sc \textsc{II} &   4 &    2.10 &    0.05 &  --1.05 &    0.01 &&
 Sc \textsc{II} &   4 &    1.20 &    0.08 &  --1.95 &  --0.02 \\
  Ti \textsc{I} &   1 &    4.36 & \nodata &  --0.59 &    0.46 &&
  Ti \textsc{I} &   3 &    3.19 &    0.03 &  --1.76 &    0.17 \\
 Ti \textsc{II} &   3 &    4.33 &    0.04 &  --0.62 &    0.44 &&
 Ti \textsc{II} &   3 &    3.44 &    0.10 &  --1.51 &    0.42 \\
   V \textsc{I} &   1 &   :3.33 & \nodata & :--0.60 &   :0.45 &&
   V \textsc{I} &   3 &    1.85 &    0.01 &  --2.08 &  --0.15 \\
  Cr \textsc{I} &  15 &    4.46 &    0.05 &  --1.18 &  --0.13 &&
  Cr \textsc{I} &  12 &    3.49 &    0.05 &  --2.15 &  --0.22 \\
 Cr \textsc{II} &   3 &    4.76 &    0.02 &  --0.88 &    0.17 &&
 Cr \textsc{II} &   2 &    3.90 &    0.02 &  --1.74 &    0.19 \\
  Mn \textsc{I} &   3 &    4.09 &    0.06 &  --1.34 &  --0.28 &&
  Mn \textsc{I} &   3 &    3.09 &    0.03 &  --2.34 &  --0.41 \\
  Fe \textsc{I} &  51 &    6.45 &    0.10 &  --1.05 &    0.00 &&
  Fe \textsc{I} &  62 &    5.57 &    0.11 &  --1.93 &    0.00 \\
 Fe \textsc{II} &  13 &    6.50 &    0.07 &  --1.00 &    0.05 &&
 Fe \textsc{II} &  14 &    5.61 &    0.12 &  --1.89 &    0.04 \\
  Co \textsc{I} &   1 &    3.94 & \nodata &  --1.05 &    0.00 &&
  Co \textsc{I} &   2 &    2.95 &    0.11 &  --1.09 &  --0.11 \\
  Ni \textsc{I} &   5 &    5.29 &    0.02 &  --0.93 &    0.13 &&
  Ni \textsc{I} &   5 &    4.22 &    0.11 &  --2.00 &  --0.07 \\
  Cu \textsc{I} &   1 &    2.53 & \nodata &  --1.66 &  --0.61 &&
  Cu \textsc{I} &   1 &    1.36 & \nodata &  --2.09 &  --0.16 \\
  Zn \textsc{I} &   2 &    3.58 &    0.03 &  --0.98 &    0.07 &&
  Zn \textsc{I} &   2 &    2.72 &    0.03 &  --1.83 &    0.10 \\
 Sr \textsc{II} &   2 &    1.99 &    0.02 &  --0.88 &    0.17 &&  
 Sr \textsc{II} &   1 &    0.69 & \nodata &  --2.18 &  --0.25 \\
  Y \textsc{II} &   5 &    1.14 &    0.05 &  --1.07 &  --0.02 &&
  Y \textsc{II} &   7 &    0.12 &    0.11 &  --2.09 &  --0.16 \\
  Zr \textsc{I} &   0 & \nodata & \nodata & \nodata & \nodata &&
  Zr \textsc{I} &   1 &    0.79 & \nodata &  --1.79 &    0.14 \\
 Zr \textsc{II} &   0 & \nodata & \nodata & \nodata & \nodata &&
 Zr \textsc{II} &   1 &    0.68 & \nodata &  --1.90 &    0.03 \\
 Ba \textsc{II} &   2 &    1.31 &    0.07 &  --0.87 &    0.18 &&
 Ba \textsc{II} &   2 &    0.22 &    0.02 &  --1.96 &  --0.03 \\
 La \textsc{II} &   1 &    0.50 & \nodata &  --0.60 &    0.45 &&
 La \textsc{II} &   1 &  --0.68 & \nodata &  --1.78 &    0.16 \\
 Ce \textsc{II} &   2 &    0.37 &    0.03 &  --1.21 &  --0.16 &&
 Ce \textsc{II} &   5 &  --0.39 &    0.18 &  --1.97 &  --0.04 \\
 Nd \textsc{II} &   3 &    0.56 &    0.06 &  --0.86 &    0.19 &&
 Nd \textsc{II} &  10 &  --0.36 &    0.04 &  --1.78 &    0.15 \\
 Eu \textsc{II} &   1 &  --0.33 & \nodata &  --0.85 &    0.20 &&
 Eu \textsc{II} &   1 &  --1.06 & \nodata &  --1.58 &    0.33 \\

\cline{1-6} \cline{8-13} \\ \\
\multicolumn{6}{c}{HD\,141531} && \multicolumn{6}{c}{HD\,142948} \\
\cline{1-6} \cline{8-13}
   C (CH)       &   2 &    6.33 &    0.20 &  --2.10 &  --0.48 &&
   C (CH)       &   2 &    7.72 &    0.20 &  --0.71 &    0.03 \\
   O \textsc{I} &   2 &    7.33 &    0.01 &  --1.35 &    0.34 &&
   O \textsc{I} &   2 &    8.43 &    0.02 &  --0.26 &    0.47 \\
  Na \textsc{I} &   2 &    4.28 &    0.05 &  --1.96 &  --0.27 &&
  Na \textsc{I} &   3 &    5.73 &    0.13 &  --0.51 &    0.22 \\
  Mg \textsc{I} &   2 &    6.30 &    0.15 &  --1.29 &    0.40 &&
  Mg \textsc{I} &   3 &    7.24 &    0.12 &  --0.36 &    0.38 \\
  Al \textsc{I} &   2 &    4.74 &    0.10 &  --1.71 &  --0.02 &&
  Al \textsc{I} &   4 &    5.94 &    0.08 &  --0.51 &    0.23 \\
  Si \textsc{I} &   5 &    6.03 &    0.10 &  --1.48 &    0.21 &&
  Si \textsc{I} &   5 &    7.07 &    0.05 &  --0.44 &    0.30 \\
   K \textsc{I} &   1 &    3.99 & \nodata &  --1.04 &    0.65 &&
   K \textsc{I} &   1 &    5.04 & \nodata &    0.01 &    0.75 \\
  Ca \textsc{I} &   4 &    4.90 &    0.03 &  --1.44 &    0.25 &&
  Ca \textsc{I} &   4 &    5.78 &    0.01 &  --0.56 &    0.18 \\
 Sc \textsc{II} &   5 &    1.40 &    0.11 &  --1.75 &  --0.06 &&
 Sc \textsc{II} &   5 &    2.57 &    0.12 &  --0.58 &    0.16 \\
  Ti \textsc{I} &   4 &    3.33 &    0.07 &  --1.62 &    0.07 &&
  Ti \textsc{I} &   4 &    4.44 &    0.09 &  --0.51 &    0.23 \\
 Ti \textsc{II} &   4 &    3.71 &    0.08 &  --1.24 &    0.46 &&
 Ti \textsc{II} &   3 &    4.40 &    0.21 &  --0.55 &    0.19 \\
   V \textsc{I} &   4 &    2.10 &    0.07 &  --1.83 &  --0.13 &&
   V \textsc{I} &   5 &    3.31 &    0.04 &  --0.62 &    0.12 \\
  Cr \textsc{I} &  12 &    3.68 &    0.06 &  --1.96 &  --0.27 &&
  Cr \textsc{I} &  13 &    4.67 &    0.15 &  --0.97 &  --0.23 \\
 Cr \textsc{II} &   2 &    4.11 &    0.02 &  --1.53 &    0.16 &&
 Cr \textsc{II} &   3 &    4.88 &    0.03 &  --0.76 &  --0.02 \\
  Mn \textsc{I} &   3 &    3.29 &    0.04 &  --2.14 &  --0.45 &&
  Mn \textsc{I} &   3 &    4.45 &    0.06 &  --0.98 &  --0.24 \\
  Fe \textsc{I} &  54 &    5.81 &    0.06 &  --1.69 &    0.00 &&
  Fe \textsc{I} &  61 &    6.76 &    0.10 &  --0.74 &    0.00 \\
 Fe \textsc{II} &  13 &    5.86 &    0.03 &  --1.64 &    0.05 &&
 Fe \textsc{II} &  13 &    6.75 &    0.06 &  --0.75 &  --0.02 \\
  Co \textsc{I} &   3 &    3.22 &    0.12 &  --1.77 &  --0.08 &&
  Co \textsc{I} &   3 &    4.36 &    0.11 &  --0.63 &  --0.13 \\
  Ni \textsc{I} &   7 &    4.42 &    0.12 &  --1.80 &  --0.11 &&
  Ni \textsc{I} &   5 &    5.62 &    0.04 &  --0.60 &    0.13 \\
  Cu \textsc{I} &   1 &    1.60 & \nodata &  --2.59 &  --0.90 &&
  Cu \textsc{I} &   1 &    3.10 & \nodata &  --1.09 &  --0.35 \\
  Zn \textsc{I} &   2 &    2.80 &    0.04 &  --1.76 &  --0.07 &&
  Zn \textsc{I} &   2 &    3.89 &    0.06 &  --0.67 &    0.07 \\
 Sr \textsc{II} &   1 &    1.00 & \nodata &  --1.87 &  --0.18 &&
 Sr \textsc{II} &   1 &    1.89 & \nodata &  --0.98 &  --0.24 \\
  Y \textsc{II} &   6 &    0.27 &    0.13 &  --1.94 &  --0.24 &&
  Y \textsc{II} &   6 &    1.33 &    0.32 &  --0.88 &  --0.14 \\
  Zr \textsc{I} &   0 & \nodata & \nodata & \nodata & \nodata &&
  Zr \textsc{I} &   0 & \nodata & \nodata & \nodata & \nodata \\
 Zr \textsc{II} &   1 &    0.75 & \nodata &  --1.83 &  --0.14 &&
 Zr \textsc{II} &   0 & \nodata & \nodata & \nodata & \nodata \\
 Ba \textsc{II} &   2 &    0.39 &    0.05 &  --1.79 &  --0.10 &&
 Ba \textsc{II} &   2 &    1.17 &    0.01 &  --1.01 &  --0.27 \\
 La \textsc{II} &   1 &  --0.56 & \nodata &  --1.67 &    0.03 &&
 La \textsc{II} &   1 &    0.56 & \nodata &  --0.54 &    0.20 \\
 Ce \textsc{II} &   4 &  --0.31 &    0.12 &  --1.89 &  --0.20 &&
 Ce \textsc{II} &   3 &    0.54 &    0.20 &  --1.04 &  --0.30 \\
 Nd \textsc{II} &  10 &  --0.20 &    0.08 &  --1.62 &    0.07 &&
 Nd \textsc{II} &   6 &    0.79 &    0.10 &  --0.63 &    0.11 \\
 Eu \textsc{II} &   1 &  --0.95 & \nodata &  --1.47 &    0.22 &&
 Eu \textsc{II} &   1 &    0.08 & \nodata &  --1.55 &    0.14 \\
\hline
\end{longtable}

\begin{longtable}[h!]{lcrcrrclcrcrrcc}
\caption{Program Star Abundances\label{tab:program-star-abundances}} \tabularnewline
\cline{1-13}
Species & $N$ & $\log\epsilon(X)$ & $\sigma$ & [X/H] & [X/Fe] && 
Species & $N$ & $\log\epsilon(X)$ & $\sigma$ & [X/H] & [X/Fe] \tabularnewline 
\cline{1-15} \tabularnewline
% This works.
\endhead
\hline
\multicolumn{13}{r}{Continued..}
\endfoot
\hline
\endlastfoot

\\
\multicolumn{6}{c}{J221821-183424} & \, & \multicolumn{6}{c}{C222531-145437} \\
\cline{1-6} \cline{8-13}
   C (CH)       &   2 &    6.55 &    0.20 &  --1.88 &  --0.30 &&
   C (CH)       &   2 &    7.15 &    0.20 &  --1.28 &  --0.05 \\
   O \textsc{I} &   2 &    7.55 &    0.04 &  --1.13 &    0.45 &&
   O \textsc{I} &   2 &    7.96 & \nodata &  --0.73 &    0.49 \\
  Na \textsc{I} &   1 &    4.75 & \nodata &  --1.49 &    0.09 &&
  Na \textsc{I} &   2 &    5.12 &    0.02 &  --1.12 &    0.10 \\
  Mg \textsc{I} &   3 &    6.37 &    0.09 &  --1.23 &    0.35 &&
  Mg \textsc{I} &   2 &    6.90 &    0.08 &  --0.70 &    0.53 \\
  Al \textsc{I} &   1 &    5.08 & \nodata &  --1.37 &    0.21 &&
  Al \textsc{I} &   4 &    5.94 &    0.10 &  --0.51 &    0.71 \\
  Si \textsc{I} &   5 &    6.32 &    0.08 &  --1.19 &    0.39 &&
  Si \textsc{I} &   5 &    7.07 &    0.15 &  --0.44 &    0.79 \\
   K \textsc{I} &   1 &    4.34 & \nodata &  --0.69 &    0.89 &&
   K \textsc{I} &   1 &    4.42 & \nodata &  --0.61 &    0.62 \\
  Ca \textsc{I} &   4 &    5.01 &    0.04 &  --1.33 &    0.25 &&
  Ca \textsc{I} &   4 &    5.57 &    0.04 &  --0.77 &    0.45 \\
 Sc \textsc{II} &   4 &    1.50 &    0.12 &  --1.65 &  --0.07 &&
 Sc \textsc{II} &   4 &    2.08 &    0.13 &  --1.07 &    0.16 \\
  Ti \textsc{I} &   0 & \nodata & \nodata & \nodata & \nodata &&
  Ti \textsc{I} &   4 &    4.10 &    0.03 &  --0.85 &    0.37 \\
 Ti \textsc{II} &   4 &    3.80 &    0.13 &  --1.15 &    0.43 &&
 Ti \textsc{II} &   2 &    4.12 &    0.13 &  --0.83 &    0.40 \\
   V \textsc{I} &   3 &    2.28 &    0.01 &  --1.65 &  --0.07 &&
   V \textsc{I} &   5 &    2.91 &    0.10 &  --1.01 &    0.22 \\
  Cr \textsc{I} &  11 &    3.80 &    0.06 &  --1.84 &  --0.26 &&
  Cr \textsc{I} &   8 &    4.24 &    0.17 &  --1.40 &  --0.17 \\
 Cr \textsc{II} &   2 &    4.07 &    0.03 &  --1.57 &    0.01 &&
 Cr \textsc{II} &   1 &    4.38 & \nodata &  --1.26 &  --0.03 \\
  Mn \textsc{I} &   2 &    3.38 &    0.03 &  --2.05 &  --0.46 &&
  Mn \textsc{I} &   3 &    3.98 &    0.05 &  --1.45 &  --0.23 \\
  Fe \textsc{I} &  52 &    5.92 &    0.09 &  --1.58 &    0.00 &&
  Fe \textsc{I} &  60 &    6.27 &    0.10 &  --1.23 &    0.00 \\
 Fe \textsc{II} &  13 &    5.94 &    0.05 &  --1.56 &    0.02 &&
 Fe \textsc{II} &  10 &    6.30 &    0.06 &  --1.20 &    0.03 \\
  Co \textsc{I} &   1 &    3.32 & \nodata &  --1.67 &  --0.09 &&
  Co \textsc{I} &   4 &    3.77 &    0.09 &  --1.22 &    0.00 \\
  Ni \textsc{I} &   5 &    4.61 &    0.14 &  --1.61 &  --0.03 &&
  Ni \textsc{I} &   7 &    5.07 &    0.09 &  --1.15 &    0.08 \\
  Cu \textsc{I} &   1 &    1.81 & \nodata &  --2.38 &  --0.80 &&
  Cu \textsc{I} &   1 &    2.72 & \nodata &  --1.47 &  --0.24 \\
  Zn \textsc{I} &   1 &    3.07 & \nodata &  --1.49 &    0.09 &&
  Zn \textsc{I} &   2 &    3.56 &    0.24 &  --1.00 &    0.23 \\
 Sr \textsc{II} &   1 &    1.39 & \nodata &  --1.48 &    0.10 &&
 Sr \textsc{II} &   1 &   :1.99 & \nodata & :--0.88 &   :0.35 \\
  Y \textsc{II} &   3 &    0.44 &    0.02 &  --1.77 &  --0.19 &&
  Y \textsc{II} &   5 &    1.78 &    0.16 &  --0.43 &    0.79 \\
  Zr \textsc{I} &   0 & \nodata & \nodata & \nodata & \nodata &&
  Zr \textsc{I} &   3 &    2.07 &    0.05 &  --0.51 &    0.72 \\
 Zr \textsc{II} &   1 &    0.97 & \nodata &  --1.61 &  --0.03 &&
 Zr \textsc{II} &   0 & \nodata & \nodata & \nodata & \nodata \\
 Ba \textsc{II} &   1 &    0.60 & \nodata &  --1.58 &    0.00 &&
 Ba \textsc{II} &   2 &    1.58 &    0.01 &  --0.60 &    0.62 \\
 La \textsc{II} &   1 &  --0.58 & \nodata &  --1.67 &  --0.09 &&
 La \textsc{II} &   2 &    0.51 &    0.02 &  --0.59 &    0.64 \\
 Ce \textsc{II} &   3 &  --0.39 &    0.06 &  --1.97 &  --0.39 &&
 Ce \textsc{II} &   5 &    0.73 &    0.15 &  --0.85 &    0.37 \\
 Nd \textsc{II} &  10 &  --0.22 &    0.07 &  --1.64 &  --0.06 &&
 Nd \textsc{II} &   8 &    0.88 &    0.13 &  --0.54 &    0.69 \\
 Eu \textsc{II} &   1 &  --0.86 &    0.11 &  --1.38 &    0.20 &&
 Eu \textsc{II} &   1 &  --0.29 & \nodata &  --0.81 &    0.42 \\

\cline{1-6} \cline{8-13} \\ \\
\multicolumn{6}{c}{J223504-152834} && \multicolumn{6}{c}{J223811-104126} \\
\cline{1-6} \cline{8-13}
   C (CH)       &   2 &    7.71 &    0.30 &  --0.72 &  --0.10 &&
   C (CH)       &   2 &    7.05 &    0.25 &  --1.38 &    0.05 \\
   O \textsc{I} &   2 &    8.50 &    0.10 &  --0.19 &    0.43 &&
   O \textsc{I}\footnote{Abundance derived from the permitted O \textsc{i} triplet instead of the forbidden [O \textsc{i}] lines, see \S\ref{sec:alpha-elements}} &   3 &    7.41 &    0.13 &  --1.28 &    0.15 \\
  Na \textsc{I} &   3 &    5.87 &    0.12 &  --0.37 &    0.26 &&
  Na \textsc{I} &   1 &    4.89 & \nodata &  --1.35 &    0.08 \\
  Mg \textsc{I} &   3 &    7.48 &    0.15 &  --0.12 &    0.51 &&
  Mg \textsc{I} &   2 &    6.51 &    0.03 &  --1.09 &    0.34 \\
  Al \textsc{I} &   3 &    6.12 &    0.09 &  --0.33 &    0.29 &&
  Al \textsc{I} &   2 &    5.13 &    0.13 &  --1.32 &    0.11 \\
  Si \textsc{I} &   5 &    7.24 &    0.10 &  --0.27 &    0.35 &&
  Si \textsc{I} &   3 &    6.42 &    0.04 &  --1.09 &    0.34 \\
   K \textsc{I} &   1 &    5.05 & \nodata &    0.02 &    0.64 &&
   K \textsc{I} &   1 &    4.50 & \nodata &  --0.53 &    0.90 \\
  Ca \textsc{I} &   4 &    6.06 &    0.03 &  --0.28 &    0.34 &&
  Ca \textsc{I} &   4 &    5.32 &    0.03 &  --1.02 &    0.41 \\
 Sc \textsc{II} &   5 &    2.65 &    0.10 &  --0.50 &    0.13 &&
 Sc \textsc{II} &   2 &    1.60 &    0.03 &  --1.55 &  --0.12 \\
  Ti \textsc{I} &   4 &    4.65 &    0.02 &  --0.30 &    0.32 &&
  Ti \textsc{I} &   0 & \nodata & \nodata & \nodata & \nodata \\
 Ti \textsc{II} &   1 &    4.67 & \nodata &  --0.28 &    0.34 &&
 Ti \textsc{II} &   4 &    3.79 &    0.09 &  --1.16 &    0.27 \\
   V \textsc{I} &   4 &    3.50 &    0.11 &  --0.43 &    0.19 &&
   V \textsc{I} &   1 &    2.45 & \nodata &  --1.48 &  --0.05 \\
  Cr \textsc{I} &   7 &    4.90 &    0.11 &  --0.74 &  --0.11 &&
  Cr \textsc{I} &  12 &    4.10 &    0.06 &  --1.54 &  --0.11 \\
 Cr \textsc{II} &   2 &    4.84 &    0.04 &  --0.79 &  --0.17 &&
 Cr \textsc{II} &   3 &    4.34 &    0.07 &  --1.30 &    0.12 \\
  Mn \textsc{I} &   3 &    4.66 &    0.04 &  --0.77 &  --0.15 &&
  Mn \textsc{I} &   2 &    3.50 &    0.01 &  --1.93 &  --0.51 \\
  Fe \textsc{I} &  63 &    6.88 &    0.12 &  --0.62 &    0.00 &&
  Fe \textsc{I} &  33 &    6.07 &    0.06 &  --1.43 &    0.00 \\
 Fe \textsc{II} &  12 &    6.87 &    0.07 &  --0.63 &  --0.01 &&
 Fe \textsc{II} &   9 &    6.04 &    0.07 &  --1.46 &  --0.03 \\
  Co \textsc{I} &   3 &    4.39 &    0.09 &  --0.60 &    0.02 &&
  Co \textsc{I} &   0 & \nodata & \nodata & \nodata & \nodata \\
  Ni \textsc{I} &   7 &    5.64 &    0.09 &  --0.58 &    0.05 &&
  Ni \textsc{I} &   2 &    4.84 &    0.04 &  --1.38 &    0.05 \\
  Cu \textsc{I} &   1 &    3.72 & \nodata &  --0.47 &    0.15 &&
  Cu \textsc{I} &   1 &    1.96 & \nodata &  --2.23 &  --0.80 \\
  Zn \textsc{I} &   2 &    4.21 &    0.03 &  --0.35 &    0.27 &&
  Zn \textsc{I} &   2 &    3.15 &    0.05 &  --1.41 &    0.02 \\
 Sr \textsc{II} &   1 &   :2.25 & \nodata & :--0.62 &   :0.00 &&
 Sr \textsc{II} &   1 &    1.64 & \nodata &  --1.23 &    0.20 \\
  Y \textsc{II} &   3 &    1.80 &    0.03 &  --0.41 &    0.21 &&
  Y \textsc{II} &   6 &    0.76 &    0.06 &  --1.45 &  --0.02 \\
  Zr \textsc{I} &   3 &    2.26 &    0.05 &  --0.32 &    0.31 &&
  Zr \textsc{I} &   0 & \nodata & \nodata & \nodata & \nodata \\
 Zr \textsc{II} &   0 & \nodata & \nodata & \nodata & \nodata &&
 Zr \textsc{II} &   0 & \nodata & \nodata & \nodata & \nodata \\
 Ba \textsc{II} &   2 &    1.65 &    0.02 &  --0.53 &    0.10 &&
 Ba \textsc{II} &   2 &    0.78 &    0.07 &  --1.40 &    0.03 \\
 La \textsc{II} &   1 &    0.76 & \nodata &  --0.34 &    0.28 &&
 La \textsc{II} &   0 & \nodata & \nodata & \nodata & \nodata \\
 Ce \textsc{II} &   3 &    0.87 &    0.13 &  --0.71 &  --0.09 &&
 Ce \textsc{II} &   2 &  --0.07 &    0.02 &  --1.65 &  --0.22 \\
 Nd \textsc{II} &   6 &    1.27 &    0.13 &  --0.15 &    0.47 &&
 Nd \textsc{II} &   1 &  --0.25 & \nodata &  --1.67 &  --0.24 \\
 Eu \textsc{II} &   1 &    0.40 & \nodata &  --0.12 &    0.50 &&
 Eu \textsc{II} &   1 &  --0.55 & \nodata &  --1.07 &    0.36 \\

\cline{1-6} \cline{8-13} \\ \\
\multicolumn{7}{c}{C2306265-085103} \\
\cline{1-6}
   C (CH)       &   2 &    7.20 &    0.20 &  --1.23 &  --0.10 \\
   O \textsc{I} &   2 &    8.02 &    0.04 &  --0.67 &    0.46 \\
  Na \textsc{I} &   2 &    5.31 &    0.01 &  --0.93 &    0.21 \\
  Mg \textsc{I} &   2 &    6.90 &    0.06 &  --0.70 &    0.44 \\
  Al \textsc{I} &   4 &    5.65 &    0.08 &  --0.80 &    0.33 \\
  Si \textsc{I} &   5 &    6.78 &    0.08 &  --0.73 &    0.40 \\
   K \textsc{I} &   1 &    4.46 & \nodata &  --0.57 &    0.56 \\
  Ca \textsc{I} &   4 &    5.56 &    0.04 &  --0.78 &    0.35 \\
 Sc \textsc{II} &   3 &    2.15 &    0.09 &  --1.00 &    0.13 \\
  Ti \textsc{I} &   4 &    4.13 &    0.03 &  --0.82 &    0.32 \\
 Ti \textsc{II} &   3 &    4.32 &    0.35 &  --0.63 &    0.51 \\
   V \textsc{I} &   4 &    2.85 &    0.06 &  --1.09 &    0.05 \\
  Cr \textsc{I} &   3 &    4.13 &    0.12 &  --1.51 &  --0.38 \\
 Cr \textsc{II} &   1 &    4.50 & \nodata &  --1.14 &  --0.01 \\
  Mn \textsc{I} &   3 &    4.10 &    0.05 &  --1.33 &  --0.20 \\
  Fe \textsc{I} &  62 &    6.37 &    0.12 &  --1.13 &    0.00 \\
 Fe \textsc{II} &  11 &    6.39 &    0.10 &  --1.11 &    0.02 \\
  Co \textsc{I} &   3 &    3.88 &    0.06 &  --1.11 &    0.02 \\
  Ni \textsc{I} &   7 &    5.11 &    0.07 &  --1.11 &    0.02 \\
  Cu \textsc{I} &   1 &    2.96 & \nodata &  --1.23 &  --0.10 \\
  Zn \textsc{I} &   2 &    3.48 &    0.15 &  --1.08 &    0.05 \\
 Sr \textsc{II} &   1 &    1.74 & \nodata &  --1.13 &    0.00 \\
  Y \textsc{II} &   4 &    1.26 &    0.26 &  --0.95 &    0.18 \\
  Zr \textsc{I} &   3 &    1.60 &    0.05 &  --0.98 &    0.16 \\
 Zr \textsc{II} &   0 & \nodata & \nodata & \nodata & \nodata \\
 Ba \textsc{II} &   2 &    0.95 &    0.10 &  --1.23 &  --0.10 \\
 La \textsc{II} &   1 &    0.07 & \nodata &  --1.03 &    0.10 \\
 Ce \textsc{II} &   2 &    0.09 &    0.02 &  --1.49 &  --0.36 \\
 Nd \textsc{II} &   7 &    0.58 &    0.21 &  --0.84 &    0.29 \\
 Eu \textsc{II} &   1 &  --0.41 & \nodata &  --0.93 &    0.20 \\
 \hline
\end{longtable}
\twocolumn

\begin{figure}
	\includegraphics[width=0.98\columnwidth]{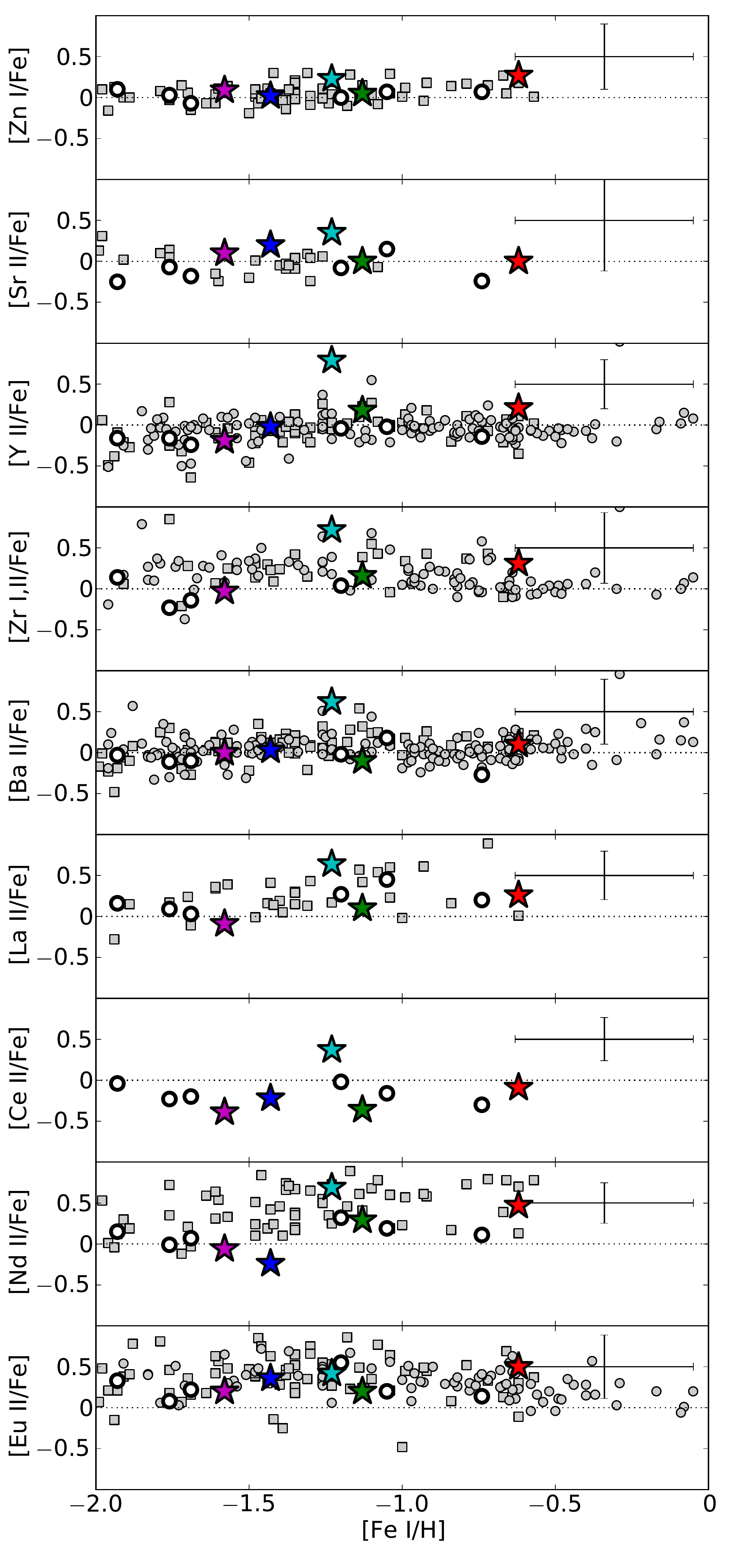}
	\caption{Element ratios for Aquarius stream stars. In the case of [Zr/Fe], {[Zr\,\textsc{i}/Fe]} is taken where available and {[Zr\,\textsc{ii}/Fe]} if no measurement was available for {Zr\,\textsc{i}}. See Table \ref{tab:program-star-abundances} for details. Standard stars are shown as open circles. Mean conservative \emph{total uncertainties} (random and systematic) for this study are shown in each panel. Filled circles and squares represent Milky Way field stars from \citet{fulbright_2000} and \citet{ishigaki;et-al_2013}, respectively. Panels have varying $y$-axis ranges to accommodate the data.}
	\label{fig:x-on-fe}
\end{figure}

\subsection{Uncertainties in Chemical Abundances}
\label{sec:chemical-abundance-uncertainties}

The uncertainties in chemical abundances are primarily driven by systematic uncertainties in stellar parameters, with a small contribution of random measurement scatter from individual lines. In order to calculate the abundance uncertainties due to stellar parameters, we have independently varied the stellar parameters by the adopted uncertainties, and measured the resultant change in chemical abundances. For lines requiring synthesis due to hyperfine structure, the difference in chemical abundances has been calculated from EWs. However, the effect of wing broadening due to hyperfine or isotopic splitting was generally small.

The individual abundance errors from varying each of the stellar parameters were added in quadrature to obtain the systematic error. To obtain the total error, we added in quadrature the systematic error with the standard error of the mean (random error). In some cases, the standard error about the mean is unrealistically small. As discussed earlier in Section \ref{sec:sodium-abundances}, we have conservatively adopted an abundance floor of 0.10\,dex for the standard deviation (i.e., $\max(0.10, \sigma(\log_{\epsilon}X)$). These resultant changes in abundances and total uncertainties are listed for all stars in Table \ref{tab:chemical-abundance-uncertainties}. These total uncertainties have been used in all figures. This provides us with an uncertainty for all abundances, in all stars, of [X/H]. Generally though, we are most interested in the uncertainty in [X/Fe]. In order to calculate this uncertainty, the correlations in uncertainties due to stellar parameters between (X, Fe) need to be considered. We have followed the description in \citet{johnson_2002} to calculate these correlations, and the overall uncertainties in [X/Fe], which are listed in Table \ref{tab:chemical-abundance-uncertainties} for all program and standard stars.

\begin{table*}
\caption{Abundance Uncertainties Due to Errors in Stellar Parameters\label{tab:chemical-abundance-uncertainties}}
\resizebox{\textwidth}{!}{%
\begin{tabular}{lrrrccc}
\hline
\hline
& & & & & \multicolumn{2}{c}{\textbf{Total Uncertainty}} \\
 \cline{6-7}
 Species &  $T_{\rm eff} +125\,{\rm K}$ & $\log{g} +0.20\,{\rm dex}$ & ${\xi_t} +0.30$\,km s$^{-1}$ & $\max(0.10, \sigma)/\sqrt(N)$ & [X/H] & [X/Fe] \\
 & $\Delta$abund. & $\Delta$abund. & $\Delta$abund. & (dex) & (dex) & (dex) \\
\hline
\\
%		& Teff+125		& logg+0.20		& vt+0.30		& Max(0.1,sigma)/sqrt(N)		& Total/H		& 
\multicolumn{7}{c}{\textbf{HD 41667}} \\
\hline
O I  		& 0.03		& 0.08		&$-$0.01		& 0.07		& 0.11		& 0.16 \\
Na I  		& 0.13		& 0.00		&$-$0.02		& 0.10		& 0.16		& 0.21 \\
Mg I  		& 0.08		& 0.01		&$-$0.01		& 0.05		& 0.10		& 0.18 \\
Al I  		& 0.10		& 0.01		&$-$0.01		& 0.06		& 0.11		& 0.18 \\
Si I  		& 0.02		& 0.03		&$-$0.02		& 0.04		& 0.06		& 0.16 \\
K I  		& 0.14		&$-$0.03	&$-$0.17		& 0.10		& 0.24		& 0.28 \\
Ca I  		& 0.13		&$-$0.01	&$-$0.10		& 0.05		& 0.17		& 0.22 \\
Sc II  		&$-$0.03	& 0.08		&$-$0.08		& 0.05		& 0.13		& 0.18 \\
Ti I  		& 0.22		& 0.01		&$-$0.01		& 0.05		& 0.22		& 0.25 \\
Ti II  		&$-$0.04	& 0.07		&$-$0.15		& 0.14		& 0.23		& 0.27 \\
V I  		& 0.25		& 0.01		&$-$0.03		& 0.05		& 0.26		& 0.28 \\
Cr I  		& 0.23		& 0.00		&$-$0.20		& 0.03		& 0.31		& 0.33 \\
Cr II  		&$-$0.07	& 0.08		&$-$0.06		& 0.07		& 0.14		& 0.20 \\
Mn I  		& 0.17		& 0.01		&$-$0.07		& 0.06		& 0.19		& 0.23 \\
Fe I  		& 0.16		& 0.01		&$-$0.07		& 0.02		& 0.17		& \nodata \\
Fe II  		&$-$0.10	& 0.08		&$-$0.04		& 0.03		& 0.14		& \nodata \\
Co I  		& 0.18		& 0.03		&$-$0.01		& 0.04		& 0.19		& 0.22 \\
Ni I  		& 0.12		& 0.03		&$-$0.01		& 0.05		& 0.13		& 0.18 \\
Cu I  		& 0.19		& 0.03		&$-$0.16		& 0.10		& 0.27		& 0.29 \\
Zn I  		&$-$0.03	& 0.06		&$-$0.09		& 0.07		& 0.14		& 0.20 \\
Sr II  		& 0.24		& 0.01		&$-$0.06		& 0.10		& 0.27		& 0.29 \\
Y II  		& 0.00		& 0.08		&$-$0.09		& 0.09		& 0.15		& 0.19 \\
Zr I  		& 0.27		& 0.01		& 0.00			& 0.07		& 0.28		& 0.30 \\
Zr II  		& 0.00		& 0.08		&$-$0.01		& 0.10		& 0.13		& 0.18 \\
Ba II  		& 0.01		& 0.06		&$-$0.21		& 0.07		& 0.23		& 0.26 \\
La II  		& 0.01		& 0.07		&$-$0.02		& 0.07		& 0.10		& 0.16 \\
Ce II  		& 0.04		& 0.08		&$-$0.04		& 0.09		& 0.13		& 0.17 \\
Nd II  		& 0.02		& 0.06		&$-$0.07		& 0.03		& 0.10		& 0.16 \\
Eu II  		&$-$0.05	& 0.04		&$-$0.06		& 0.10		& 0.13		& 0.20 \\
\hline
\end{tabular}}
\parbox{\textwidth}{Table \ref{tab:chemical-abundance-uncertainties} is published for all standard and program stars in the electronic edition. A portion is shown here for guidance regarding its form and content.}
\end{table*}

\section{Discussion}
\label{sec:discussion}

\subsection{Stellar Parameter Discrepancies with \\ \citet{wylie-de-boer;et-al_2012}}
We now seek to investigate the nature of the Aquarius stream and in particular, the globular cluster origin suggested by \citet{wylie-de-boer;et-al_2012}. Before proceeding, we now compare our stellar parameters with those of \citet{wylie-de-boer;et-al_2012} for the four stars in common. \citet{wylie-de-boer;et-al_2012} deduce their stellar parameters ($T_{\rm eff}$, $\log{g}$, $\xi_t$, [M/H]) by minimizing the $\chi^2$ difference between the observed spectra and synthetic spectra from the \citet{munari;et-al_2005} spectral library. For the four stars common to both samples, the stellar parameters reported in \citet{wylie-de-boer;et-al_2012} differ from our values listed in Table \ref{tab:stellar-parameters}.  In general, effective temperatures between the two studies agree within the uncertainties. The only aberration is {J223811-104126}, where we find an effective temperature of {5190\,K}, $\sim$450\,K cooler than the {5646\,K} found by \citet{wylie-de-boer;et-al_2012}. Similarly, \citet{williams;et-al_2011} report a hotter effective temperature of {5502\,K} from low-resolution spectra. This is the largest discrepancy we find in any of our standard or program stars.

Photometric temperature relationships support our spectroscopic temperature for J223811-104126. The \citet{ramirez;melendez_2005} relationship for giants suggests an effective temperature of 5240\,K, which is 50\,K warmer than our spectroscopically-derived temperature. Furthermore, the metallicity-independent $J-K$ colour-$T_{\rm eff}$ relationship for giants by \citet{alonso;et-al_1999} yields an effective temperature of 5215\,K, 25\,K warmer than our spectroscopic temperature. As a test, we set the temperature for {J223811-104126} to be 5600\,K -- within the temperature regime reported by \citet{williams;et-al_2011} and \citet{wylie-de-boer;et-al_2012}. The slopes and offsets in abundance with excitation potential and REW were large: ${m_{\rm Fe\,\textsc{i}} = -0.099}$\,dex\,eV$^{-1}$, 0.162\,dex, ${m_{\rm Fe\,\textsc{ii}} = -0.133}$\,dex\,eV$^{-1}$, $-0.033$\,dex respectively, and in doing so we could not find a representative solution for this temperature. 

\citet{williams;et-al_2011} and \citet{wylie-de-boer;et-al_2012} find {J223811-104126} to be a sub-giant/dwarf, with a surface gravity {$\log{g} = 4.16$} and {4.60} respectively. We note that the \citet{williams;et-al_2011} and \citet{wylie-de-boer;et-al_2012} effective temperatures for J223811-104126 are 150-300\,K hotter than the \citet{casagrande;et-al_2010} $J-K$ photometric temperature calibration for dwarfs and sub-giants. We find the surface gravity for {J223811-104126} to be {$\log{g} = 2.93 \pm 0.30$\,dex}, placing this star at the base of the red giant branch.

With the exception of {J223811-104126}, our surface gravities are largely in agreement with \citet{wylie-de-boer;et-al_2012}. The only other noteworthy difference is for {J221821-183424}, where we find a lower gravity of {$\log{g} = 0.88 \pm 0.30$} and \citet{wylie-de-boer;et-al_2012} find {$\log{g} = 1.45 \pm 0.35$}. Given the difference in the $S/N$  between these studies, this difference is not too concerning. \citet{wylie-de-boer;et-al_2012} calculate the microturbulence from empirical relationships derived by \citet{reddy;et-al_2003} for dwarfs and \citet{fulbright_2000} for giants. These relationships are based on the effective temperature and surface gravity. Our published microturbulent velocities agree excellently with the values presented in \citet{wylie-de-boer;et-al_2012}, again with the exception of {J223811-104126}, where the difference in $v_{t}$ is directly attributable to the offsets in other observables.

Of all the stellar parameters, metallicities exhibit the largest discrepancy between the two studies. In the \citet{wylie-de-boer;et-al_2012} study, after the stellar parameters ($T_{\rm eff}$, $\log{g}$, $v_{t}$, and an initial [M/H] estimate) were determined through a $\chi^2$ minimisation, the authors synthesised individual {Fe\,\textsc{i}} and Fe\,\textsc{ii} lines using \textsc{moog}. \citet{castelli;kurucz_2003} stellar atmosphere models were employed (K. Freeman, private communication, 2013) -- the same ones used in this study -- albeit the interpolation schemes will have subtle differences. The median abundance of synthesised {Fe\,\textsc{i}} lines was adopted as the overall stellar metallicity, and scaled relative to the Sun using the \citet{grevesse;sauval_1998} Solar composition.

%This offset translates into a factor of at least 15 electrons per pixel.
The study of \citet{wylie-de-boer;et-al_2012} is of slightly lower resolution ($\mathcal{R} = 25,000$ compared to $\mathcal{R} = 28,000$ presented here), but with a much lower $S/N$ ratio: ${\sim{}25\,{\rm pixel}^{-1}}$ compared to {$>$100\,pixel$^{-1}$} achieved here. The line list employed in the \citet{wylie-de-boer;et-al_2012} utilized astrophysical oscillator derived from a reverse solar analysis on the Solar spectrum. However, there are very few transitions listed in their line list: a maximum of 14 {Fe\,\textsc{i}} lines and 3 {Fe\,\textsc{ii}} lines were available. For contrast, our analysis is based on 63 {Fe\,\textsc{i}} and 13 {Fe\,\textsc{ii}} clean, unblended lines. 

We first suspected that the discrepancy in derived metallicities could be primarily attributed to the difference in line lists. In order to test this hypothesis, we re-analysed our data using the \citet{wylie-de-boer;et-al_2012} line list and their stellar parameters. Excitation or ionization equilibria could not be achieved using any stellar parameters from \citet{wylie-de-boer;et-al_2012} within their quoted uncertainties. We find different metallicities for each star in common with differences up to 0.42\,dex. Alternatively, if the \citet{wylie-de-boer;et-al_2012} line list is employed and we solve for stellar parameters (see Section \ref{sec:stellar-parameter-derivation}), we obtain stellar parameters closer to \emph{our} existing measurements tabulated in Table \ref{tab:stellar-parameters}, which are also distinct from the \citet{wylie-de-boer;et-al_2012} values. From the four stars common between these studies, using the \citet{wylie-de-boer;et-al_2012} line list and our stellar parameters\footnote{Here we are referring to a separate test in which we found the stellar parameters by excitation and ionization equilibria using the \citet{wylie-de-boer;et-al_2012} line list, \emph{not} those listed in Table \ref{tab:stellar-parameters}.}, we observe a metallicity dispersion of $\sigma$([Fe/H]) = 0.32\,dex. This is contrast to the $\sigma$([Fe/H]) = 0.10\,dex reported by \citet{wylie-de-boer;et-al_2012} from six Aquarius stream stars. The discrepancy might be explainable by different methods of determining stellar parameters (e.g. $\chi^2$-minimisation compared to excitation and ionization equilibria), as well as the difference in quality of the spectra between these two studies.

We note that given the small numbers of Fe lines used by \citet{wylie-de-boer;et-al_2012}, even subtle changes to the stellar parameters produced large variations to both the individual and mean Fe abundances. Furthermore, one {Fe\,\textsc{i}} transition at $\lambda$6420 in the \citet{wylie-de-boer;et-al_2012} line list was either not detected at the {3$\sigma$} level -- even though the $S/N$ at this point exceeds {115\,pixel$^{-1}$} in every observation -- or it was blended with a stronger neighbouring transition.

\subsection{The Aquarius Stream Metallicity Distribution}

Given the overall data quality and the limited Fe line list used for analysis, it appears the Aquarius stream stars conspired to present a tight metallicity distribution of {$\sigma(\mbox{[Fe/H]}) = 0.10$\,dex} in the \citet{wylie-de-boer;et-al_2012} analysis (in our analysis we find {$\sigma(\mbox{[Fe/H]}) = 0.33$\,dex}). When viewed in light of enhanced [Ni/Fe] and [Na/Fe] abundance ratios, \citet{wylie-de-boer;et-al_2012} interpreted this chemistry as a signature of a globular cluster origin for the Aquarius stream. Our study of high-resolution spectra with high $S/N$ reveals a much broader metallicity distribution for the stream than previously reported. With just 5 stars we find the metallicity varies from {$\mbox{[Fe/H]} = -0.63$} to {$-1.58$}. Although this is a small sample, we find the mean abundance and standard deviation to be {$\mbox{[Fe/H]} = -1.20 \pm 0.33$}. 

If the metallicity dispersion were smaller, as found by \citet{wylie-de-boer;et-al_2012}, a globular cluster scenario may be plausible. Classical globular clusters typically exhibit very little dispersion in metallicity. An intrinsic [Fe/H] dispersion of 0.33\,dex -- ignoring error contribution -- is substantially larger than that seen in any globular cluster, with the exception of the unusual system $\omega$-Centauri. In that cluster the total abundance range is about $\Delta$[Fe/H] $\sim$ 1.4\,dex: from $-$2 to $-$0.6 \citep[e.g.][]{marino;et-al_2011}, and many sub-populations have been identified \citep[e.g.,][]{johnson_pilachowski_2010}. 

Other clusters with established intrinsic [Fe/H] dispersions include M54 -- a nuclear star cluster of the Sagittarius dSph -- where $\sigma_{\rm int}$([Fe/H])$ = 0.19$ \citep{carretta;et-al_2010}, and M22, where the interquartile range in [Fe/H] is $\sim{}$0.24 dex \citep{da_costa;et-al_2009,marino;et-al_2009,marino;et-al_2011}. There are a few clusters where the intrinsic dispersion is $\sim$0.10, namely NGC 1851 \citep{carretta;et-al_2011}, NGC 5824 \citep{da_costa;et-al_2014}, and NGC 3201 \citep{simmerer;et-al_2013}. These globular clusters are outliers, and even amongst these unusual systems they largely do not match the abundance spread observed in the Aquarius stream \citep[e.g., see Figure 4 in][]{simmerer;et-al_2013}. In fact, the Aquarius stream metallicity distribution -- on its own -- is large enough to be reconcilable with dSph galaxies like Fornax \citep[e.g., ][]{letarte;et-al_2010}. Similarly, the mean Aquarius stream metallicity and the $\log(L)$, $\langle$[Fe/H]$\rangle$ relation of \citet{kirby;et-al_2011} also suggest a relatively luminous system with $L_{\rm tot} \sim 10^{7.5}L_\odot$ \citep{kirby;et-al_2011}.  However, the Aquarius stream stars exhibit very different abundance ratios to Fornax. For example, [Ba/Y] (e.g., heavy/light $s$-process) abundance ratios in the Aquarius stream vary between $-$0.24 and +0.19, significantly lower than the [Ba/Y] $\geq$ 0.5 level generally observed in the present day dSphs \citep{venn;et-al_2004}.

\subsection{The Na-O Relationship}

Extensive studies of stars in globular clusters have revealed variations in light element abundances, most notably an anti-correlation in sodium and oxygen content \citep[see][and references therein]{norris;da_costa_1995,carretta;et-al_2009a}. This chemical pattern has been identified in every well-studied globular cluster, although the magnitude and shape of the anti-correlation varies from cluster to cluster. The direct connection between Na and O abundances requires an additional synthesis mechanism for Na, at least for the Na content that exceeds the Na in the primordial population.

\begin{figure*}
	\includegraphics[width=\textwidth]{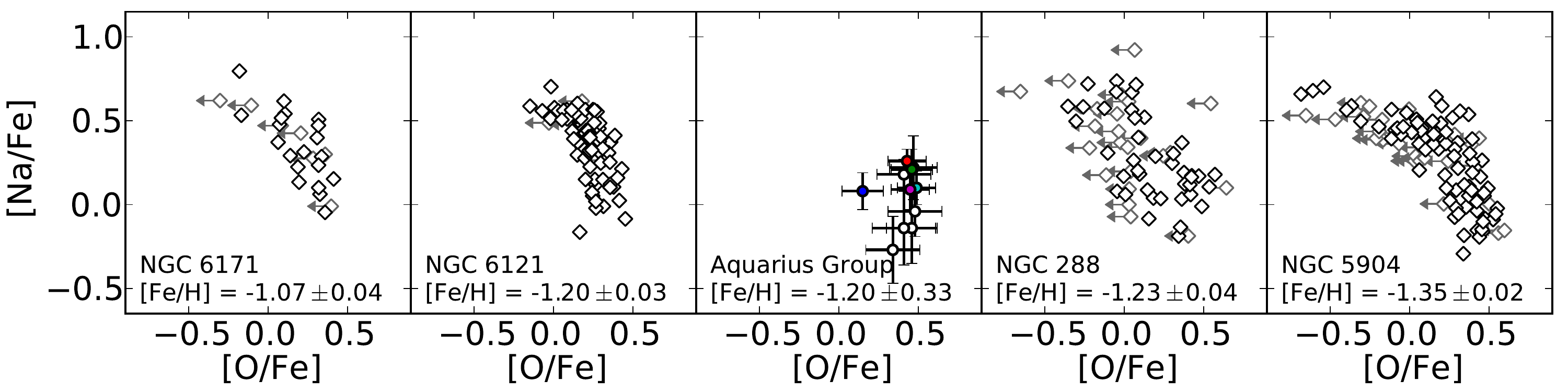}
	\caption{Oxygen and sodium abundances for 4 classical globular clusters with mean metallicities similar to the Aquarius stream \citep{carretta;et-al_2009a}. All clusters demonstrate a {Na-O} anti-correlation. [O/Fe] and [Na/Fe] abundances from this study for the 5 Aquarius stream stars and standard stars (open circles) are shown in the middle panel. Colors for Aquarius stars is as per Figure \ref{fig:toombre}.}
	\label{fig:o-na-clusters}
\end{figure*}

Sodium is primarily produced through carbon burning in massive stars by the dominant $^{12}\mbox{C}(^{12}\mbox{C}, p)^{23}\mbox{Na}$ reaction. The final Na abundance is dependent on the neutron excess of the star, which slowly increases during carbon burning due to weak interactions \citep{arnett;truran_1969}. Massive stars ($>10 M_\odot$) deliver their synthesized sodium to the interstellar-medium through {SN\,\textsc{ii}} explosions. Because the eventual {SN\,\textsc{ii}} explosion is devoid of any significant $\beta$-decay processes, the neutron excess of the ejected material is representative of the pre-explosive abundance. The ejected material eventually condenses to form the next generation of stars, which will have a net increase in their neutron excess with respect to their predecessors. Thus an overall increase in the total sodium content \textit{and} Na-production rate between stellar generations can be expected. The sodium content also becomes important for production of nickel during the {SN\,\textsc{ii}} event (see Section \ref{sec:na-ni-relationship}) because $^{23}$Na is the only stable isotope produced in significant quantities during C-burning.

Oxygen depletion is likely the result of complete CNO burning within the stellar interior. The nucleosynthetic pathways that produce the {Na-O} anti-correlation are well understood to be proton-capture nucleosynthesis at high temperatures \citep{prantzos;et-al_2007}. However, the temperatures required to produce these patterns are not expected within the interiors of globular cluster stars. While the exact mechanism for which these conditions occur remains under investigation, we can describe the abundance variation as an external oxygen depletion (or dilution) model with time. Through comparisons with existing globular clusters, we can make inferences on the star-formation history of a system by measuring sodium and oxygen abundances in a sample of its stars. 

\begin{figure}
	\includegraphics[width=\columnwidth]{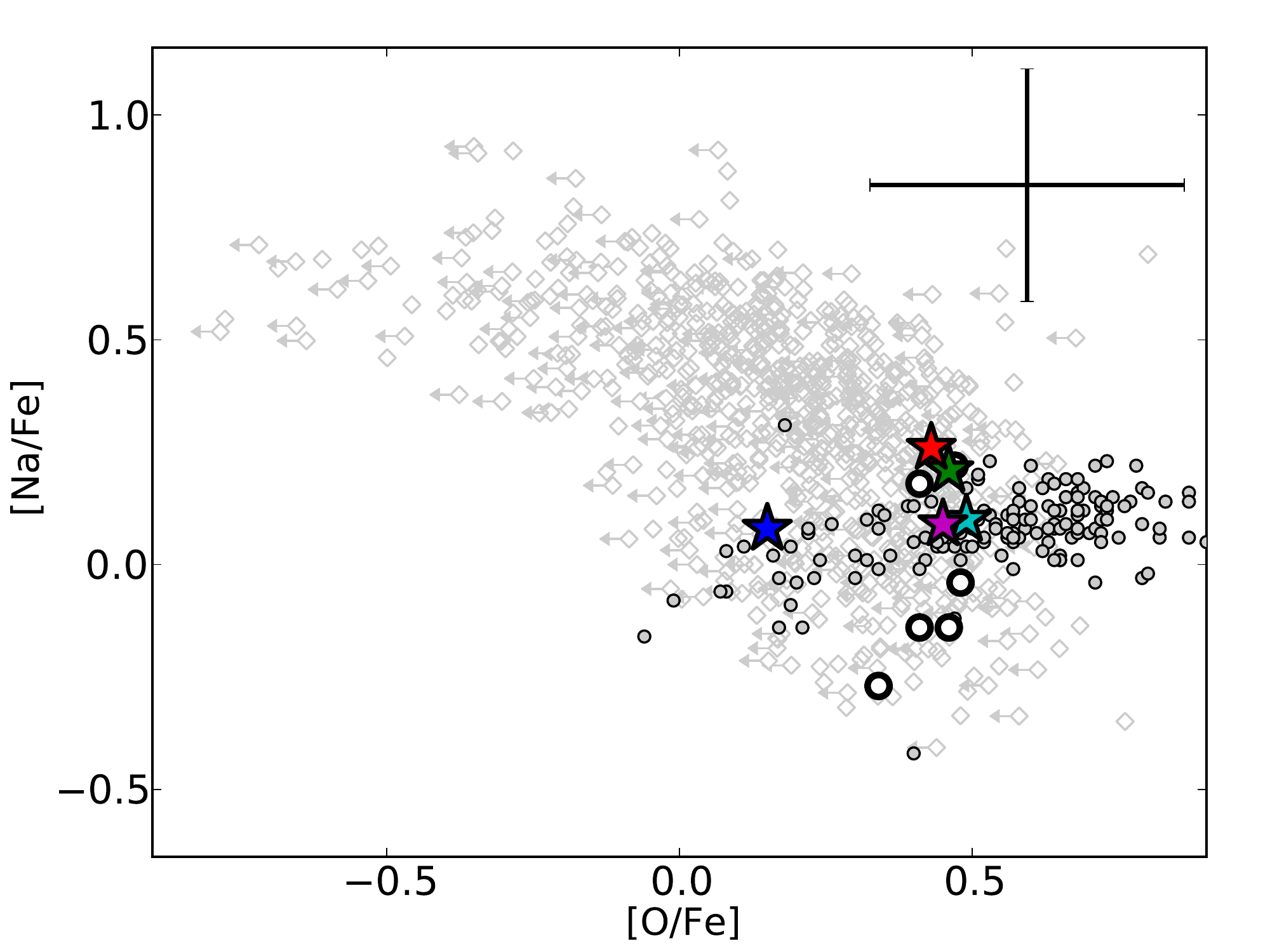}
	\caption{Oxygen and sodium abundances for disk/halo stars from \citet{reddy;et-al_2006} are shown as grey circles, and globular cluster stars from \citet{carretta;et-al_2009a} are shown as diamonds. The Aquarius stream stars are also shown -- following the same colors in Figure \ref{fig:toombre} -- illustrating how their [O/Fe], [Na/Fe] content is not dissimilar from Galactic stars. Although some standard stars (open circles) have lower [Na/Fe] abundances than the \citet{reddy;et-al_2006} sample, our values are consistent with \citet{ishigaki;et-al_2013}, who did not measure O abundances.}
	\label{fig:o-na}
\end{figure}

\citet{wylie-de-boer;et-al_2012} measured sodium and oxygen abundances for four of their six Aquarius stream members. These abundance measurements exist for only three stars common to this study and \citet{wylie-de-boer;et-al_2012}, as the data quality for {J223811-104126} in the \citet{wylie-de-boer;et-al_2012} was too low to permit oxygen measurements. We have measured sodium and oxygen abundances for all of our stars, which are plotted in Figures \ref{fig:o-na-clusters} and \ref{fig:o-na}\footnote{Although the \citet{reddy;et-al_2006} sample primarily consists of dwarfs/subgiants and we are observing primarily giants/subgiants, this does not affect our interpretation.}. These figures employ the corrected [O/Fe] value for {J223811-104126} rather than a conservative upper limit (see Section \ref{sec:alpha-elements}).

The \citet{wylie-de-boer;et-al_2012} measurements show two stars with solar levels of [Na/Fe] -- identical to field star abundances for their metallicity -- and two stars with slightly enhanced sodium content: {J223504-152834} and {J232619-080808}. We also find {J223504-152834} to be slightly sodium-enhanced, whereas the second star in their study, {J232619-080808}, is not in our sample. We find the additional star not present in the \citet{wylie-de-boer;et-al_2012} sample, {C2306265-085103}, to be enhanced to almost the same level of {J223504-152834} with {$\mbox{[Na/Fe]} = 0.26$}. The sodium-enhanced stars are not enhanced significantly above the total uncertainties, and they do not exhibit depletion of oxygen: their chemistry is not representative of a {Na-O} anti-correlation. 

In the Aquarius sample we observe no intrinsic dispersion above the measurement uncertainty in [O/Fe] or [Na/Fe] (Figure \ref{fig:o-na}). A dispersion of 0.14\,dex is observed for [O/Fe] (or 0.03\,dex when J223811-104126 is excluded), which is only marginally larger than the mean total uncertainty of $\sigma({\rm [O/Fe]}) = 0.12$\,dex. Similarly for [Na/Fe], a dispersion of $\sigma({\rm [Na/Fe]}) = 0.08$\,dex is observed, when taking the uncertainties into account, is consistent with zero dispersion. We also see no significant variation in [C/Fe] outside the uncertainties, or relationship between [C/Fe]-[Na/Fe] \citep[e.g., see][]{yong}.

Now we consider the possibility that the Aquarius stars did originate in a globular cluster. Given the negligible dispersions present in [(O, Na, C)/Fe] (among other elements), the stars would be considered as members the primordial component, which comprises $\sim$33\% of the total population for any globular cluster \citep{carretta;et-al_2009a}. The likelihood of randomly observing five globular cluster stars that all belong to the primordial component is 0.4\%. If the primordial component was a larger fraction (e.g., 40\% or 50\%), this probability raises marginally, to 1\% and 3\% respectively. Given the dispersion in overall metallicity though, such a globular cluster would be an unusual object.

%Such inferences must be made with careful consideration. In addition to the normal care afforded for measuring elemental abundances from high-resolution spectroscopic data, attention must be given to telluric absorption, contamination from Ni\,\textsc{i}, as well as non-LTE and 3D effects when determining oxygen abundances. Furthermore, when characterising the oxygen depletion rate -- the strength of the {Na-O} anti-correlation in a globular cluster -- it is vital to sample, where possible, stars belonging to all three components \citep[primordial, intermediate and extreme, see][]{carretta;et-al_2009a}. 

%If only a primordial sample of stars is observed, their [Na/Fe] and [O/Fe] abundances will be, by definition, indistinguishable from field stars of a similar metallicity. In such a scenario any inferred anti-correlation could equally be explained by small abundance variations or observational uncertainties. 

If the Aquarius stream is the result of a disrupted globular cluster, a large part of the picture must still be missing. Almost all of the Aquarius stream stars studied to date (either in this sample or the \citet{wylie-de-boer;et-al_2012} study), would be unambiguously classified as belonging to a ``primordial'' component \citep[see][]{carretta;et-al_2009a}, with chemistry indistinguishable from field stars. In this scenario any inferred anti-correlation is equally explainable by observational uncertainties. Identifying more Aquarius stream members belonging to a postulated intermediate component with strong oxygen depletion, or perhaps members of an extreme component, would be convincing evidence for a {Na-O} anti-correlation and a globular cluster origin. Three stream stars identified to date (including two from this sample) might tenuously be classified as members of an intermediate population, with only a slight enhancement in sodium and no oxygen-depletion. Recall our [Na/Fe] abundance ratios appear systematically higher in our standard stars when compared to the literature sources listed in Table \ref{tab:stellar-parameters}. Thus, if the strength of any {Na-O} relationship is to be used to vet potential disrupted hosts for the Aquarius stream, many more stream members will need to be identified and observed spectroscopically with high-resolution and high $S/N$. In the absence of such data, no evidence exists for a {Na-O} anti-correlation in the Aquarius stream.

\subsection{The Al-Mg Relationship}

Although not ubiquitous to every system, many globular clusters exhibit an anti-correlation between aluminium and magnesium. This is perhaps unsurprising, given the nucleosynthetic pathways for these elements. In addition to the CNO cycle operating during hydrogen burning, the {Mg-Al} chain can also operate under extreme temperatures \citep[$T \sim 8 \times 10^{6}\,$K;][]{arnould;et-al_1999}. Aluminium is produced by proton-capture onto magnesium, beginning with $^{25}$Mg to $^{26}$Al. The relative lifetime of $\beta$-decay to proton-capture allows for the production of unstable $^{27}$Si through proton-capture. Seconds later, the isotope decays to $^{27}$Al, completing the $^{27}$Si path of the Mg-Al chain. The alternative process from $^{26}$Al involves $\beta$-decay to $^{26}$Mg.

%The Mg-Al cycle can explain the observed anti-correlation observed in some globular clusters, and some AGB models can predict such a pattern (e.g., \citet{ventura;et-al_2011} but see nucleosynthesis yields from \citealt{karakas;lattanzio_2007} and \citealt{karakas_2010}). However, at present the observed Mg isotope ratios are in tension with AGB model predictions \citep{yong;et-al_2006,da_costa;et-al_2013}. Thus, while the pathways for creating these chemical patterns are understood, the temperatures required to produce this chain are much higher than expected for low-mass stellar interiors, so the exact site and requisite conditions are lacking a full description.

\citet{wylie-de-boer;et-al_2012} published magnesium and aluminium abundances for five stars in their Aquarius sample. No inverse correlation is present in their data; their abundances are indistinguishable from field stars. The [(Mg, Al)/Fe] abundance ratios tabulated in Table \ref{tab:program-star-abundances} are generally in agreement with the \citet{wylie-de-boer;et-al_2012} sample, and we also find no {Mg-Al} anti-correlation. 

\begin{figure}
	\includegraphics[width=\columnwidth]{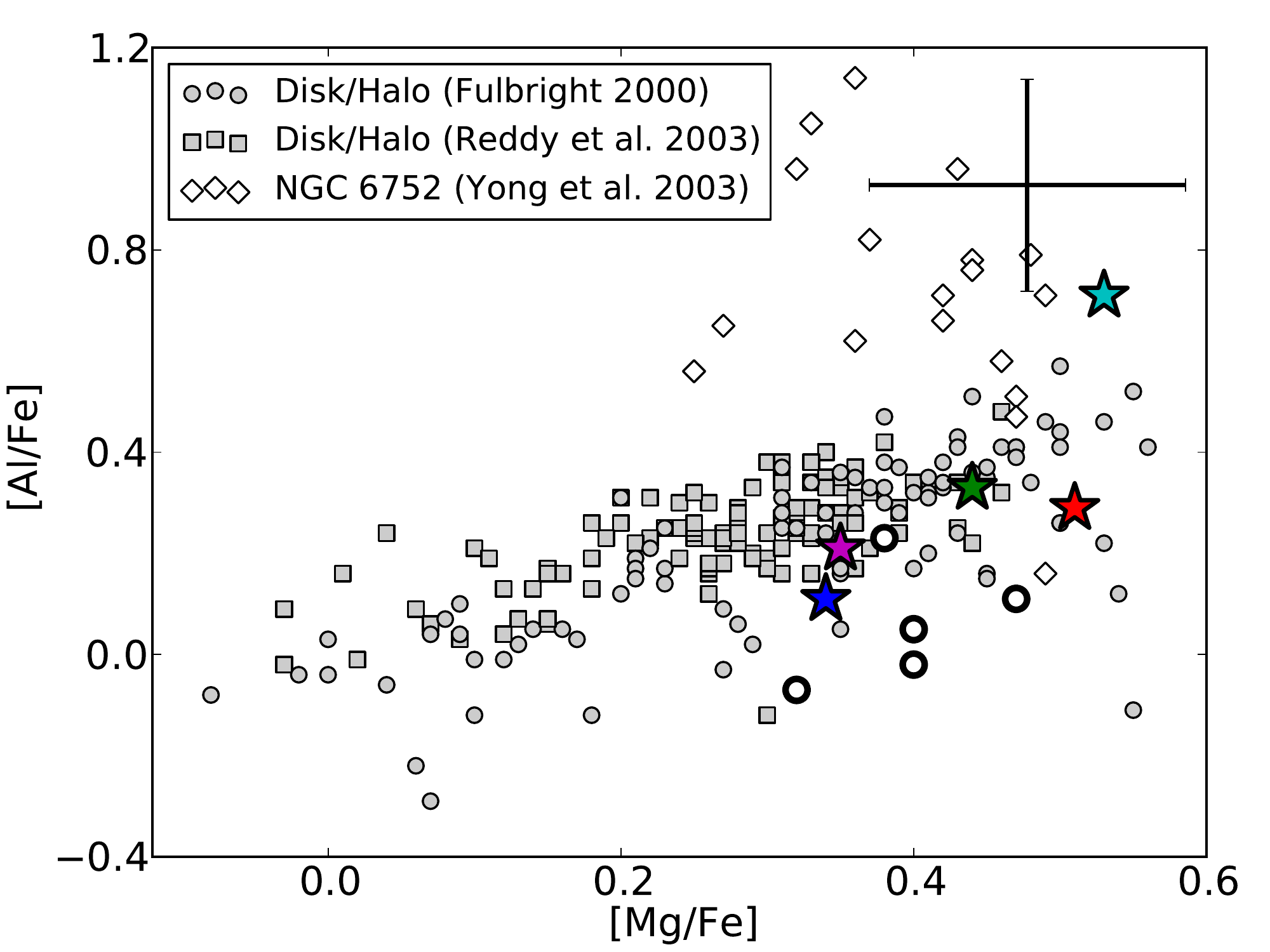}
	\caption{Magnesium and aluminium abundances for Aquarius stream stars, as well as Milky Way halo/disk stars from \citet{reddy;et-al_2003} and \citet{fulbright_2000}. Aquarius stars are colored as described in Figure \ref{fig:toombre}.}
	\label{fig:mg-al}
\end{figure}

However it {\it is} surprising that we find such a strong positive relationship in [Mg/Fe] and [Al/Fe], with a best-fitting slope of ${{\rm [Al/Fe]} = 2.08\times\mbox{[Mg/Fe]} - 0.57}$. If we exclude the chemically peculiar star {C222531-145437}, the slope decreases to {${\rm [Al/Fe]} = 0.96\times\mbox{[Mg/Fe]} - 0.16$}, a near 1:1 relationship. Even when a {Mg-Al} anti-correlation is not detected in globular clusters, there is generally more scatter in [Al/Fe] at near-constant [Mg/Fe] \citep[e.g., see Figure \ref{fig:mg-al} or ][]{carretta;et-al_2009a}. This is because Mg is much more abundant than Al, requiring only a small amount of Mg atoms to be synthesized to Al before the differences in Al abundance become appreciable, whilst the observed Mg abundance could remain within the uncertainties.

No classical globular clusters exhibit a positive correlation, and nor is such a pattern expected in globular clusters. However a positive relationship between magnesium and aluminium can result from {SN\,\textsc{ii}} contributions to the local interstellar medium. Intermediate-mass ($\gtrsim$4$M_\odot$) AGB models can also contribute towards a positive correlation between aluminium and magnesium. Under extreme temperatures ($T \gtrsim 300 \times 10^{6}\,$K), substantial $^{25}$Mg and $^{26}$Mg are produced by $\alpha$-capture onto $^{22}$Ne by the $^{22}$Ne($\alpha$,n)$^{25}$Mg and $^{22}$Ne($\alpha$,$\gamma$)$^{26}$Mg reactions respectively \citep[e.g.,][]{karakas;et-al_2006}. Depending on uncertain numerical details of stellar modelling , the third dredge-up can mix significant quantities of $^{25}$Mg and $^{26}$Mg into the photosphere, even more than the quantity of $^{26}$Al produced through the {Mg-Al} cycle. Therefore a positive relationship between magnesium and aluminium can occur if there has been significant contributions from intermediate-mass AGB stars, however this should also produce a Na-O anti-correlation \citep[][but see results in \citealt{ventura;et-al_2011}]{karakas;lattanzio_2003}. The strong {Mg-Al} relationship observed provides additional chemical evidence against a globular cluster scenario for the Aquarius stream, and further suggests the chemistry is indicative of Milky Way disk stars.
 
%An unrealistically high fraction of intermediate-mass AGB stars would be required to produce the same chemical signature of a single {SN\,\textsc{ii}} event. Given the efficiency of chemical mixing following supernovae, the observed positive {Mg-Al} correlation in the thick disk is likely the result of {SN\,\textsc{ii}} events. Distinguishing between these processes observationally requires careful measurements of magnesium isotope abundances $^{24}$Mg (indicating supernovae mixing), $^{25}$Mg and $^{26}$Mg (suggesting significant AGB contribution), which is not possible given our $S/N$ or spectral resolution. 

\subsection{The Na-Ni Relationship}
\label{sec:na-ni-relationship}

\begin{figure}
	\includegraphics[width=\columnwidth]{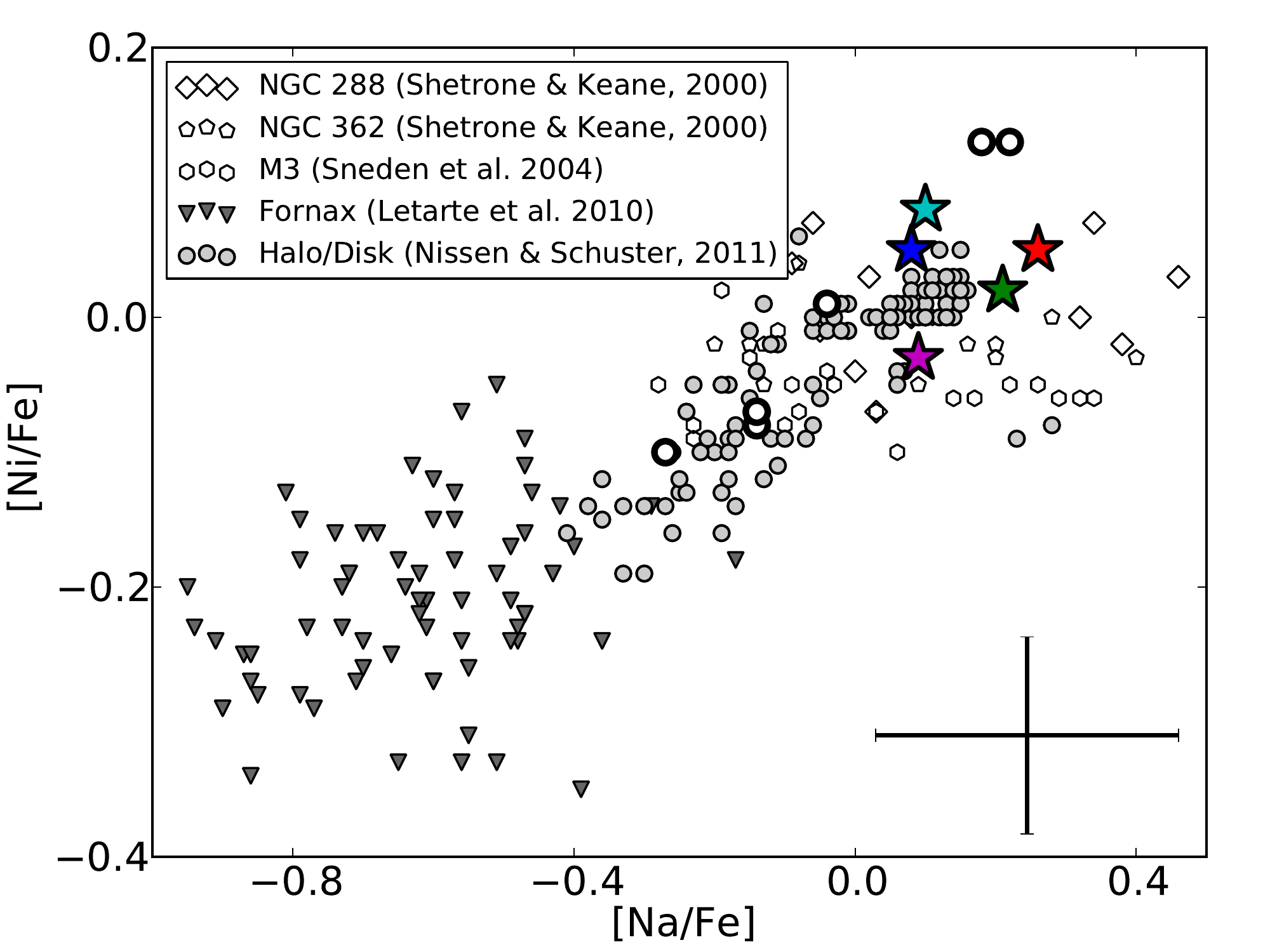}
	\caption{[Na/Fe] and [Ni/Fe] for Aquarius stream stars and for globular cluster, dSph, and field (halo/disk) stars. Aquarius targets are colored as per Figure \ref{fig:toombre}. Stars from the most representative dSph galaxy, Fornax, are shown as downward triangles ($\blacktriangledown$). Fornax has been chosen as it lies closest to the luminosity that one would expect for an Aquarius host system, given its overall metallicity, metallicity dispersion, and the $\log(L)-\langle$[Fe/H]$\rangle$ relationship \citep{kirby;et-al_2011}. The \citet{nissen;schuster_2011} plotted sample included halo stars, as well as low- and high-$\alpha$ disk members. As noted by \citet{nissen;schuster_1997,nissen;schuster_2011}, a positive Na-Ni relationship exists for field stars, with dSph members exhibiting strong depletion in both elements and globular cluster stars consistently showing an enhancement. Sodium and nickel content for Aquarius members indicate a dSph accretion origin is unlikely.}
	\label{fig:na-ni}
\end{figure}

Detailed chemical studies of nearby disk stars have noted a correlation with [Na/Fe] and [Ni/Fe] abundance ratios (Figure \ref{fig:na-ni}. This relationship was first hinted in \citet{nissen;schuster_1997}, where the authors found eight stars that were under-abundant in [$\alpha$/Fe], [Na/Fe] and [Ni/Fe]. Interestingly, the authors noted that stars at larger Galactocentric radii were most deficient in these elements. \citet{fulbright_2000} saw a similar signature: stars with low [Na/Fe] were only found at large {($R_{\rm GC} > 20$\,kpc)} distances. \citet{nissen;schuster_1997} proposed that since the outer halo is thought to have been largely built up by accretion, then the {Na-Ni} pattern may be a chemical indicator of merger history within the galaxy.

With additional data from \citet{nissen;schuster_2011}, the {Na-Ni} relationship was found to be slightly steeper than originally proposed. The pattern exists only for stars with {$-1.5 < \mbox{[Fe/H]} < -0.5$}, and is not seen in metal-poor dSph stars \citep{venn;et-al_2004}, providing a potentially useful indicator for investigating chemical evolution. However, it is crucial to note that although there are only a few dSph stars in the $-1.5 < \mbox{[Fe/H]} < -0.5$ metallicity regime with [Na/Fe] and [Ni/Fe] measurements, they agree reasonably well with the Galactic trend. 

The correlation between sodium and nickel content is the nucleosynthetic result of neutron-capture in massive stars. As previously discussed, the total Na abundance is controlled by the neutron excess, which limits the production of $^{58}$Ni during {SN\,\textsc{ii}} events. When the inevitable supernova begins, the core photodissociates into neutrons and protons, allowing the temporary creation of $^{56}$Ni before it decays to $^{56}$Fe. A limited amount of $^{54}$Fe is also formed, which is the main source of production for the stable $^{58}$Ni isotope through $\alpha$-capture. The quantity of $^{54}$Fe (and hence $^{58}$Ni) produced is dependent on the abundance of neutron-rich elements during the explosion. As $^{23}$Na is a relatively plentiful neutron source with respect to other potential sources (like $^{13}$C), the post-supernova $^{58}$Ni abundance is driven by the pre-explosion $^{23}$Na content. Thus, through populations of massive stars undergoing C-burning, a positive correlation between sodium and nickel can be expected. 

Stars originating in dSph galaxies and globular clusters have very different chemical enrichment environments. Consequently, both types of systems exhibit chemistry that reflects their nucleosynthetic antiquity. Stars in dSphs do not demonstrate enhanced sodium or nickel content with respect to iron, as there has been a relatively small lineage of massive stars undergoing supernova. In contrast, globular cluster stars do have elevated [Na/Fe] and [Ni/Fe] signatures. This sharp contrast between dSph and globular cluster star chemistry is highlighted in Figure \ref{fig:na-ni}. Given the extended star formation within the Milky Way disk, globular cluster stars and disk stars are indiscernible in the Na/Ni plane: they both show an extended contribution of massive stars. The most that can be inferred from the Na and Ni abundances of Aquarius stream stars is that their enrichment environment is less like a dSph galaxy, and more representative of either a globular cluster, or the Milky Way disk. 

\subsection{The Chemically Peculiar Star C222531-145437}

In almost every element with respect to iron, {C222531-145437} is distinct from the other Aquarius stream stars. It is over-abundant in light  and neutron-capture elements, with a high barium abundance of {$\mbox{[Ba/Fe]} = 0.68$}. This value is well in excess of the halo ({$\mbox{[Ba/Fe]} \sim 0.0$}) -- and our other Aquarius stream stars -- which vary between $-$0.02 to {0.15\,dex}.

\begin{figure}
	\includegraphics[width=\columnwidth]{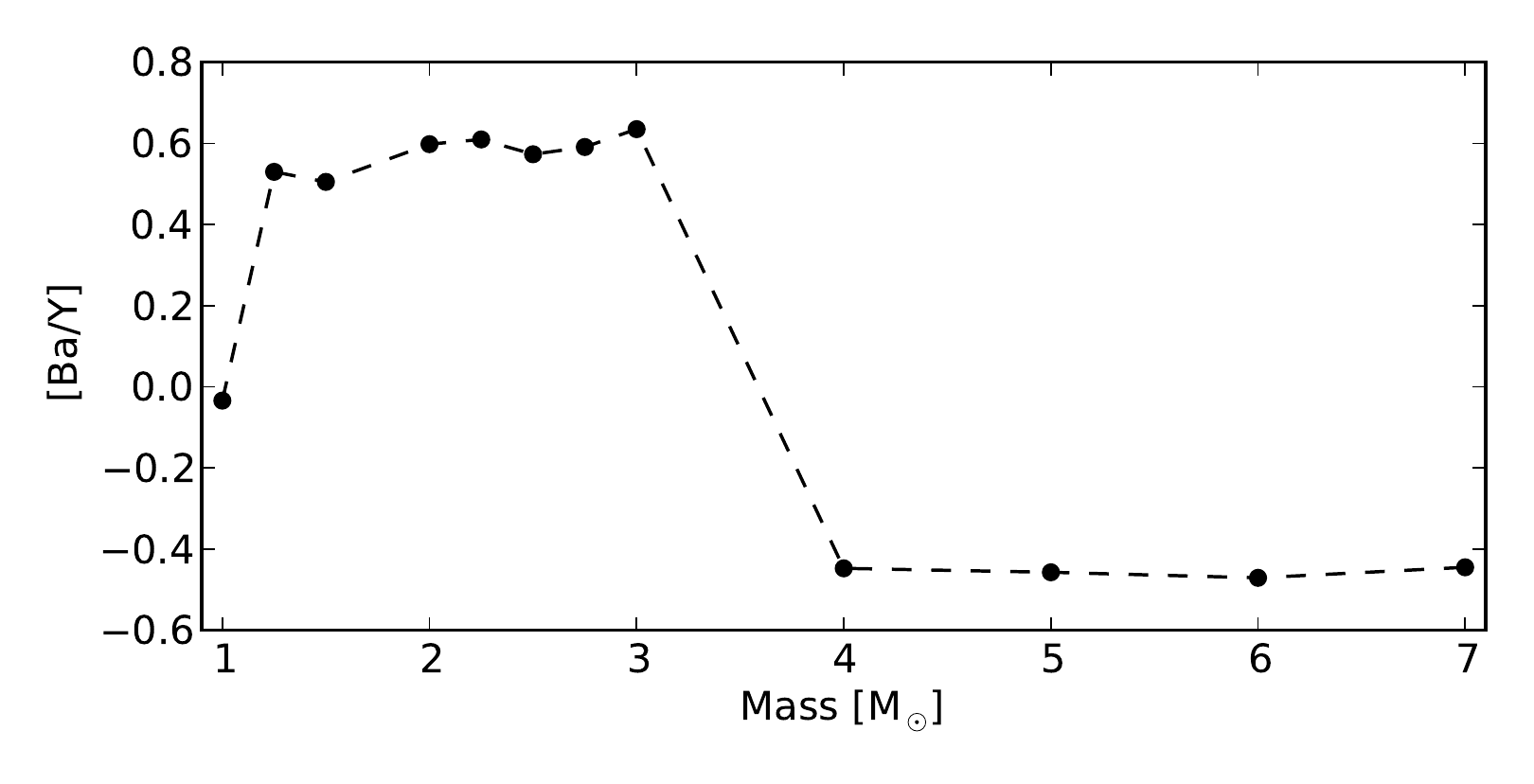}
	\caption{Distribution of the final surface abundance of [Ba/Y] with initial mass for each of the AGB models calculated for $Z~=~0.001$ (Fishlock et al., in preparation). The ratio of [Ba/Y] can be used as an indicator of the initial mass of the AGB companion where the low-mass AGB models show a higher [Ba/Y] ratio compared to the intermediate-mass AGB models.}
	\label{fig:ba-y}
\end{figure}

Here we discuss the possibility that an unseen companion has contributed to the surface abundances of C222531-145437. Although no radial velocity variations were observed between exposures, we do not have a sufficient baseline to detect such variation. The abundances of heavy elements produced by AGB stars have a high dependence on the initial metallicity and mass. Low-mass ($\lesssim{}3M_\odot$) AGB stars of low metallicity produce high fractions of heavy $s$-process elements compared to their light $s$-process counterparts \citep{busso;et-al_2001}. As such, [Ba/Y] is a useful indicator for considering contributions from a low-mass AGB companion. For C222531-145437, ${{\rm [Ba/Y]} = -0.17}$, which is much lower than expected if a low-mass AGB star was responsible for the heavy element enhancements ([Ba/Y] $\sim$ 0.5 as shown in Figure \ref{fig:ba-y}; see also \citealt{cristallo;et-al_2009}). If mass transfer from an AGB companion has occurred very recently, non-negligible amounts of technetium, produced by the AGB star, remain visible in the companion's photosphere before $^{99}$Tc decays over $\sim$2\,Myr \citep{brown;et-al_1990,van_eck_jorrissen_1999, uttenthaler;et-al_2011}. We saw no technetium absorption at $\lambda$4049, $\lambda$4238 or $\lambda$4297 in the spectrum of C222531-145437. Intermediate-mass (3-5$M_\odot$) AGB stars also cannot explain the abundances for C222531-145437: using recently computed intermediate-mass AGB $s$-process yields for 3-5$M_\odot$ for a star of [Fe/H] $\approx -1.2$ (Fishlock et al., in preparation) the resulting surface abundances do not match the observations (Figure \ref{fig:ba-y}). Therefore, we find no reason to suspect the heavy element enhancement in C222531-145437 is the result of mass transfer from an AGB companion.

Stars in $\omega$-Centauri show large over-abundances of $s$-process elements compared to the Galaxy \citep{norris;da_costa_1995,stanford;et-al_2010,johnson_pilachowski_2010}. M22 also hosts an $s$-process rich population \citep{marino;et-al_2011}. Like the Aquarius co-moving group, both clusters are relatively close to the Sun: {5.2\,kpc} and {3.2\,kpc}, respectively. M22 has a mean metallicity of {$\mbox{[Fe/H]} \sim -1.7$} and a range between {$-2.0 < \mbox{[Fe/H]} < -1.6$\,dex}, making an association between {C222531-145437} and M22 unlikely. Similarly, {C222531-145437} is unlikely to be associated with the metal-rich Argus association \citep[IC 2391;][]{de_silva;et-al_2013}, which also shows large enhancement in $s$-process abundances. Other groups have identified field stars enriched in $s$-process elements, which have generally been associated as tidal debris from $\omega$-Centuari \citep{wylie-de-boer;et-al_2010,majewski;et-al_2012}. The high $s$-process abundance ratios and overall metallicity of {C222531-145437} ({$\mbox{[Fe/H]} = -1.22$}) suggests this star may also be a remnant from the tidal disruption process. This is illustrated in Figure \ref{fig:omega-cen-barium}. In contrast to the comparison field stars, {C222531-145437} also demonstrates a high [Eu/Fe] abundance ratio of +0.42, which is also consistent with studies of $\omega$-Centauri stars \citep[e.g., see Figure 10 of ][]{johnson_pilachowski_2010}.

\begin{figure}
	\includegraphics[width=\columnwidth]{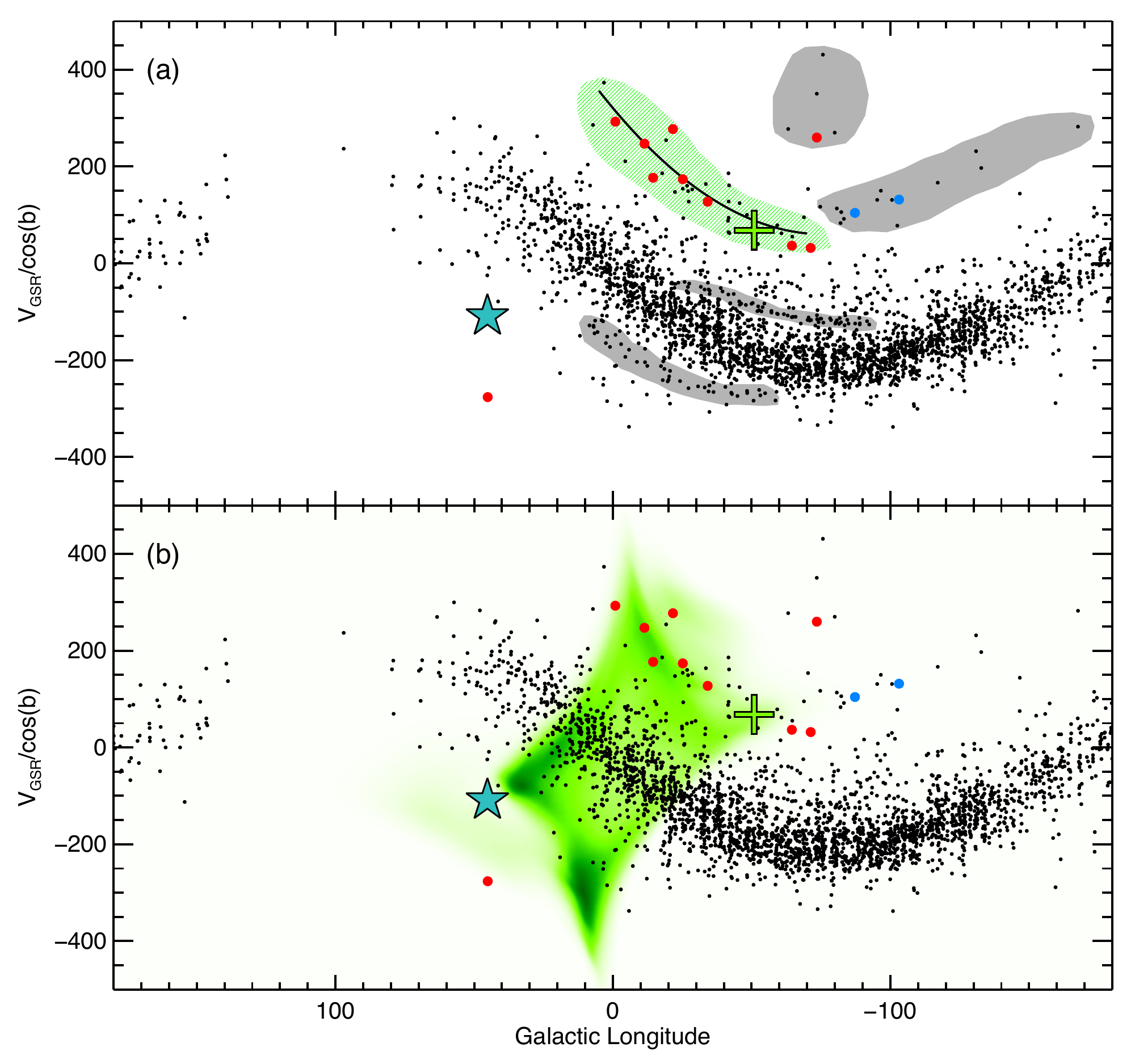}
	\caption{Panel (a) shows the distribution of giant stars in the GGSS \citep{majewski;et-al_2012} in Galactic longitude and $V_{\rm GSR}/cos(b)$ after excluding stars with $|b| > 60^\circ$. Stars from the GGSS sample believed to be $\omega$-Centauri tidal debris are shown in green shading. Red points are stars from the GGSS sample with abundances that follow the $\omega$-Centauri [Ba/Fe]--[Fe/H] pattern. Blue points are those with high-resolution spectra that do not follow this trend. Grey shading highlights other potential halo substructures from their study. Panel (b) shows the probability distribution of $\omega$-Centauri tidal debris from 4,050 simulations.  The $\omega$-CenCentauri core is shown as a green cross and the cyan star represents {C222531-145437}, falling almost precisely where a relatively high probability of $\omega$-Centauri tidal debris is expected.}
	\label{fig:omega-cen-tidal-debris}
\end{figure}

\begin{figure}
	\includegraphics[width=\columnwidth]{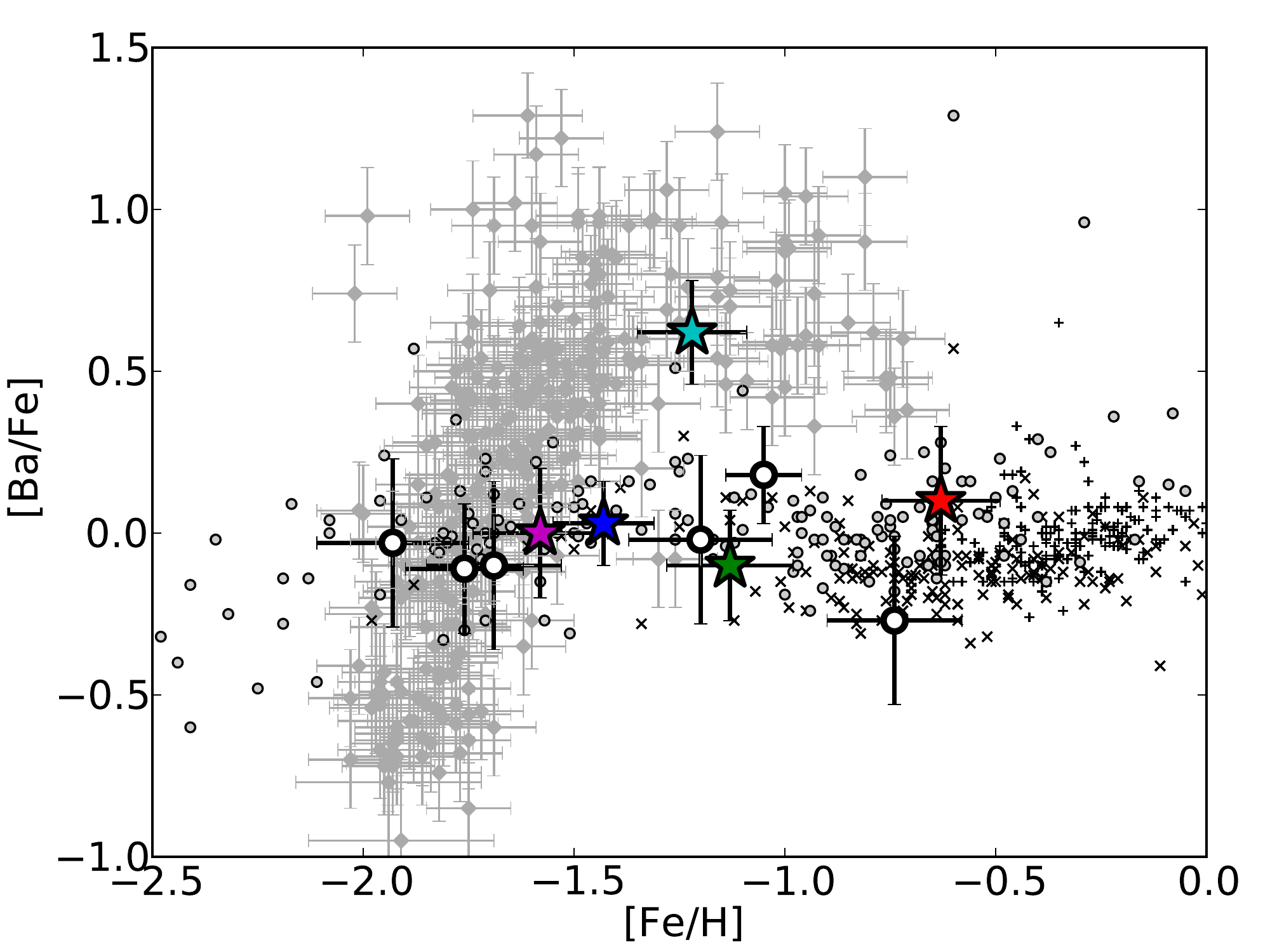}
	\caption{[Fe/H] and [Ba/Fe] for halo/disk stars (black) from \citet{fulbright_2000,reddy;et-al_2003,reddy;et-al_2006} and $\omega$-Centauri RGB stars (grey) from \citet{francois;et-al_1988,smith;et-al_2000,marino;et-al_2011}. Similar trends are observed for other heavy elements in $\omega$-Centauri members. Aquarius stars are colored as per Figure \ref{fig:toombre}.}
	\label{fig:omega-cen-barium}
\end{figure}

$\omega$-Centauri has a retrograde orbit with low inclination. Many groups simulating this orbit have predicted retrograde tidal debris to occur near the Solar circle \citep{dinescu_2002,tsuchiya_2003,tsuchiya_2004,bekki;freeman_2003}. Subsequent searches for $\omega$-Centauri debris in the Solar neighbourhood have led to tantalising signatures of debris. From over 4,000 stars targeted by \citet{da_costa;coleman_2008} in the vicinity of the cluster's tidal radius, only six candidate debris members were recovered, consistent with tidal stripping occurring long ago. Using data from the Grid Giant Star Survey (GGSS), an all-sky search looking for metal-poor giant stars, \citet{majewski;et-al_2012} identified 12 stream candidates. In addition, \citet{majewski;et-al_2012} performed 4,050 simulations in order to predict likely locations for $\omega$-Centauri tidal debris. The results of their simulation are replicated in Figure \ref{fig:omega-cen-tidal-debris}, where the location of {C222531-145437} is also shown. The velocity and position of {C222531-145437} align almost precisely where \citet{majewski;et-al_2012} predict a high probability of tidal debris. More interestingly, the angular momentum and orbital energy for {C222531-145437} (Figure \ref{fig:orbits}) matches excellently for $\omega$-Centauri cluster stars as well as its previously identified tidal remnants \citep{wylie-de-boer;et-al_2010}. The chemical and phase-space information strongly suggests that {C222531-145437} is associated with the remnants of tidal stripping that occurred as the proto-$\omega$-Centauri fell into the Galaxy \citep{bekki;freeman_2003}.

In the Aquarius stream discovery paper, \citet{williams;et-al_2011} attempted to exclude possible known progenitors for the Aquarius stream. On the basis of metallicity, distance, proper motions, transverse velocities and orbital energies, the authors were able to exclude all known Milky Way satellites with the notable exception of $\omega$-Centauri. Although the Aquarius stream metallicity distribution is not dissimilar from a known sub-population in $\omega$-Centauri, the individual chemical abundances are quite distinct. The strong $s$-process enhancement with overall metallicity is not observed in the rest of our sample. Thus, with the exception of {C222531-145437}, the Aquarius members do not have a chemistry that is synonymous with $\omega$-Centauri tidal debris. It will be most interesting to learn how many other members of the Aquarius stream are tidal remnants of $\omega$-Centauri, given the frequency of these objects is quite low \citep[e.g., see][]{da_costa;coleman_2008,majewski;et-al_2012}.

\subsection{Disrupted Disk/Halo Stars -- Signature of a Disk-Satellite Interaction?}

Since the Aquarius stream is kinematically coherent, it has been assumed that the moving group has been accreted onto the Milky Way from a tidally disrupted satellite.  The chemical abundances presented in this study do not favour an accretion scenario from a globular cluster or a dSph; there is conflicting evidence for either hypothesis. As it stands, the moving group  appears chemically indistinguishable from thick disk/halo stars. These results force us to consider other scenarios that may replicate the observations.

The Aquarius stream has an unusually wide intrinsic velocity distribution. Generally a stellar stream is considered kinematically `cold' when its velocity dispersion is $\lesssim 8$\,km s$^{-1}$. We find the velocity dispersion from five members to be $\sim$30\,km\,s$^{-1}$, consistent with \citet{williams;et-al_2011}. Hypotheses invoked to explain the Aquarius moving group must account for the high velocity dispersion.
 
There are other moving groups in the Milky Way that were initially considered as tidal tails from disrupted satellites but are no longer regarded as accretion events. We now list some examples. The Hercules moving group is significantly offset from the bulk of the velocity distribution observed in the field. Members of the Hercules group exhibit a wide range of metallicities and ages \citep{bensby;et-al_2007, bovy;hogg_2010}. Furthermore, Hercules group stars have [X/Fe] abundance ratios at a given [Fe/H] that are {\it not} substantially different from the thin or thick disk. The Hercules group kinematics are well replicated in simulations by stars in the outer disk resonating with the bar in the central region of the Milky Way \citep{dehnen_2000,fux_2001}, and strong predictions are made for disk velocity distributions that would lend further weight to this hypothesis \citep{bovy_2010}. Similarly, the Canis Major stellar over-density was also first considered to be an accretion feature from the postulated Canis Majoris dSph galaxy \citep{martin;et-al_2004}. However, \citet{momany;et-al_2004} demonstrate that the star counts, proper-motions, photometry and kinematics of the ``accreted feature'' can be easily explained by the warp and flare in the outer thick disk. The Monoceros ring \citep{newberg;et-al_2002,juric;et-al_2008} is perhaps another example of such an occurrence, as similar features naturally emerge as a consequence of galaxy-satellite interactions \citep{purcell;et-al_2011}, which has prompted considerable discussion \citep{lopez-corredoira;et-al_2012}. It is clear that not all kinematic groups are attributable to accretion events; in many scenarios a Galactic origin is more likely, and simpler.

We hypothesise that the Aquarius group is the result of displaced stars from a perturbation in the thick disk. That is, the stars are Galactic in origin but have been displaced by a disk-satellite interaction. Minor mergers can significantly disrupt the host galaxy \citep{villalobos;helmi_2008}, producing extended spatial and kinematic structure in the process. \citet{minchev;et-al_2009} proposed that such a perturbation would cause a Galactic ``ringing'' effect in the neighbourhood surrounding the merger site, analogous to the resulting compression wave propagating outwards from a stone falling in water. Stars move closer together in the wave peak, a signature which is observable in the velocities and orbital motions of nearby stars. This signature is most prominent in the $U$--$V$ velocity plane as concentric circles \citep{gomez;et-al_2012}, and dissolves over time (a few Gyr, depending on the mass of the perturber). After the $U$--$V$ velocity signature dissipates, a clear signature in angular momentum and orbital energy ($L_Z$, $E$) persists for long periods following the merger \citep[e.g., see][]{gomez;et-al_2012}.

Through Milky Way-Sagittarius simulations, \citet{purcell;et-al_2011} found that these disk-satellite interactions can explain ringing perturbations within the disk. Additionally, \citet{widrow;et-al_2012} and \citet{gomez;et-al_2012b} independently observed these phenomena -- a {\it``wavelike perturbation''}, as Widrow described -- in the SDSS and SEGUE catalogues. More recently, \citet{gomez;et-al_2013} proposed that these patterns were induced by the Sagittarius dSph interacting with the disk. Their simulations reproduce the observed north-south asymmetries and vertical wave-like structure, and show that the amplitude of these oscillations is strongly dependent on Galactocentric distance. Combined with the oscillating vertical motions with the $U$--$V$ velocity pattern, corrugated waves are observed as a result of the interaction.

The stars in these oscillations should exhibit a wide range of ages, metallicities and a large spread in velocity dispersion. Thus, resultant oscillations following a disk-satellite interaction can satisfactorily explain the existence of the Aquarius moving group. We do not observe a distinct coherence in the $U$--$V$ velocity plane in our data, but the angular momentum and orbital energies for Aquarius members qualitatively reproduces the theoretically predicted pattern by \citet{gomez;et-al_2012} in a retrograde direction. The extent and gradient of this $L_Z$--$E$ signature is dependent on the mass of the perturber and the time since infall. Although our sample size is minute -- and the sample size would still be small even if all Aquarius members had reliable orbits -- the fact that we see no $U$--$V$ velocity coherence (Figure \ref{fig:toombre}) is consistent with the observed $L_Z$--$E$ pattern: signatures in the $L_Z$--$E$ plane (Figure \ref{fig:orbits}) become more extended over time as the $U$--$V$ signature dissipates. This is consistent with a disk-satellite interaction occurring in the disk approximately a few Gyr ago.

The Aquarius moving group resides at a an intermediate latitude ($b \approx -55^\circ$) and with a radial distance of up to $\sim$5\,kpc for some stars, the stars are slightly out of the plane. This is not inconsistent with a disk-satellite interaction, as similar features in the Galactic field star population naturally emerge. \citet{gomez;et-al_2013} find that a significant fraction of the total energy goes into vertical perturbations. While the mean vertical distance $\langle{}Z\rangle$ in their simulations are near zero, this is an average of disk particles at all plane heights -- positive and negative -- and the dispersions around $\langle{}Z\rangle$ are very large (F. G{\'o}mez, private communication, 2013). Moreover, \citet{gomez;et-al_2013} were only able to reliably track particles up to $|Z| \approx 1.4$\,kpc due to a finite number of particles in each cell volume.

If the Aquarius group is a feature of a disk-satellite interaction, the perturber must have a mass on the order of a large globular cluster or a dSph satellite to produce the residual pattern in orbital energy and angular momenta. The Sagittarius dSph galaxy is an obvious candidate, but $\omega$-Centauri is also a possible perturber. On the basis on position, velocities, chemical abundances and orbit, we identify {C222531-145437} was highly likely stripped from $\omega$-Centauri in the past. Thus, it is plausible that $\omega$-Centauri has disrupted Galactic stars as it passed through the plane, adding to any other oscillating modes rippling through the disk, resulting in what we now observe as the Aquarius stream.

\section{Conclusions}
\label{sec:conclusions}
We have presented a detailed chemical and dynamical analysis for 5 members of the recently discovered Aquarius stream from data taken with the MIKE spectrograph on the Magellan Clay telescope. Hereafter we solely refer to the discovery as a moving group instead of a stellar stream, as we find no evidence that the group is a tidal tail of a disrupted satellite. The main conclusions are as follows:

\begin{itemize}
\item The Aquarius stream is not mono-metallic. A wide spread in metallicities is observed, with [Fe/H] ranging from $-0.63$ to $-1.58$ in just 5 members. The mean of the sample is [Fe/H] $= -1.20$ and the dispersion is $\sigma({\rm [Fe/H]}) = 0.33$\,dex.

\item No Na-O anti-correlation is observed in the Aquarius group. Two members have slightly enhanced levels of sodium with respect to iron. If the candidates were \textit{known globular cluster members}, they would be classified as belonging to either the primordial component, or at most, tenuous membership could be argued for the lower envelope of the intermediate group.

\item We find no evidence that the Aquarius group is the result of a disrupted classical globular cluster. The large [Fe/H] variation severely limits the number of possible parent hosts, and both the extreme and intermediate component of the Na-O anti-correlation have not been observed. A strong positive Mg-Al relationship is observed, reminiscent of Milky Way field stars. In total, high-resolution spectra exists for more than half of the stream.

\item The moving group shows an $\alpha$-enhancement of ${[\alpha/{\rm Fe}] = +0.40}$\,dex, similar to the Milky Way, and distinct to that typically observed in stars in dSph galaxies with comparable metallicities.

\item Aquarius members are enhanced in [Na/Fe] and [Ni/Fe] to levels typically observed in either the thick disk or globular clusters. These levels of [Na/Fe] and [Ni/Fe] enhancement are not observed in stars from dSph galaxies. Low [Ba/Y] abundance ratios are also observed in the Aquarius group, in conflict with chemistry of present day dSph galaxies. Thus, on the basis of [(Na, Ni, $\alpha$)/Fe] and [Ba/Y] abundance ratios, it is unlikely the Aquarius moving group is the result of a tidally disrupted dSph galaxy.

\item One of the candidates, {C222531-145437}, has an abundance pattern that is clearly distinct from the other Aquarius members, most notably in barium where {[Ba/Fe] = 0.68}. We exclude the possibility that the abundance variations have resulted from an AGB companion.

\item The position and velocity of {C222531-145437} agrees excellently where simulations by \citet{majewski;et-al_2012} predict large amounts of $\omega$-Centauri tidal debris, and the orbital energy and angular momenta are consistent with the $\omega$-Centauri cluster. The chemical and phase-space information suggests that {C222531-145437} is a rare tidal debris remnant from the globular cluster $\omega$-Centauri. Removing C222531-145437 from the Aquarius sample does not extinguish or diminish any of the aforementioned conclusions.

\item While no evidence exists for an accreted origin, and the Aquarius group members are indistinguishable from thick disk/halo stars, we hypothesise the moving group is the result from a disk-satellite interaction. We see no coherent pattern in the $U$--$V$ plane from Monte-Carlo simulations, but the orbital energies and angular momenta for the Aquarius group qualitatively reproduces patterns predicted by \citet{gomez;et-al_2012}. This is consistent with a minor merger in the Milky Way thick disk occurring perhaps up to a few gigayear ago. Given the location and velocity of the Aquarius group, and the identification of {C222531-145437} as a star tidally stripped from $\omega$-Centauri, it is plausible that the Milky Way-$\omega$Cen interaction sufficiently perturbed outer disk/halo stars to produce what we now observe as the Aquarius group.
\end{itemize}

It is clear that not all moving groups are tidal tails of disrupted satellites, and that the structure of the Milky Way is indeed complex. While we find no chemical evidence that the Aquarius group is a tidal tail from a disrupted satellite, we propose the members are Galactic in origin, and the group is a result of a disk-satellite interaction. Thus, although the Aquarius group has not been accreted onto the Galaxy, it certainly adds to the rich level of kinematic substructure within the Milky Way.

\section*{Acknowledgements}
We thank the anonymous referee for constructive comments which improved this paper. We are indebted to David Nidever and Steven Majewski for kindly replicating a plot from their $\omega$-Centauri simulations (Figure \ref{fig:omega-cen-tidal-debris} in this text) which helped to place {C222531-145437} in context of the cluster's tidal debris. A. R. C. acknowledges the financial support through the Australian Research Council Laureate Fellowship LF0992131, and from the Australian Prime Minister's Endeavour Award Research Fellowship, which facilitated his research at MIT. S.K. and G. Da C. acknowledge research support from the Australian Research Council through Discovery Project grant DP120101237, while A. A-B. acknowledges research support from the same source through Super Science Fellowship program FS110200016, and through FONDECYT Chile (3100013). A. F. acknowledges support from NSF grant AST-1255160. Australian access to the Magellan Telescopes was supported through the National Collaborative Research Infrastructure Strategy of the Australian Federal Government. This publication makes use of data products from the Two Micron All Sky Survey, which is a joint project of the University of Massachusetts and the Infrared Processing and Analysis Center/California Institute of Technology, funded by the National Aeronautics and Space Administration and the National Science Foundation. \\

%\facilities{Magellan:Clay}
\label{lastpage}

\newcommand{\actaa}{{Acta Astronomica}}
\newcommand{\aj}{{AJ}}
\newcommand{\araa}{{ARA\&A}}
\newcommand{\apj}{{ApJ}}
\newcommand{\apjl}{{ApJ}}
\newcommand{\apjs}{{ApJS}}
\newcommand{\ao}{{Appl.~Opt.}}
\newcommand{\apss}{{Ap\&SS}}
\newcommand{\aap}{{A\&A}}
\newcommand{\aapr}{{A\&A~Rev.}}
\newcommand{\aaps}{{A\&AS}}
\newcommand{\azh}{{AZh}}
\newcommand{\baas}{{BAAS}}
\newcommand{\jrasc}{{JRASC}}
\newcommand{\memras}{{MmRAS}}
\newcommand{\mnras}{{MNRAS}}
\newcommand{\pra}{{Phys.~Rev.~A}}
\newcommand{\prb}{{Phys.~Rev.~B}}
\newcommand{\prc}{{Phys.~Rev.~C}}
\newcommand{\prd}{{Phys.~Rev.~D}}
\newcommand{\pre}{{Phys.~Rev.~E}}
\newcommand{\prl}{{Phys.~Rev.~Lett.}}
\newcommand{\pasp}{{PASP}}
\newcommand{\pasj}{{PASJ}}
\newcommand{\qjras}{{QJRAS}}
\newcommand{\skytel}{{S\&T}}
\newcommand{\solphys}{{Sol.~Phys.}}
\newcommand{\sovast}{{Soviet~Ast.}}
\newcommand{\ssr}{{Space~Sci.~Rev.}}
\newcommand{\zap}{{ZAp}}
\newcommand{\nat}{{Nature}}
\newcommand{\iaucirc}{{IAU~Circ.}}
\newcommand{\aplett}{{Astrophys.~Lett.}}
\newcommand{\apspr}{{Astrophys.~Space~Phys.~Res.}}
\newcommand{\bain}{{Bull.~Astron.~Inst.~Netherlands}}
\newcommand{\fcp}{{Fund.~Cosmic~Phys.}}
\newcommand{\gca}{{Geochim.~Cosmochim.~Acta}}
\newcommand{\grl}{{Geophys.~Res.~Lett.}}
\newcommand{\jcp}{{J.~Chem.~Phys.}}
\newcommand{\jgr}{{J.~Geophys.~Res.}}
\newcommand{\jqsrt}{{J.~Quant.~Spec.~Radiat.~Transf.}}
\newcommand{\memsai}{{Mem.~Soc.~Astron.~Italiana}}
\newcommand{\nphysa}{{Nucl.~Phys.~A}}
\newcommand{\physrep}{{Phys.~Rep.}}
\newcommand{\physscr}{{Phys.~Scr}}
\newcommand{\planss}{{Planet.~Space~Sci.}}
\newcommand{\procspie}{{Proc.~SPIE}}
\newcommand{\ieeesigprocm}{{IEEE~Signal~Processing~Magazine}}
\newcommand{\icarus}{{Icarus}}

\end{document}